%% file: courseai2arxiv.tex
\documentclass[12pt]{article}

\usepackage[margin=1in]{geometry}
\usepackage{setspace}
\usepackage{amsmath}
\usepackage{amssymb}
\usepackage{graphicx}
\usepackage{booktabs}
\usepackage{float}
\usepackage{hyperref}
\hypersetup{
    colorlinks=true,
    linkcolor=black,
    citecolor=black,
    urlcolor=blue
}
\usepackage{natbib}
\usepackage{caption}
\usepackage{subcaption}
\usepackage{threeparttable}
\usepackage{placeins}%
\usepackage{rotating}
\usepackage{pdflscape}
\usepackage{lmodern}
\usepackage[T1]{fontenc}
\usepackage{microtype}
\usepackage{xcolor}
\usepackage{tabularray}
\usepackage{codehigh}
\usepackage[normalem]{ulem}
\UseTblrLibrary{booktabs}
\UseTblrLibrary{siunitx}
\usepackage{comment}
\usepackage{fancyhdr}
\usepackage{etoolbox}
\makeatletter
\fancypagestyle{lscapepage}{%
  \fancyhf{}%
  \fancyfoot[C]{%
    \setlength{\unitlength}{1pt}%
    \begin{picture}(0,0)%
      \put(\strip@pt\dimexpr.5\paperwidth-1in+\footskip\relax,%
           \strip@pt\dimexpr.5\paperheight-1in+\footskip\relax){%
        \rotatebox{90}{\makebox[0pt][c]{\thepage}}}%
    \end{picture}%
  }%
}
\makeatother
\AtEndEnvironment{landscape}{\thispagestyle{lscapepage}}

\NewTableCommand{\tinytableDefineColor}[3]{\definecolor{#1}{#2}{#3}}

\newcommand\blfootnote[1]{%
  \begingroup
  \renewcommand\thefootnote{}\footnote{#1}%
  \addtocounter{footnote}{-1}%
  \endgroup
}

\title{\onehalfspacing{
Generative AI Availability, Grades, and Student Satisfaction at a Large University
}}

\vspace{-60mm}
\author{
  James M. Zumel Dumlao\textsuperscript{1}, 
  Meng Wang\textsuperscript{2}, 
  Zhonghan Xie\textsuperscript{2}, 
  Junyao Hu\textsuperscript{3}, \and
  Ivan Bar\textsuperscript{1}, 
  George Chaney III\textsuperscript{1}, 
  Henry Gold\textsuperscript{1}, 
  Misha Teplitskiy\textsuperscript{1}\thanks{Address correspondence to: 
  \href{mailto:tepl@umich.edu}{tepl@umich.edu}.
  }
}

\date{October 14, 2026}
\begin{document}
\maketitle
\vspace{-1cm}

\begin{center}
\{\textsuperscript{1}School of Information,
\textsuperscript{2}Marsal Family School of Education,
\textsuperscript{3}Ford School of Public Policy\} 

at the University of Michigan
\end{center}
\begin{abstract}
\singlespacing{\noindent The availability of generative AI (GenAI) has raised concerns that college students substitute GenAI for cognitive effort, earning high grades without learning. If this ``GenAI substitution hypothesis'' is true, grades should rise disproportionately in GenAI-susceptible courses---those relying more on assessments like take-home problem sets and essays rather than in-class exams. Student satisfaction (self-reported understanding and interest) may also shift. We test this hypothesis using syllabus and administrative data from a large U.S. university (2016–2025; 138,386 students; 72,730 course offerings). We measure GenAI susceptibility by extracting assessment types from syllabi with a human-validated LLM pipeline. We then use a difference-in-differences design comparing courses before and after ChatGPT’s release, modeling COVID-19 effects as persistent or transient. Effects on grades and satisfaction are small and sensitive to COVID modeling choices, with no detectable differences by students’ prior academic preparation. Exploratory analyses show that the link between pre-college test scores and grades has not detectably weakened in more susceptible courses, and that instructors have been making courses less susceptible since before ChatGPT. Our findings temper concerns that GenAI inflates grades and reduces satisfaction, partly because instructors may adapt their pedagogy.}
\end{abstract}

\blfootnote{The authors thank Annaliese Paulson, Joshua Ashkinaze, members of the Computational Social Science Working Group and the Michigan Behavioral and Experimental Economics (BEE Lab) at the University of Michigan, and attendees of the 2026 International Conference on Computational Social Science and the Causal Inference in Higher Education Research Seminar (CIERS) for their invaluable feedback. This work would not have been possible without administrative support from Lisa Emery, Richard Richter, and the Undergraduate Research Opportunity Program at the University of Michigan.}

\newpage

\section{Introduction}
\label{sec:intro}

    Generative artificial intelligence (GenAI) is rapidly transforming higher education institutions. Many students have adopted GenAI tools to help with studying and assessments. One 2023 study found that over half of students were regularly using GenAI technology \citep{coffeyStudentsOutrunningFaculty2023}, and more recent estimates find 42\% of students use GenAI weekly \citep{mowreader_report_2025}. While it is accepted that GenAI use is rampant, its effects on students' learning and satisfaction in higher education courses are not yet well-established.

Although GenAI could complement learning and augment human performance \citep{barkhuff_situated_2026, lau_barriers_2026}, a key concern is that students are offloading the cognitive effort necessary for learning (see  \citet{gerlichAIToolsSociety2025} and \citet{liuAIAssistanceReduces2026} on ``cognitive offloading'').  GenAI tools are capable of producing outputs that often outperform, and are difficult to distinguish from, human work \citep{noyZhangExperimentalEvidence2023, dellacqua_navigating_2026, finnie-ansley_robots_2022}. There is widespread worry that students may substitute their own effort with these tools to receive high grades on ``GenAI-susceptible'' assessments, like homework and essays \citep{shirkyOpinionStudentsHate2025}. We call this concern the ``GenAI substitution hypothesis'' for higher education. Assessments largely determine final grades, which signal students' ability to themselves, future instructors, and employers \citep{barGradeInformationGrade2009}. Grades on assessments are also important for instructors to identify gaps in understanding and improve courses. Consequently, if this GenAI substitution hypothesis is true, grades could lose their signaling value, and all of these outcomes might be undermined. 

    Additionally, GenAI use for coursework may erode students' satisfaction with academic subjects through two mechanisms. First, GenAI may alter students’ engagement with assessments. Because assessment structure influences students’ motivation and participation \citep{solomonTwoPointSystemMethod1979, azorlosaEffectAnnouncedQuizzes2011}, reduced engagement on assessments could weaken students' reduced interest in subjects and understanding of course material. Second, GenAI may reshape students’ social engagement by replacing traditional help-seeking behaviors. Recent studies suggest that students increasingly turn to GenAI rather than peers or instructors for academic support \citep{hou_all_2025, hou_effects_2024, gu2026policeguidehighereducation}, potentially weakening learning communities. The resulting social isolation could have negative consequences on student satisfaction, as was similarly documented during the COVID-19 pandemic \citep{birdNegativeImpactsShift2022}.

We contribute new evidence to this debate using administrative data from a large university in the United States. We combine syllabus, student academic records, and course evaluation data over the fall and winter terms from fall 2016 through fall 2025. The resulting panel comprises 138,386 unique students and 72,730 offerings across 6,357 unique courses. We define a course's GenAI susceptibility as its final grade weighting of assignments that can be completed with GenAI tools and out of the instructor's view, such as open-notes exams, take-home problem sets and essays, and participation quizzes. In contrast, in-class closed-book exams, presentations, and live skills performances are less susceptible to GenAI because they require live human demonstration of mastery. We develop an LLM-based pipeline for reconstructing assignment types and their weighting in the final grade from syllabi, and found high accuracy when validated against 525 human consensus labels. We then deploy this pipeline on 36,357 syllabi and calculate the proportion of the final grade determined by GenAI susceptible assignments in each course offering.
    
    To identify the effects of GenAI availability, we take a difference-in-differences approach, comparing outcomes before and after ChatGPT's introduction in November 2022 in courses with higher and lower GenAI susceptibility. If the GenAI substitution hypothesis is true, grades would rise disproportionately in courses that weight susceptible assessments more heavily. If it is false, grades would rise similarly across courses regardless of susceptibility, or not rise at all. To account for common trends, like overall grade inflation or changes to student composition, we use course, semester, and, where possible, student fixed effects.

    Importantly, the COVID-19 pandemic greatly disrupted higher education \citep{asgeirsdottirHowHomeExams2026, birdNegativeImpactsShift2022}, complicating the analysis. In 2020 and 2021, most courses shifted to online lectures, more susceptible assessment structures, and likely more lenient grading standards. The return to in-person, less susceptible courses was uneven leading up to ChatGPT's release, leaving no clear separation between grades under COVID-19 and grades after GenAI availability. To account for this, we define 2020–2022 as the COVID-19 disruption window. We then measure the effects of GenAI availability separately from COVID effects under two assumptions, that COVID effects were either fully transient or fully persistent. Because pandemic effects plausibly lingered at the time of ChatGPT's release, the estimates under each assumption act as bounds on the true effect. We do not privilege one assumption over the other in our interpretation of the results, drawing conclusions only when the sign and statistical significance of estimates align.

    We find no robust effect of the introduction of ChatGPT on final grades in more susceptible courses relative to less susceptible ones. The grade estimate is statistically indistinguishable from zero when COVID effects are treated as transient and positive but small when they are treated as persistent. Withdrawal and failure rates show no significant change under either assumption. Continuous difference-in-differences analyses (see \citet{callawayDifferenceinDifferencesContinuousTreatment2024}) show non-positive and non-significant grade effects comparing every level of course susceptibility to non-susceptible courses. These findings align with robustness analyses simulating violations of the parallel trends assumption (see \citet{rambachanMoreCredibleApproach2023}).
    
   Null average grade effects could mask underlying heterogeneous effects, so we also estimate GenAI availability's effects within grade distributions (i.e., share of A's versus share of D's) and between students based on prior academic performance measured six ways. GenAI did not increase grades at the bottom of the distribution. We do find an increase in the share of A's and B's in more susceptible courses after ChatGPT, but these results are not robust across COVID recovery assumptions. We also find that grade effects are flat across prior academic performance terciles, robust across our various proxies and between the COVID assumptions. Together, we conclude that our non-robust average grade effects do not mask heterogeneity in grade effects by contemporaneous or previous performance. Although absence of evidence should not be conflated with evidence of absence, we argue that our estimated effect sizes are also economically small. We are able to rule out average grade effects larger than 0.11 grade points on a 4.0 scale, less than half of a letter-grade step (e.g., from B- to B). 

    What might explain the lack of robust evidence of grade disruption following GenAI availability? We conduct two exploratory analyses to give suggestive evidence and gesture toward future inquiry. First, we examine instructors' role in preserving grade distributions. Comparing syllabi of the same courses across semesters, we find that courses are returning to pre-COVID assessment structures after GenAI susceptibility peaked during the COVID-affected era. Although median susceptibility post-ChatGPT is still higher than pre-COVID, this endogenous response (i.e., making classes less susceptible) is one way instructors could be preserving grade signals. Second, we examine the possibility that, while distributions are mostly preserved, where students fall on the distribution may have changed. Within each course, we regress grades on students' pre-college standardized test scores and ask whether, after ChatGPT's release, that relationship has weakened in more susceptible courses. We find no detectable change, suggesting that grades in susceptible courses still sort students by earlier performance much as they did before. However, these analyses are suggestive rather than conclusive and do not rule out other ways in which GenAI may affect student learning.

    The estimated effects of GenAI availability on student satisfaction-related course evaluations depend on how COVID is modeled. We find null effects on self-reported understanding in either scenario, and changes in self-reported interest and relative workload after ChatGPT are not distinguishable from pandemic-era changes. In any case, the non-negative results for understanding and interest are evidence against the erosion of student satisfaction after GenAI became available.

    This study makes four contributions to the emerging literature on GenAI in higher education. First, we contribute to research on the effects of GenAI availability on student performance. Two prior studies have approached the GenAI substitution hypothesis: \citet{hausmanGenerativeAIsImpact} reports data from an Israeli university, while \citet{chirikovAIGradeInflation2026} uses data from a public university in Texas. Both find evidence of GenAI substitution, with grades increasing in more susceptible courses after the release of ChatGPT. Placed on a common scale, our estimates are generally smaller than theirs but mostly not statistically distinguishable from them. The three studies agree more on magnitudes than on interpretation. We argue that effects of this size, a fraction of the step between adjacent letter grades, do not yet amount to the erosion of grade signals that the prior studies emphasize. Second, the more established literature on GenAI's productivity effects in the workplace has examined differential impacts for workers of low and high previous performance, finding evidence of equalizing in some contexts and widening in others \citep{brynjolfsson_generative_2025, noyZhangExperimentalEvidence2023, otis_uneven_2026}. We extend this question to higher education and provide the first analysis of grade effects by students' prior academic performance, finding neither pattern. Third, while prior research has considered the role of instructors in adapting to GenAI availability \citep{hausmanGenerativeAIsImpact}, we provide early descriptive, behavioral evidence of how assessment structures have changed in recent years. Fourth, we examine for the first time the effects of GenAI availability on student satisfaction at a university-wide level, measured through course evaluations. While most prior work on GenAI's effects has focused on performance, student experiences provide a complementary perspective on the implications of GenAI availability and have potential downstream consequences for students' course selection, major choice, and career trajectories. 

\section{Background and Related Literature}
\label{sec:lit}

First, this section develops the ``GenAI substitution hypothesis,'' which relates the availability of GenAI to student outcomes under different assessment structures and considers heterogeneous performance effects. Second, we give an overview of the empirical literature on GenAI's performance effects in the workplace. Finally, we review the emerging empirical literature on GenAI's performance effects in the classroom, concluding with a close look at previous studies most similar to ours.

\subsection{GenAI Substitution Hypothesis}
\label{sec:lit-assessments}
In both popular and academic conversations on GenAI in higher education, many have expressed concerns that students are increasingly relying on GenAI tools to study and produce work, which is harming learning and making it more difficult to measure learning \citep{whitfordBrownProfessorSuspects, shirkyOpinionStudentsHate2025, hausmanGenerativeAIsImpact, chirikovAIGradeInflation2026, gu2026policeguidehighereducation}. We label this concern the ``GenAI substitution hypothesis,'' which posits that students receive higher grades with access to GenAI tools, relative to when they do not have access, because they may substitute away some of their own effort using GenAI. Here, we consider the mechanisms driving this hypothesis and its empirical signature. 

Early widely-available GenAI tools were capable of producing outputs of decent quality and difficult to distinguish from human outputs, especially writing and code \citep{Kadoma2025-tm, finnie-ansley_robots_2022}. GenAI capabilities have grown rapidly in recent years. These tools can now manage projects, perform complex reasoning, and draw on vast informational resources from its training data and web search. Arguably, GenAI tools are now capable of producing convincing and satisfying outputs for most assessments.

Still, conversations about the GenAI substitution hypothesis implicitly acknowledge that some types of assessments are more susceptible to GenAI use than others. For example, \citet{shirkyOpinionStudentsHate2025} and \citet{whitfordBrownProfessorSuspects} both discuss closed-book, invigilated exams as GenAI-proof. As GenAI capabilities have developed, the remaining difference between more and less ``GenAI-susceptible'' assessments is in their modality.  Instructors may explicitly restrict GenAI use or require demonstration of mastery without access to GenAI tools. Oral and closed-book exams disallow the use of many external resources, GenAI tools included. Presentations and live skills demonstrations require the student to internalize information before performing the task in real-time. Thus, we consider in-person closed-book exams, presentations, and live demonstrations non-susceptible assessments. Susceptible assessments, on the other hand, are those where GenAI tools may directly substitute for student ability, both when students are permitted to use GenAI or when use is not monitored. These include take-home exams, homework and problem sets, projects, and essays. While many such assessments are done asynchronously, it is also possible for synchronous assessments to be susceptible (e.g., in-class open-notes quiz on laptop). 

While conversations about the role of GenAI education discuss both learning and grades, the real effects of GenAI on learning are unclear and evidence usually centers on grade performance. Learning is not a perfectly observable process. Assessments and grades act as imperfect measures \citep{elliot_approach_1999, LINNENBRINK2000195, rosovskyEvaluationAcademyAre2002}. But this distinction between susceptible and non-susceptible assessments motivates an empirical signature for the GenAI substitution hypothesis. If GenAI tools are used to develop students' mastery of a subject (i.e., as a complement), then effects on grades should be similar in courses regardless of whether they rely on susceptible or non-susceptible assessments. However, full substitution of students' effort and skill is only possible for susceptible assessments. So, if substitution is happening, then we may observe an increase in grades in courses relying more on susceptible assessments relative to other courses. Our study tests for the latter scenario, seeking to measure performance increases in excess of learning benefits from GenAI use.

\subsubsection{Heterogeneous Grade Effects of GenAI}
Whether or not GenAI has had effects on average grades, the GenAI substitution hypothesis could be true for subsets of students. Most empirical work on the economic consequences of GenAI has been set in the workplace. Although studies of GenAI in the workplace consistently find average productivity gains \citep{dellacqua_navigating_2026}, effects differed based on workers' previous performance. Among professional workers, introducing GenAI had an equalizing effect, helping previously worse performers most \citep{brynjolfsson_generative_2025, noyZhangExperimentalEvidence2023, cui_effects_2026}. However, in an experiment involving entrepreneurs in Kenya, \citet{otis_uneven_2026} find an opposite dynamic: Previously high performers did better with GenAI, while previously low performers did worse. 

This question of heterogeneous performance effects may be extended to higher education. In college computer science courses, \citet{prather_widening_2024} find small-scale, mixed-methods evidence that the benefits of GenAI use may depend on previous metacognitive ability. Their findings align with a widening, rather than an equalizing, effect. \citet{lehmannCorneliusStingAIMeetsClassroom2024} corroborate a widening effect in their lab experiments. However, university-scale research on whether any GenAI performance increases differ by previous student performance is lacking.

\subsubsection{Could GenAI Impact Student Satisfaction?}
Beyond performance-based metrics like grades, an additional concern is that GenAI may weaken students' satisfaction with their education. To our knowledge, there are no quantitative observational studies examining how course GenAI-susceptibility has affected students' satisfaction. While qualitative data can provide a proof of existence for this relationship, quantitative data can help distinguish idiosyncratic effects from systematic ones.

We posit that, if GenAI availability affects satisfaction, then we will observe a change in course evaluations related to students' self-reported subject understanding, interest, and relative workload in more susceptible courses. This might happen in two ways. First, the aforementioned cognitive offloading \citep{gerlichAIToolsSociety2025, liuAIAssistanceReduces2026} that occurs when students substitute their abilities with GenAI capabilities might result in students feeling like they have less of an understanding or interest in the subject, and that the GenAI-susceptible coursework has a relatively lighter workload than less-susceptible coursework. Second, students may turn to GenAI tools rather than their peers or instructors. This weakening of the social fabric of the classroom is documented in qualitative studies \citep{hou_all_2025, hou_effects_2024, gu2026policeguidehighereducation}. During the COVID-19 pandemic, virtual learning and the increased friction of accessing social resources was associated with feelings of isolation and poorer retention \citep{birdNegativeImpactsShift2022}. This change in help-seeking behavior may increase feelings of understanding, while decreasing subject interest fostered in community with other learners. Prior studies show that having regular objectives to complete can enhance student motivation \citep{elliot_approach_1999, liuMeasuringLearningOutcomes2012} and class participation \citep{solomonTwoPointSystemMethod1979, azorlosaEffectAnnouncedQuizzes2011}. However, little work exists on the consequences of obviating assessments with access to GenAI.

\subsection{Empirical Literature}
\subsubsection{Small-scale Experiments}
The education setting raises a tension that is less emphasized in the workplace. Education aims to build human capital, not just to produce outputs. A small experimental literature speaks directly to this tension (see \citet{chenEffectGenerativeArtificial2025} for a review). AI tutors designed to scaffold problem solving show learning gains on later unassisted tasks \citep{nieGPTSurpriseCoding2024, bastaniGenerativeAIGuardrails2025}, while access to GenAI tools during graded work tends to raise performance while available and lower it once removed, sometimes below the level of those who never had access \citep{lehmannCorneliusStingAIMeetsClassroom2024, darvishiImpactAIStudentAgency2024, bastaniGenerativeAIGuardrails2025}. Qualitative and journalistic work documents a pattern of rapid, heterogeneous adoption \citep{ellisAITeachingNext2025, coffeyStudentsOutrunningFaculty2023, ammariHowStudentsReally2025}, further complicating predictions about the effects of GenAI on the educational system. 

Such experiments provide a valuable but partial picture of how GenAI is transforming higher education. They mostly measure effects of GenAI usage, constraining whether and how students use GenAI, rather than observing effects of availability, in which students can choose to use the tools and potentially adapt usage for their particular learning situation. Thus, these experiments measure effects of GenAI use in contexts not necessarily relevant for the future of higher education and the economy. Quasi-experimental evidence based on administrative data complements this literature by examining how academic outcomes change in real-world educational settings. 

\subsubsection{Large-scale Observational Studies of Grades}

Quasi-experimental evidence at a large-scale remains scarce, but a small literature has begun to form, focusing thus far on average grade effects and distributional grade effects (i.e., whether there are more A grades or fewer failing within courses).%
\citet{hausmanGenerativeAIsImpact} use administrative records for roughly 36,000 students at an Israeli university. They define a course as ``AI-compatible'' if most of the final grade comes from take-home exams and assignments, and they compare within-student performance across AI-compatible and incompatible courses before and after ChatGPT. They estimate that AI availability raises average grades in AI-compatible courses by roughly 0.6 to 1.3 points on a 100-point scale. Looking at grade distributions, they find that the largest grade gains were concentrated near the 25th percentile, corresponding to a compression of the grade distribution and a reduction in failure rates. \citet{hausmanGenerativeAIsImpact} compute AI-compatibility at the course offering level rather than the course level and do not include course fixed effects, so a course can switch treatment groups from year to year in their primary analysis. Their estimates become statistically insignificant in the most restrictive matching models where only courses with consistent treatment status are included. Treating AI compatibility as continuous rather than binary also yields positive estimates of similar magnitude when allowing courses to change in intensity over time, but null estimates if intensity is anchored to a pre-GenAI year. Therefore, the statistical and economic significance of their estimates depend on modeling choices. Their strictest analyses seem not to support their conclusion that GenAI has resulted in grade inflation.

\citet{chirikovAIGradeInflation2026} uses data from a large public university in the United States, a setting closer to ours. Instead of measuring assessment format, he measures the share of a course's required tasks that involve writing and coding, extracted from pre-ChatGPT syllabi. He estimates that GPA rose by 0.12 points on the 4-point scale per unit of exposure, which is the change for a course moving from no writing or coding tasks to only such tasks. Regarding grade distributions, \citet{chirikovAIGradeInflation2026} finds that the share of A grades rose by about 13 percentage points and there was no change in the share of C or D/F grades. Hence, prior work diverges on whether performance gains were experienced most at the top or bottom of the grade distribution. \citet{chirikovAIGradeInflation2026} anchors his exposure measure to a pre-ChatGPT (Fall 2022) version of each course and restricts analysis to fall offerings with at least 20 students that appear in every year of his 2018-2025 panel, which removes complicated seminar courses and accounts for cyclicality in offerings. These choices yield cleaner comparisons but exclude graduate courses, those offered in the winter or every other year, and small seminars, which are perhaps more common in humanities departments. His task-exposure measure is validated through spot checks \citep{chirikovHowInstructorsRegulate2026}, while our assessment structure susceptibility measure is validated against a human-coded sample. 

Both studies read their results as evidence of GenAI substituting for student effort on unsupervised work, raising measured grades without an analogous gain in underlying skill. However, each study is of just one institution, so generalizability may be limited due to differences in setting, including geographically specific effects of the COVID-19 pandemic. This warrants a closer look at the GenAI substitution hypothesis and whether it generalizes across university settings. Additionally, neither study examines heterogeneity of GenAI effects by prior academic preparation or effects on student satisfaction. 

\section{Data and Measurement}
\label{sec:data}

The setting of this study is a large flagship public university in the Midwest United States (``University'') from 2015 to 2025. The University shares characteristics with both elite private universities and less-prestigious state universities, making it an informative setting for understanding GenAI impacts on U.S. higher education. Throughout the paper, ``year'' refers to calendar years, ``term'' to academic terms (fall, winter, and the smaller spring/summer terms), ``course'' to unique subject-catalog-number combinations (e.g., ``ECON 101''), and ``offering'' to individual sections of a course, which may have different instructors and syllabus policies. We combine three data sources, described below, since no single database in our setting contains all of the information needed.

\subsection{University Course Syllabi and LLM Annotations}
\label{sec:data-syllabi}

Course syllabi codify course design and policies and have been described as ``roadmaps'' and ``contracts'' \citep{tongWhatAreWe2025}. We use offering-level syllabi to measure course susceptibility to student GenAI use (see Section~\ref{sec:methods}). In 2013, one of the colleges in the University established an online archive of course syllabi, hosted on a closed platform with single sign-on. Instructors voluntarily upload syllabi, with some retroactive uploads going back to 2000.  Although some of the highest-enrollment programs are not represented (e.g., business administration, computer science, nursing), the archive also includes records from other colleges and departments.

We scraped metadata and PDF files for 44,876 syllabi, the entire archive as of February 2026, and applied an LLM-based annotation pipeline to parse the corpus at scale (Section~\ref{sec:methods-llm}). 36,357 of these syllabi are for offerings between 2015 and 2025. Six syllabi involved sensitive content that raised model guardrails and were excluded from the analytic sample.

\subsection{Student Demographics and Academic Career Data}
\label{sec:data-admin}

We combine syllabus data with deidentified administrative records on student demographics and academic careers, including course enrollments and grades. The data are at the term-offering-student level from as early as 1986, with more even coverage beginning in the late 1990's. Final grade data is displayed as a letter grade rather than as a percentage, and a standard grade point scale is associated with each letter grade (see Appendix \ref{app:grade-pts}).

\subsection{Course Evaluation Data}
\label{sec:data-evals}

Near the end of each term, students are invited to voluntarily evaluate their courses. The standard course evaluation survey includes required five-point Likert-scale questions and additional instructor-selected questions, each with open-ended text boxes. Starting in 2016, each evaluation included eight standard questions regarding instructor efficacy and course experiences. From 2008 to 2016, similar questions were included with different wording. Before 2008, evaluations were conducted on paper and are not digitally accessible.

We obtained aggregated course evaluation records from 2008 to 2025 using metadata from the syllabus archive. This search yielded 43,025 evaluations at the offering-instructor level. In many unmatched cases, particularly at the graduate level, departments did not request evaluations. This data includes median scores and counts of each five-point Likert response for each of the eight standard questions. We focus on the questions relating to self-reported outcomes related to student learning, interest in subject, and relative workload (see Appendix \ref{app:eval-qs} for question wordings). For consistency in question wording, we only include evaluation data from fall 2016 and on in the analysis presented here.

\subsection{Analytic Sample}
\label{sec:data-sample}
To maximize coverage across the three administrative datasets and comparability between courses over the years, we restrict analysis to fall and winter terms. The assembled panel spans fall 2016 through fall 2025 and includes 138,386 unique students and 72,730 offerings across 6,357 unique courses. 

Under our difference-in-differences approach with course and semester fixed effects, only courses that appear in both the pre- and post-ChatGPT periods add information to regression results. Finally, since we anchor course GenAI susceptibility to a course's 2019 offering (Section~\ref{sec:measure-exposure}), courses without a 2019 syllabus are excluded from the main regressions.

Seminar courses often have different topics, instructors, and assessment structures, but share a course code (e.g., ENGLISH 710). To keep them in the sample, we identified seminar courses via their titles using a set of keywords like ``seminar'' and ``special topics.'' For these courses, we include the lead instructor's name alongside the course code and treat each combination as a persistent course over time. 

The resulting balanced analytic sample comprises 1,220,330 student-offering observations, with 122,663 unique students and 38,758 offerings across 1,391 courses. Further descriptives of this data are in Appendix~\ref{app:descriptives}. While these restrictions help with internal validity, they also make the analytic sample unrepresentative of the full course catalog. This could have consequences for the generalizability of our results. Appendix~\ref{app:archive} compares the analytic sample with every graded course in the University's administrative records over the same window.

\section{Empirical Strategy}
\label{sec:methods}

Our empirical strategy exploits the public release of OpenAI's ChatGPT in November 2022. This aligns with previous studies on the same question \citep{hausmanGenerativeAIsImpact, chirikovAIGradeInflation2026}. While GenAI tools were available prior to ChatGPT, the release represents an unexpected and widespread shock. Crucially, courses varied in their susceptibility to GenAI use for direct substitution of student ability in a plausibly exogenous way before ChatGPT. 

Taking a difference-in-differences approach, we compare outcomes in more and less GenAI susceptible courses before and after the Fall 2022 semester. It should be noted that there have been shifts in the technologies' capabilities, students' adoption \citep{coffeyStudentsOutrunningFaculty2023}, and instructors' policies and grade structures \citep{chirikovHowInstructorsRegulate2026} since the release of ChatGPT. Each of these are potential mechanism modifying the effects of course susceptibility on outcomes post-ChatGPT. Below we describe the measures and methods underlying our empirical approach.

\subsection{Measures}
\label{sec:measures}

\subsubsection{Treatment: Course GenAI Susceptibility}
\label{sec:measure-exposure}

We define a course's GenAI susceptibility as the proportion of the
final grade allocated to susceptible assessments. Susceptible assessments are operationally defined as those completed independently from instructor observation: open (-notes or -book)/take-home exams (``\textit{open\_exam}''), homework, papers, and projects. Such assessments could in principle be fully automated using GenAI. Whereas, non-susceptible assessments are those with an in-person, proctored, or real-time component: in-class closed (-notes and -book) exams, presentations, and live
skill demonstrations (``\textit{performance}''). Although students could use GenAI in preparation for these assessments, the real-time nature of those assessments precludes full automation. In cases where it is not explicit whether exams were open or closed, we categorize the weighting as an unknown exam. Unknown exams are assumed non-susceptible for our analyses, as that is the traditional default. For projects with a presentation component, unless the weighting of each component is specified, we split the final grade weight equally between susceptible and non-susceptible. In summary, we calculate GenAI susceptibility as $1 - (closed/unknown\_exam + presentation + performance + 0.5*project\_present)$, relying more on the measuring of non-susceptible assessments due to greater accuracy in those categories (see Section \ref{sec:methods-llm} and Appendix \ref{app:llm-pipeline}).

For the main specifications, we anchor course GenAI susceptibility to the course's 2019 offering (\texttt{Susceptibility}), taking an average across sections if multiple exist. Because instructors can change assessment weights after the shock, a time-varying treatment measure would mechanically absorb part of the instructor response into the treatment definition, whereas a fixed anchor preserves that response as an object of study. Reclassification is not hypothetical here. Under a contemporaneous classification, courses graded almost entirely on proctored work lose about two-thirds of their observations per semester after 2022, as courses shift toward unproctored assessment. We use 2019 specifically, rather than the last pre-AI year 2022, for two reasons. First, 2019 is the last clean pre-COVID year. Anchoring to 2022 would fold pandemic-era assessment changes into the treatment, and we show in Section~\ref{sec:results-grades} and Appendix~\ref{app:robustness-2019} that the COVID-affected anchors inflate the transient-COVID grade estimate. Second, we argue that a course's susceptibility is relatively persistent, so the 2019 value is representative of the course's stable assessment design. We document year-to-year persistence in course susceptibility in Appendix Table~\ref{tab:async_persistence}. The mean correlation of courses' susceptibility in every other year with that of 2019 is 0.73. Using 2019 as both the treatment anchor and the event-study reference keeps the two choices internally consistent and the pre-period uncontaminated.

GenAI susceptibility is continuous, ranging from 0 to 1, in concept. In practice the variable concentrates around multiples of five (e.g., 55\%, 90\%). We use the continuous measure throughout rather than comparing binarized treatment and control groups, including in the event studies, where the coefficient of interest is the per-unit effect of $\text{Susceptibility}_c$ at each semester relative to the reference. A binary design would also be thin in our setting. Only 14 courses in the balanced sample graded less than 10\% of their 2019 final grade on unproctored work, the threshold \citet{hausmanGenerativeAIsImpact} use to define AI-incompatible courses.

\subsubsection{Outcomes: Student Performance and Satisfaction}
\label{sec:measure-student}

Student outcomes are measured at the term-offering-student level. We examine average final grade value (\texttt{GPA}), withdrawal rate (\texttt{withdraw}), and failure rate (\texttt{fail}). Final grades range from A+ to E with corresponding values of 4.0 and 0.0, respectively (see Appendix \ref{app:grade-pts}). Withdrawals carry no grade points and are excluded from the grade outcome, while incompletes are scored as 0.0. Per university policy, students receive course credit for grades of D- (0.7 value) or higher. Therefore, we define failure as not receiving a D- grade or higher. For grade distribution analyses, we construct grade threshold measures similar to those used by \citet{chirikovAIGradeInflation2026}. We use binary indicators for grades of (\texttt{at\_least\_A}), (\texttt{at\_least\_B}), (\texttt{at\_least\_C}), and (\texttt{at\_least\_D}), and for the complement of the last (\texttt{not\_at\_least\_D}). As in \citet{chirikovAIGradeInflation2026}, the thresholds use every enrollment as the denominator, so a withdrawal or non-letter grade counts as not clearing any threshold. An increase in the likelihood of (\texttt{at\_least\_A}) would indicate grade inflation at the top-end of the distribution, while an increase in the likelihood of (\texttt{at\_least\_D}) would indicate fewer grades at the bottom of the distribution.

We proxy student satisfaction with course evaluations. Course evaluation outcomes are median scores measured at the term-offering level: student self-reported understanding of the subject matter after taking the course (\texttt{understand}), interest in the subject after taking the course (\texttt{interest}), and perceived workload relative to their other courses (\texttt{workload}). For the relative workload question, the original Likert scale associated higher values with a lower relative workload. We reverse-coded this variable in regressions for interpretability; lower estimates represent lower relative workload ratings. 

We interpret the understanding and interest course evaluations as related to student satisfaction with the course. Relative workload acts as a sanity check on claims of student disengagement with assessments, comparing evaluations between courses of similar assessment structures before and after the introduction of ChatGPT. 

\subsubsection{Mediator: Prior Academic Preparation}
\label{sec:measure-abilityrank}

For one heterogeneity analysis we measure students' prior academic preparation with a course-residualized, within-cohort first-term GPA rank (\texttt{AbilityRank}). For simplicity, we exclude graduate students and a small minority of undergraduates who did not enter in a fall semester for this measure only.

We build this measure in three steps. First, to account for regular grade inflation trends and changes in student composition, we rank each student against their own entering cohort rather than pooling cutoffs across the panel. First-year and transfer cohorts are measured separately. Second, to disentangle rank and selection into courses, we residualize each first-term grade against its course-by-term mean before averaging and ranking. This strips the two channels through which a naive rank correlates with GenAI susceptibility: Students who select into leniently graded or susceptible courses early would otherwise rank higher, and field or course difficulty correlates with susceptibility. Residualizing compares a student against coursemates in the same offering, so the rank reflects relative performance rather than course choice. Third, because the proxy is a pre-treatment ability measure, we restrict the heterogeneity sample to cohorts that entered before the public release of ChatGPT, for whom first-term performance is observed pre-treatment. The raw within-cohort rank is retained as an interpretable robustness comparison and yields the same null results (Appendix~\ref{app:ability-robustness}).

One cohort requires special handling. The Fall 2020 entering cohort's first-term GPA is distorted due to changes in assessment modality and a university-wide grading policy that records low grades as ``No Record COVID'' by default. Both forces push genuinely lower-performing Fall 2020 entrants toward the middle of their cohort's rank. Because the 2020 to 2022 cohorts supply most of the post-AI observation mass, the resulting misclassification would concentrate in the post period. \texttt{AbilityRank} uses the official first-term GPA, which omits these pass/no-record grades. We confirm in Appendix~\ref{app:ability-robustness} that grade effects remain flat across terciles when we drop the Fall 2020 cohort, when we recompute first-term GPA from the instructor letter grades recorded for pass/no-record elections, and when we use an external high school GPA or standardized test-score rank.

\subsection{LLM Annotation Pipeline}
\label{sec:methods-llm}

Syllabi are unstructured and highly heterogeneous documents, and large language models (LLMs) are well suited for reconstructing structured labels from natural language at scale. We use the term ``reconstruction'' rather than ``extraction'' because LLMs do not directly extract information, but rather generate labels that are likely to match the document.  

Our pipeline converts each syllabus PDF to Markdown using \texttt{pymupdf4llm}, then reconstructs the final grade weighting for each assessment category via GPT-5.1 through a university-managed API gateway approved for moderately sensitive data. The normalized reconstructed weights feed the continuous treatment variable (\texttt{Susceptibility}). We validate the pipeline against a stratified human-annotated sample of 525 syllabi (two trained coders per syllabus with consensus resolution) and find strong recovery. \texttt{Susceptibility} is reconstructed with mean absolute error of 0.063 on the 0--1 scale. Appendix~\ref{app:llm-pipeline} reports the validation results by assessment category. 

\subsection{Model Specifications}
\label{sec:methods-regression}

We estimate three progressive specifications that build from a simple ordinary least-squared (OLS) baseline to our main two-way fixed effects (TWFE) design with an explicit COVID-window adjustment. We use linear models throughout for simplicity. Some outcomes are bounded discrete variables, but the continuous treatment is also bounded in $[0,1]$, so linear regression remains a sensible approximation of the average treatment effect within the observed data range. Standard errors are clustered at the course level. $\text{Susceptibility}_c$ is measured once per course and treatment timing varies only with calendar year, so course-level clustering is the natural level and accommodates arbitrary within-course serial correlation across years and sections.

\paragraph{From OLS to DiD.}

The OLS baseline regresses each outcome on our measure of course GenAI susceptibility ($\text{Susceptibility}_c$), a post-AI indicator $\text{Post}^{\text{AI}}_t$, and their interaction. OLS gives raw associations but is subject to selection bias since high- and low-susceptibility courses differ in many time-invariant ways (see Table \ref{tab:balance}). The two-way fixed effects (TWFE) specification adds course fixed effects ($\delta_c$) to absorb time-invariant course characteristics and semester fixed effects ($\tau_t$) to absorb common shocks, including university-wide grade inflation trends. For student-level outcomes, we further include student fixed effects ($\gamma_s$). The main effects of $\text{Susceptibility}_c$ and $\text{Post}^{\text{AI}}_t$ are collinear with the fixed effects and are dropped; the coefficient on $\text{Susceptibility}_c \times \text{Post}^{\text{AI}}_t$ is the difference-in-differences (DiD) estimator, identified within courses (and within students for student-level outcomes) over time.

\paragraph{Modeling the COVID disruption.}
\label{sec:covid-modeling}

The COVID-19 pandemic represents a significant disruption to higher education that overlaps the pre-period of our DiD design. From early 2020 through mid-2022, courses moved online, assessment weighting shifted, and recovery to pre-pandemic conditions was slow and incomplete for reasons unrelated to GenAI. We address this by adding an explicit COVID-affected period to the basic DiD. However, there is no a priori way to settle whether or how the effect of the pandemic contaminates the post-ChatGPT period for three reasons. First, the institutional return to normal was gradual and federated rather than a single clean switch university-wide. Second, students enrolled in Fall 2022 experienced COVID-19 lockdowns in high school, with likely effects on their college preparedness. Third, the persistence of the disruption is visible in the outcomes themselves. For example, the withdrawal rate event study presented in Figure \ref{fig:es-withdraw-fail} suggests that Fall 2022 is not a fully recovered, pre-GenAI baseline. The average grade event study in Figure \ref{fig:es-grade} instead dips in early COVID and returns to zero by 2022. The pandemic disruption decays slowly while the GenAI shock ramps up, with no clean post-COVID, pre-AI window in between. All together, the entire 2020 to 2022 period is arguably COVID-affected without clear evidence of whether an equilibrium emerged prior to introduction of ChatGPT.

Because the correct model cannot be data-determined, we take a ``multiverse'' approach. We estimate the disruption both as a temporary window that closes after 2022 and as a persistent level shift, and we report the GenAI effect as a range bounded by the two. Define the indicators
\[
  \text{COVID Year}_t \;=\; \mathbf{1}\{2020 \leq t \leq 2022\},
  \qquad
  \text{Post}^{\text{AI}}_t \;=\; \mathbf{1}\{t \geq 2023\}.
\]
The two indicators are mutually exclusive. $\text{COVID Year}_t$ takes value one only during the three calendar years 2020 through 2022, and $\text{Post}^{\text{AI}}_t$ takes value one for 2023 and later.

Given this approach, our main specification for student-level outcomes is
\begin{equation}
\begin{split}
  y_{sct} \;=\; & \gamma_s + \delta_c + \tau_t \\
    & + \beta_1 \, \text{Susceptibility}_{c} \times \text{Post}^{\text{AI}}_t \\
    & + \beta_2 \, \text{Susceptibility}_{c} \times \text{COVID Year}_t \\
    & + \varepsilon_{sct},
\end{split}
\label{eq:main}
\end{equation}
estimated on the full panel of fall 2016 to fall 2025. For course-level outcomes, student fixed effects are excluded. 

The first coefficient of interest is the \emph{transient-COVID estimate} $\beta_1$, the per-unit response of the outcome to GenAI susceptibility in the post-AI period (2023 onward) relative to the pre-COVID baseline (2016 to 2019), with the COVID window itself absorbed by $\beta_2$. The second is the \emph{persistent-COVID estimate} $\beta_1 - \beta_2$. Each main table reports it in a ``GenAI net COVID'' row with delta-method standard errors beneath the regression coefficients.  The two estimates are endpoints of a range where we propose the true effects lie. Suppose a share $\lambda$ of the COVID-era differential persists into the post-AI period. The GenAI effect is then $\beta_1 - \lambda\beta_2$, which equals the transient-COVID estimate at $\lambda = 0$ and the persistent-COVID estimate at $\lambda = 1$. We assume $\lambda \in [0,1]$, meaning that post-ChatGPT conditions lie somewhere between the pre-COVID and COVID-era states. Courses likely recovered only partially by 2023 (Appendix~\ref{app:descriptives-susceptibility}), which is consistent with this assumption. Our estimates do not cover an overshoot beyond the pre-COVID state ($\lambda < 0$) or an amplification of the COVID-era differential ($\lambda > 1$). Which endpoint is larger depends on the sign of $\beta_2$ and varies by outcome, so we do not label either one a lower or an upper bound. We draw substantive conclusions only where both estimates agree. When stating which effects the data rule out, we use the larger of the two upper confidence limits. Because that limit is convex in $\lambda$, it also holds for every intermediate degree of persistence.

\paragraph{Identification and parallel trends.}

The key identifying assumption is parallel trends, that high- and low-susceptibility courses would have continued on their existing trajectories after 2022 had ChatGPT not been released. This assumption cannot be directly tested, and estimates may be read as descriptive if deemed unrealistic.\footnote{As \citet{rothPretestCautionEventStudy2022} cautions, pre-treatment trends cannot tell us much about the validity of the parallel trends assumption. They can, however, help us estimate bounded effects under various scenarios of how parallel trends might be violated \citep{rambachanMoreCredibleApproach2023}, which we discuss in Appendix \ref{app:honestdid} } 

For visualization, we estimate the continuous event study at the term level, replacing $\text{Post}^{\text{AI}}_t$ with a full set of semester-by-treatment interactions and normalizing to the Fall 2019 reference semester $k_0$:
\begin{equation}
y_{sct} \;=\;  \gamma_s + \delta_c + \tau_t + \sum_{k \neq k_0} \lambda_k \, \text{Susceptibility}_c \times \mathbf{1}\{\text{sem}_t = k\} + \varepsilon_{sct}.
\label{eq:event-study}
\end{equation}
Aligned with the main regressions, student fixed effects are excluded for course-level outcomes. The coefficients $\lambda_k$ trace the per-unit response of each outcome to GenAI susceptibility across semesters. 

\paragraph{Interpreting a continuous treatment.}

Our susceptibility ``treatment'' is theoretically continuous, although it most often appears as counts in practice. Continuous treatment difference-in-differences carries additional assumptions and procedures not required for the traditional binary version, as documented by \citet{callawayDifferenceinDifferencesContinuousTreatment2024}. A crucial assumption is that of parallel trends for every dosage of the treatment. In other words, the counterfactual trends of courses with higher susceptibility levels must match the observed trends of those with lower susceptibility levels for causal interpretation to hold. Thus, selection into susceptibility dosage is a threat to precise identification in our setting.

One defense that our 2019-anchored treatment design affords is that courses could not have selected into assessment structure in anticipation of GenAI. However, courses likely differ by susceptibility in ways that are correlated with responses to GenAI. For example, differences in pedagogical tradition and tolerance for students' external resource usage across subjects. To support the credibility of causal interpretation, Appendix~\ref{app:contdid} reports non-positive and non-significant effects at every dosage of susceptibility, which are identified under standard, more relaxed parallel trends assumptions.

\paragraph{Heterogeneous effects by prior academic preparation.}

For the final grade outcome we also test whether the effect of GenAI susceptibility varies by students' prior academic preparation. An average of zero is consistent either with no effect anywhere or with countervailing effects that cancel in aggregate, for example a substitution-driven gain for weaker students offset against a loss for stronger ones, and only a split by prior preparation can distinguish these cases. We estimate Equation~\eqref{eq:main} separately within each tercile of \texttt{AbilityRank} (see Section~\ref{sec:measure-abilityrank}); the resulting event studies appear in Figure~\ref{fig:es-gpa-het}. For robustness, we also reconduct the analysis with a non-course-residualized version of \texttt{AbilityRank} and a battery of alternative ability measures (Appendix~\ref{app:ability-robustness}).

\section{Results}
\label{sec:results}

\subsection{Student Outcomes}
\label{sec:results-students}

\subsubsection{Average Grade Effect} 
\label{sec:results-grades}

In estimating average effects of GenAI availability on grades, we begin with a simple OLS model without accounting for COVID disruptions. The first column of Table~\ref{tab:student_outcomes} show a positive relationship between course GenAI susceptibility and grades. The GenAI period was also associated with higher average grades in general. But the interaction effect of 0.039 grade points (SE = 0.037) is not statistically significant, signaling that GenAI availability was not associated with higher grades in more susceptible courses. The second column presents a difference-in-differences estimate with fixed effects, which is also null.

The third column accounts for COVID disruptions, and the effect of course GenAI susceptibility on average final grade depends on how the pandemic is modeled. Under the fully transient assumption, we estimate GenAI availability increased final grade value by 0.010 (SE = 0.045, n.s.) points on average in fully susceptible courses. Under the fully persistent assumption, we estimate a 0.071 (SE = 0.024, $p < 0.01$) average grade point increase. Thus, assuming some lingering COVID effect, we estimate a GenAI effect in the range of $[0.010, 0.071]$. The transient-COVID estimate is indistinguishable from zero, while the persistent-COVID estimate is positive and statistically significant. Because the estimates disagree on statistical significance, we conclude that evidence of GenAI availability's effect on average grades is non-robust. Beside statistical significance, the estimated bounds are small. We rule out increases larger than 0.109 grade points per unit of susceptibility at the 5\% level (one-sided), well under half a step between adjacent letter grades (see Appendix \ref{app:prior-comparison}).

\input{tables_and_figures_v2/table_student_outcomes}

\begin{figure}[htbp]
\centering
\includegraphics[width=0.7\linewidth]{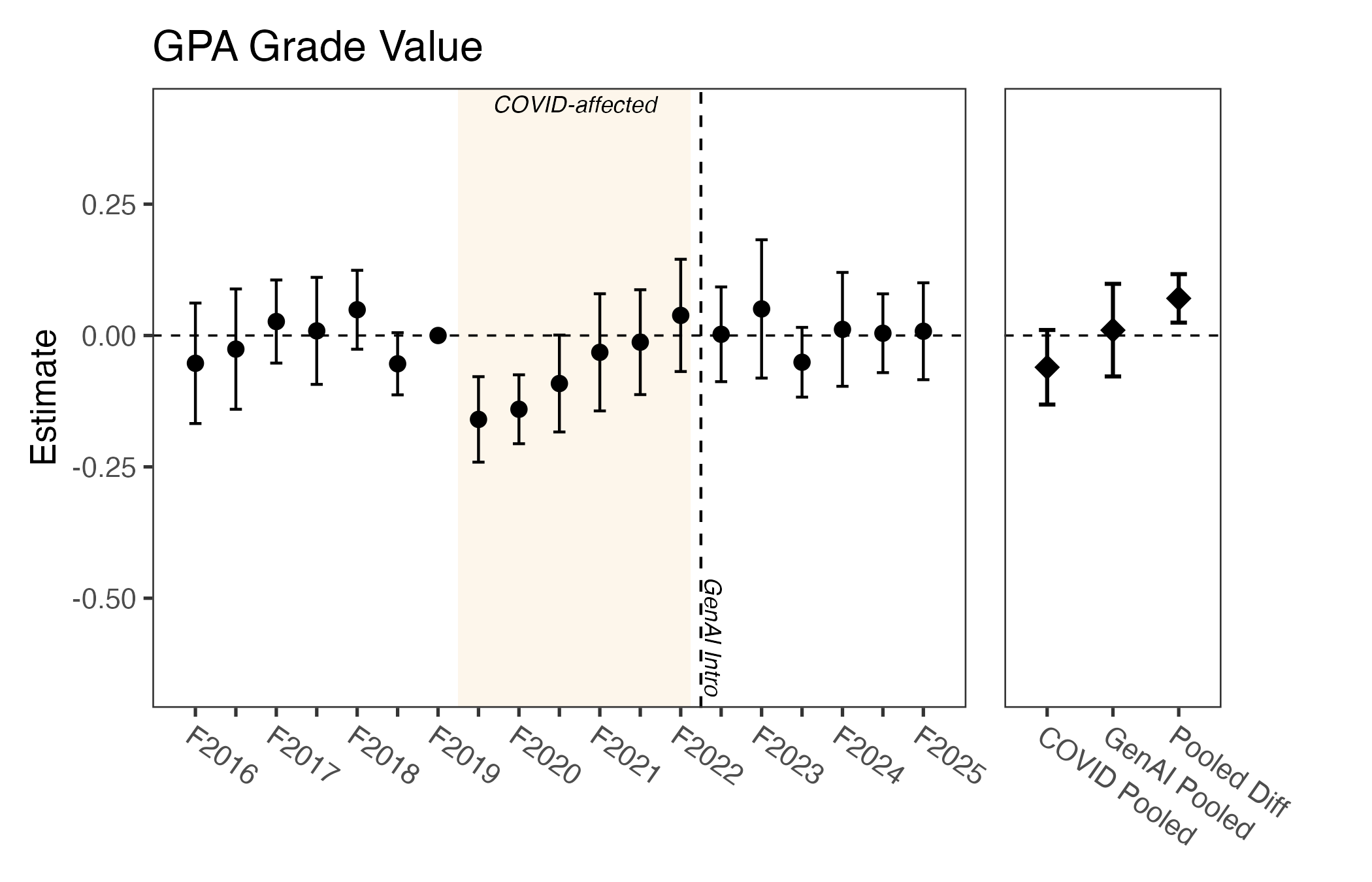}
\caption{Event Study: Average Grade. Left panel: coefficients on course susceptibility interacted with semester indicators (Equation~\ref{eq:event-study}) for final grade value, normalized to Fall 2019. Right panel: pooled coefficients for the COVID-affected terms (shaded band) and the post-GenAI terms from Equation~\ref{eq:main}, and their difference with a delta-method standard error. The post-GenAI coefficient is the transient-COVID estimate and the difference is the persistent-COVID estimate. Bars are 95 percent confidence intervals. Coefficients are per unit of susceptibility, the difference between fully susceptible and fully non-susceptible courses. Regressions include student, course, and semester fixed effects; standard errors are clustered by course.}
\label{fig:es-grade}
\end{figure}

Figure~\ref{fig:es-grade} plots the event-study dynamics of the grade effects of GenAI availability semester-by-semester, relative to Fall 2019. Qualitatively, the effects are consistent with zero in the pre-COVID period, become negative during early COVID, and return toward 0 before GenAI introduction. The differential is largest in Winter 2020 ($-0.160$) and has closed by Winter and Fall 2022 ($-0.013$ and 0.038).

We conduct three robustness checks for this average grade results. First, we re-estimate the main specification using four alternative definitions of course GenAI susceptibility (Appendix~\ref{app:robustness-2019}). The transient-COVID estimate grows as the definition includes more pandemic-era assessment change, which supports anchoring susceptibility to 2019. Second, we follow \citet{rambachanMoreCredibleApproach2023} in estimating bounds of the true effect under various non-parallel trends scenarios (Appendix~\ref{app:honestdid}). For a linear counterfactual difference in trends of arbitrary slope, the confidence set for the average post-ChatGPT grade effect still covers zero and excludes effects larger than about 0.06 grade points (4.0 scale) per unit of susceptibility. Third, causal interpretation of our continuous treatment difference-in-differences estimates requires stronger assumptions than traditional binary treatment versions. We follow \citet{callawayDifferenceinDifferencesContinuousTreatment2024} in estimating effects at every dosage of susceptibility compared to non-susceptible courses, relying on standard parallel trends assumptions rather than stricter assumptions for TWFE regression with a continuous treatment variable (Appendix~\ref{app:contdid}). We find grade effects that are non-positive and non-significant at every level of susceptibility. The slope of the dose-response curve, which our main estimates summarize, is also non-significant at every level. However, it should be noted that confidence intervals for both analyses are wide, rather than tight around zero.

\subsubsection{Grade Distribution}
\label{sec:results-grade-dist}
The non-robust average grade effect could conceal opposing heterogeneous effects that cancel in aggregate. Following \citet{chirikovAIGradeInflation2026}, we estimate our main difference-in-differences specification modeling COVID effects on indicators for clearing each letter-grade threshold, from earning at least an A down to earning at least a D. Table~\ref{tab:grade_distribution} in Appendix~\ref{app:additional-main} reports the full set of estimates.

The probability of earning at least an A rises by 0.107 under the transient-COVID estimate (SE = 0.046, $p < 0.05$). However, the persistent-COVID estimate is 0.030 (SE = 0.018, n.s.), and the event study in Figure~\ref{fig:es-thresholds} shows a long rising trend in the likelihood of an A grade before ChatGPT's release. Similarly, our estimate of B-grade or better probability is not stable across the COVID assumptions. The transient-COVID estimate is negative and not statistically significant, while the persistent-COVID estimate is positive and significant (0.030, SE = 0.011, $p < 0.01$). We therefore interpret these shifts at the top of the grade distribution as non-robust.

At the bottom of the distribution, the probability of earning at least a D and its complement (earning less than a D or no grade) are null under both COVID assumptions. Thus, we detect no robust effects of GenAI availability on the ceiling or floor of the grade distribution.

\subsubsection{Heterogeneity by Prior Academic Preparation} \label{sec:results-grades-het}

\begin{figure}[htbp]
\centering
\includegraphics[height=0.65\paperheight]{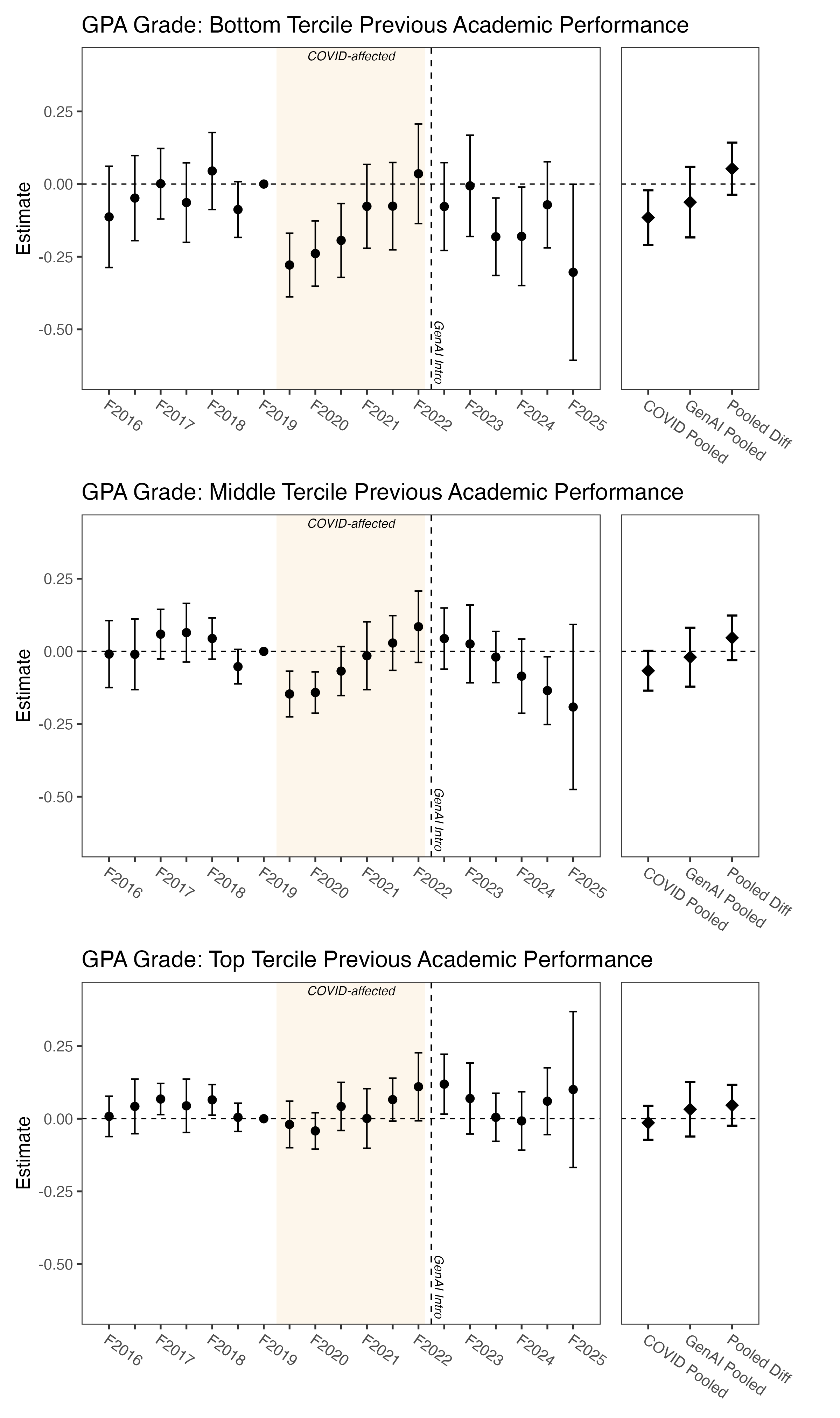}
\caption{Event Study: Average Grade by \texttt{AbilityRank} Tercile. Left panels: coefficients on course susceptibility interacted with semester indicators (Equation~\ref{eq:event-study}) for final grade value, estimated within each tercile and normalized to Fall 2019. Right panels: pooled coefficients for the COVID-affected terms (shaded band) and the post-GenAI terms from Equation~\ref{eq:main}, and their difference with a delta-method standard error. The post-GenAI coefficient is the transient-COVID estimate and the difference is the persistent-COVID estimate. Bars are 95 percent confidence intervals. Coefficients are per unit of susceptibility, the difference between fully susceptible and fully non-susceptible courses. Regressions include student, course, and semester fixed effects; standard errors are clustered by course.}
\label{fig:es-gpa-het}
\end{figure}

Countervailing heterogeneous effects by prior academic preparation could also explain our non-robust average grade effects. Estimating the main specification separately within each \texttt{AbilityRank} tercile (Table~\ref{tab:het_ability_terciles} in Appendix~\ref{app:additional-main}) yields transient-COVID estimates of $-0.062$ (SE = 0.062) for the bottom tercile, $-0.020$ (SE = 0.052) for the middle tercile, and 0.032 (SE = 0.048) for the top tercile. The persistent-COVID estimates are 0.053 (SE = 0.046), 0.047 (SE = 0.039), and 0.046 (SE = 0.036). No estimate is significant at the 5\% level. The estimates are economically small, statistically null, and, crucially, relatively flat across the terciles. Figure~\ref{fig:es-gpa-het} shows the corresponding event studies. No tercile shows separation in grades between more and less susceptible courses after the release of ChatGPT. The same flat, null pattern holds under five other proxies for prior academic preparation (Appendix~\ref{app:ability-robustness}). Only one of the 30 alternative-proxy estimates is significant at the 5\% level, which is attributable to multiple comparisons.

\subsubsection{Withdrawal and Failure Rates}
\label{sec:results-withdrawal-fail}

The last six columns of Table~\ref{tab:student_outcomes} report results for withdrawal and failure rates. Focusing on the main specification with COVID modeling, the transient-COVID withdrawal estimate is $-0.007$ (SE = 0.005, n.s.) and the persistent-COVID estimate is 0.010 (SE = 0.006, n.s.). Both point estimates are near-zero and statistically insignificant. 

Failure rates have similarly null GenAI availability effects: 0.001 (SE = 0.002, n.s.) under the transient-COVID estimate and $-0.002$ (SE = 0.002, n.s.) under the persistent-COVID estimate. Thus, more susceptible courses experienced no differential change in failure rates after the release of ChatGPT. These findings align with our results from the grade distribution analysis. The withdrawal and failure event study plots appear in Appendix Figure~\ref{fig:es-withdraw-fail}.

\subsection{Student Satisfaction} 
\label{sec:results-evals}

Table~\ref{tab:course_evals} (Appendix~\ref{app:additional-main}) reports results for the three course evaluation outcomes, and their corresponding event study plots are in Appendix Figure~\ref{fig:es-evals}. The OLS models for understanding, interest, and relative workload all show that courses did not significantly differ based on GenAI susceptibility or post-GenAI. Estimates of the interaction between susceptibility and the GenAI period are null for understanding and interest, but are negative and significant for relative workload, suggesting that work in more susceptible courses was perceived as easier compared to work in less susceptible courses once GenAI became widely available. However, the basic difference-in-differences estimates (second column for each outcome) are not statistically significant at a 5\% level for all three outcomes.

Accounting for COVID effects, the transient-COVID estimates are positive for understanding at 0.069 (SE = 0.037, n.s.) and interest at 0.147 (SE = 0.062, $p < 0.05$), and negative for relative workload at $-0.270$ (SE = 0.120, $p < 0.05$). However, the persistent-COVID estimate for understanding is $-0.010$ (SE = 0.042, n.s.), for interest is $-0.044$ (SE = 0.051, n.s.), and for relative workload is $-0.034$ (SE = 0.068, n.s.). Thus, estimates depend on the pandemic recovery assumption. 

Because the estimates for interest and workload disagree across COVID assumptions, we cannot attribute those shifts to GenAI rather than to the pandemic's recovery. Self-reported understanding shows no significant change  under either assumption. These outcomes are more noisy than the student-level ones, partially because they are collected at the course level. Because the regressions use student rows, the estimates weight each offering by its enrollment; unweighted offering-level estimates lead to the same conclusions (Appendix~\ref{app:additional-main}). Rather than precise zero estimates, we conclude that our data shows no robust evidence of changes in student satisfaction after ChatGPT.

\section{Potential Mechanisms Driving Small, Non-robust Grade Effects}
\label{sec:interpreting-null}

This section presents two exploratory analyses that seek to enrich our understanding of the non-robust grade effects found in the main analyses. The first investigates instructors' adjustments to course GenAI susceptibility after ChatGPT. The second investigates \emph{who} receives high versus low grades, which may have changed even if grade distributions remain unchanged. 

\subsection{Are Instructors Using Less Susceptible Assessments?}
\label{sec:null-instructors}

Because we anchor susceptibility to each course's 2019 offerings, our estimates capture differences between courses' pre-COVID assessment designs. Instructors shifting toward less susceptible assessment structures after ChatGPT is a potential mechanism that could explain null grade effects. \citet{hausmanGenerativeAIsImpact} considered this possibility and conducted a survey of faculty in their university setting. Finding that 40 percent of faculty made no course changes in response to student GenAI use, they remark that ``a more complete faculty response could potentially greatly reduce or even eliminate these grade effects'' \citep[][p. 39]{hausmanGenerativeAIsImpact}. Here, we use observed course syllabi to document how instructors in our setting adapted assessment structure over time.

\begin{figure}[htbp]
\centering
\includegraphics[width=0.85\linewidth]{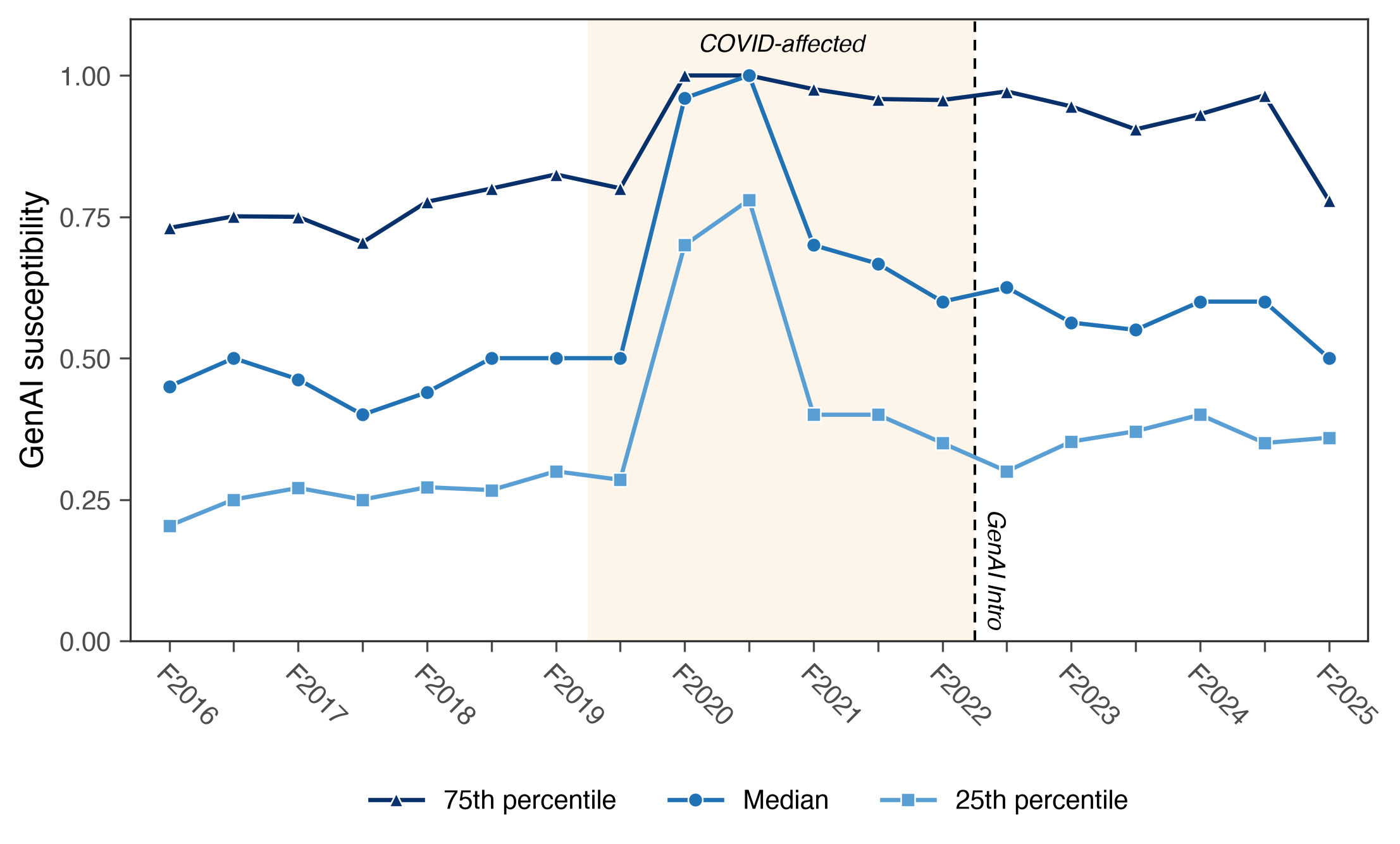}
\caption{GenAI Susceptibility by Semester. Lines show the 25th, 50th, and 75th percentiles of contemporaneous susceptibility, the share of the final grade allocated to susceptible assessments. Percentiles are computed across analytic-sample courses with a syllabus in each fall and winter semester, averaging over sections and weighting by enrollment, so the lines describe the susceptibility faced by the typical enrolled student. The shaded band marks the COVID-affected terms (Winter 2020 to Fall 2022), and the dashed line marks ChatGPT's release. The set of courses varies across semesters. Syllabus coverage falls in Winter and Fall 2025 (56 and 53 percent of enrolled course-semesters, against 68 to 80 percent in 2021 to 2024), consistent with incomplete uploading of recent terms. Restricting to the 158 courses with a syllabus in every academic year (48 percent of weighted enrollment) gives the same shape at somewhat lower levels. Weighted percentiles can jump when a single large course changes; the drop in the 75th percentile in Fall 2025 reflects one such course.}
\label{fig:susceptibility-percentiles}
\end{figure}

Figure~\ref{fig:susceptibility-percentiles} displays the susceptibility measure across the analytic sample, weighted by enrollment, using every syllabus available (not just 2019 versions). Before COVID, the median enrolled student took courses where half or less of the final grade was determined through susceptible assessments. The median spikes to nearly 100 percent in Fall 2020 and Winter 2021, presumably in response to COVID. Leading up to ChatGPT's release in Fall 2022, the median fell to roughly 60 percent. In the GenAI period, the median ranged from 50 to 63 percent with no visible break, and it stayed above its pre-COVID level through Winter 2025. Restricting the sample to only those courses for which we have a syllabus in every year yields the same pattern.

We therefore see no evidence that susceptible assessment has been abandoned en masse, at least in syllabus final grade weightings. Most of the reduction in susceptibility was during pandemic recovery and medians did not change much immediately after GenAI availability. Although there is a visible decline in the most susceptible courses in 2025, syllabus coverage is lower for these semesters.

Formal grade weighting on the syllabus is only one way instructors might respond. They could hold weights fixed while changing what the weighted work asks of students, for example by raising expectations on take-home assignments or by adding in-class components within a homework or paper grade. They could also adjust grades after the fact, through informal curving or shifting norms about what earns an A. Explicit GenAI syllabus policies are a further channel \citep{chirikovHowInstructorsRegulate2026}. More research on the role of instructors in responding to GenAI seems appropriate in this regard.

\subsection{Did Who Gets Which Grades Change?}
\label{sec:null-makeup}

While it is possible that instructors may preserve grade distributions, GenAI availability could still make that distribution less meaningful if grades do not sort students by skill level. This is difficult to test empirically, as assessments and grades are the main way skill level might be observed. However, performance on standardized tests gives some comparable measure of prior performance. Here, we analyze whether GenAI availability has changed how well grades are predicted by students' performance on pre-college standardized tests. Within each course-semester, we regress final grade value on students' SAT or ACT scores, converted to a common scale (Appendix~\ref{app:ability-robustness}). Our outcome is the slope of that regression, the change in grade points per 100 SAT points. Appendix~\ref{app:grade-signal} details the measure and specification, and Table~\ref{tab:grade-signal} reports the estimates.

\input{tables_and_figures_v2/table_grade_signal}

The slope did not detectably change in more susceptible courses after ChatGPT's release. Before COVID, every 100 SAT points predicted 0.21 more grade points, comparing students within a course-semester. Taking the difference between fully susceptible and fully non-susceptible courses, the transient-COVID estimate is $-0.007$ (SE = 0.017, n.s.) and the persistent-COVID estimate is $-0.016$ (SE = 0.010, n.s.). One-sided tests at the 5\% level rule out declines larger than 0.035 under either estimate, about a sixth of the pre-COVID slope. We do find that standardized test scores have become less predictive of grades in general, although we do not find greater declines in more susceptible courses, suggesting that GenAI substitution is not the cause.

Changes in admissions policy at our University complicate the comparison. Students entering in 2021 or later could apply without submitting standardized test scores. Coverage is 94 percent of enrollments before COVID but only 69 percent after ChatGPT's release. Among these later entrants, students without scores earned lower first-term grades than students with scores (a GPA of 3.53 versus 3.68). The second column of Table~\ref{tab:grade-signal}, therefore, keeps only students who entered the university by Fall 2020. This removes the change in who reports a score, at the cost of a post-ChatGPT period made up mostly of upper-level students. The transient-COVID point estimate becomes positive, but neither estimate is statistically significant. This exploratory evidence suggests who gets which grades in more susceptible courses has not changed with GenAI availability relative to less susceptible courses.

\section{Discussion}
\label{sec:discussion}

Our analyses reveal small and non-robust effects of generative AI (GenAI) availability on grades and student satisfaction. Once the pandemic shock is separated from the GenAI shock, more GenAI-susceptible courses show no statistically significant change in withdrawal or failure after the release of ChatGPT. The average grade effect is non-robust: near zero when COVID effects are treated as transient and positive when they are treated as persistent. We also show that heterogeneity in the grade distribution, which prior work observed, is not robustly present in our data. Furthermore, we find no statistically significant grade effect heterogeneity by prior academic preparation. 

These results may at first appear inconsistent with those from prior work. Appendix~\ref{app:prior-comparison} details an analysis comparing our grade estimates to those of \citet{hausmanGenerativeAIsImpact} and \citet{chirikovAIGradeInflation2026} on a common scale (Table~\ref{tab:prior_comparison}). Our point estimates are generally less than or equal to the prior ones, but are mostly statistically indistinguishable from them. Overall, it is important to consider the economic significance of the effects from this emerging body of work. Using the larger average grade estimate, one-sided tests at the 5\% level rule out increases above 0.11 grade points for a fully susceptible course. On a 4.0 scale, this is less than half a step between adjacent letter grades, such as going from a B- to a B. For comparison, average grades rose by 0.14 points between the pre-COVID and post-ChatGPT semesters in our analytic sample, continuing a pre-pandemic trend of about 0.03 points per year. During the pandemic window they averaged 0.14 points above their pre-COVID level. Our upper estimates in fully susceptible courses are of similar size to grade inflation from secular trends and COVID leniency; for courses of average susceptibility, the upper estimate is about 40 percent of these other effects.

Regarding course evaluations, we do not find robust evidence that the introduction of ChatGPT affected self-reported subject understanding, interest, or relative workload. Increases in interest and a decrease in workload in the post-GenAI period could plausibly be attributed to persistent COVID effects. This evidence tempers concerns in the qualitative literature that susceptibility to direct GenAI substitution could erode student engagement and satisfaction. 

Instructor adaptation, and responses from other institutional actors, may play an important role in preserving grade signals in the wake of GenAI. Our exploratory analysis of post-GenAI course susceptibility finds no evidence that instructors broadly abandoned susceptible assessments. Courses have been becoming less susceptible since the early-COVID disruptions but still remain above pre-COVID baselines. Our finding aligns with \citet{hausmanGenerativeAIsImpact} that most instructors have not resorted to ``GenAI-proofing'' courses by leaning more heavily on non-susceptible assessments. Still, instructor response could explain why grades and grade distributions appear unchanged.

We also explore whether, in spite of similar looking grade distributions, who gets which grade may have changed. Using a regression analysis of how SAT and ACT scores predict grades over time, we find that the relationship has not detectably weakened in more susceptible courses after ChatGPT. This suggests that GenAI availability has not changed the way students are sorted in the grade distribution as predicted by their previous performance on standardized tests. 

\subsection{Limitations}
\label{sec:limitations}

There are important caveats in the interpretation of our results. First, regarding the grade effects, final grades may not perfectly reflect student performance on individual assessments. For example, instructors may institute a final grade curve that reflects relative performance, rather than students' actual grade on exams. We assume that the final grade weightings observed in syllabi matches how final grades are determined, but future work could focus on assessment-level performance. Furthermore, our difference-in-differences strategy does not rule out effects that affect susceptible and non-susceptible courses similarly, such as broader learning or augmentation effects.

Second, effects of GenAI availability on student satisfaction may be less visible in responses to course evaluation surveys than in other measures. The questions on these surveys were not designed to capture the effects of GenAI, and don't directly ask about satisfaction. Also, response rates and representativeness fluctuate across offerings, so selection effects could drive our results. 

Third, our LLM-based pipeline for reconstructing susceptible assessment weighting from syllabi performs well but does make errors. Our human validation helps gauge the performance of the pipeline, and the descriptive patterns in assessment changes during COVID provide some face validity to the measurement. Still, measurement error increases statistical noise in our inference, increasing the likelihood of null findings. Future work could theoretically perform the assessment categorization and weight extraction by hand or improve the automatic reconstruction.

Fourth, we rely on an unconventional difference-in-differences strategy to separate the COVID effects from the GenAI effects. The pandemic occupies the middle of our pre-period, and it is widely documented that COVID-19 had ubiquitous impacts on higher education in the United States and globally. Our modeling choices are well-motivated, but a strict reading is that no window in this setting is clean enough to support a conventional difference-in-differences design.

Last, susceptibility measures are drawn from a syllabus archive maintained by only one of the colleges at our University, and upload is voluntary in most cases. This has implications for the generalizability of our results, no matter how internally valid. For example, instructors who expend more effort in teaching may disproportionately select into uploading their syllabi. These instructors might also be more inclined to restructure assessments in response to GenAI, which would bias our estimates downward. To provide some evidence on the representativeness of our sample, we compare features of courses in the analytic sample to those in the archive but not in the sample and those in the catalog that are not in the archive (Appendix~\ref{app:archive}). As previously acknowledged, much of the missing courses in our sample are from high-enrollment programs in colleges outside the one administering the archive (e.g., Business School). The analytic sample skews toward large undergraduate, in-person, lecture courses, arguably the population of greatest policy interest.

\subsection{Contributions}
\label{sec:contributions}

Our study makes five contributions to literature. First, we add to the nascent large-scale observational literature on GenAI substitution in higher education. Our estimates for average grades and the grade distribution are generally smaller than those of the two prior studies on the topic, though mostly not statistically distinguishable from them (Appendix~\ref{app:prior-comparison}). Together, the three studies suggest that GenAI-driven grade inflation is modest where it appears and is not a universal feature of post-ChatGPT higher education. Whether it appears seems to depend on a number of factors including whether COVID effects persist, endogenous responses to GenAI use, and measurement and modeling choices. Second, we provide what is, to our knowledge, the first quasi-experimental estimate of how the effects of GenAI availability vary with students' pre-treatment academic preparation. Third, we offer early descriptive evidence on how instruction and grading responded, which has received little attention in prior work. Fourth, we expand the outcome set beyond grades to course evaluations of subject understanding, interest, and relative workload. Additionally,  to support follow-on research, the annotation codebook and LLM prompts for the course susceptibility measurement pipeline are shared in Appendices \ref{app:codebook} and \ref{app:prompts}, respectively.

\section{Conclusion}
\label{sec:conclusion}

Higher education decision makers face many critical choices in the wake of widely available GenAI tools, and the stakes are high. 
Higher education is central to increasing human capital and economic growth \citep{romerEndogenousTechnologicalChange, jonesFutureEconomicGrowth2022}, and the grade signals produced through assessments are part of how that human capital is measured and allocated. A major concern with GenAI availability is that students are not engaging in the cognitive effort necessary to develop mastery when they can access these tools. Our study finds no robust evidence for this ``GenAI substitution hypothesis,'' as grades and student satisfaction have not substantively changed after the release of ChatGPT in courses that rely more on GenAI-susceptible assessments, relative to less susceptible courses. 

The evidence (or lack thereof) which we present should dampen alarm about GenAI, grades, and student satisfaction, rather than dismiss concerns outright. On one hand, rushed policy like shifting to fully non-susceptible assessments could harm students' learning and experience \citep{gu2026policeguidehighereducation}, potentially introducing a cure worse than the disease. On the other hand, grades guide instruction, sort students into majors and careers, and inform employers \citep{barGradeInformationGrade2009}, so whether GenAI eventually erodes them matters even where no erosion is yet visible. Ultimately, higher education institutions must adapt to a world with GenAI, maintaining evidence-based pedagogy and continuing to monitor whether, when, and how exactly grade signals and student satisfaction degrade in more susceptible courses.

\newpage
\bibliography{all_refs}

\newpage
\appendix

\section{Letter Grades and Grade Point Scale}
\label{app:grade-pts}

The academic career data (Section~\ref{sec:data-admin}) record final letter grades, which we convert to grade points using the scale in Table~\ref{tab:grade-pts}. Incompletes are scored as 0.0. Withdrawals carry no grade points and are excluded from the average grade outcome, but they remain in the withdrawal outcome and in the denominator of the grade threshold indicators.

\begin{table}[h]
    \centering
    \caption{Grade Point Values by Letter Grade}
    \label{tab:grade-pts}
    \begin{tabular}{lc}
        \hline
        Letter Grade & Grade Points \\
        \hline
        A+/A & 4.0 \\
        A- & 3.7 \\
        B+ & 3.3 \\
        B & 3.0 \\
        B- & 2.7 \\
        C+ & 2.3 \\
        C & 2.0 \\
        C- & 1.7 \\
        D+ & 1.3 \\
        D & 1.0 \\
        D- & 0.7 \\
        E & 0.0 \\
        I (incomplete) & 0.0 \\
        W (withdrawal) & -- \\
        \hline
    \end{tabular}
\end{table}

\FloatBarrier
\section{Course Evaluation Questions}
\label{app:eval-qs}

The standard course evaluation includes eight required five-point Likert-scale questions, each with an open-ended prompt. We use the questions on subject understanding, subject interest, and workload relative to the student's other courses. Their exact wordings are:
\begin{itemize}
    \item \texttt{understand}: ``This course advanced my understanding of the subject matter.''
    \item \texttt{interest}: ``My interest in the subject has increased because of this course.''
    \item \texttt{workload}: ``As compared with other courses of equal credit, the workload for this course was (Much Lighter, Lighter, Typical, Heavier, Much Heavier).''
\end{itemize}

For the understanding and interest questions, a higher score indicates greater agreement with the statement. On the original workload scale, a higher score indicated a lighter relative workload. We reverse-code it so that a higher score indicates a heavier relative workload. A negative estimate on this outcome therefore means that students rated more susceptible courses as relatively less demanding after ChatGPT's release.

\FloatBarrier
\section{Data Descriptives}
\label{app:descriptives}

The descriptives in this appendix use a balanced panel that spans the full 2015--2025 timeline of the syllabus data: all terms, restricted to courses observed both in 2015--2019 and in 2023--2025. It is broader than the regression sample, which keeps fall and winter terms from fall 2016 onward and requires a 2019 anchor (Section~\ref{sec:data-sample}). Table~\ref{tab:summary_stats} reports summary statistics for this panel at the student-offering level. Susceptibility is measured from voluntarily uploaded syllabi, so it is missing for courses without a 2019 syllabus. Course evaluations begin in 2016 rather than 2015, and some courses, particularly at the graduate level, did not request evaluations (Section~\ref{sec:data-evals}). Outcome coverage is otherwise nearly universal, with grades missing only where students withdrew or elected non-graded credit. Final grades average 3.49 on the 4.0 scale. Withdrawal (2.4\%) and failure (0.8\%) are rare, so even changes of a fraction of a percentage point would be large in relative terms.

\input{tables_and_figures_v2/table1_summary_stats}

Table~\ref{tab:top_subjects} lists the ten subjects with the greatest coverage by unique courses, course-term offerings, and student enrollments. Large introductory sequences in chemistry, mathematics, psychology, and economics dominate the enrollment ranking. The course and offering rankings instead reflect the voluntary nature of the syllabus archive. Departments differ in how systematically they upload. Some archive syllabi as a matter of departmental practice, while others leave uploads to individual instructors. Industrial and Operations Engineering (IOE) illustrates the pattern. It ranks seventh in course-term offerings and eighth in unique courses but does not appear in the enrollment top ten, which reflects a consistent upload practice rather than department size. This selection determines which courses carry susceptibility data, but it does not by itself threaten identification, because treatment and outcomes are compared within course over time.

\input{tables_and_figures_v2/table_top_subjects}

Table~\ref{tab:balance} compares pre-COVID (fall 2016 to fall 2019) means for courses in the bottom ($\leq 0.335$) and top ($> 0.665$) absolute terciles of the 2019-anchored susceptibility measure. The two groups differ substantially at baseline. High-susceptibility courses award grades 0.45 points higher than low-susceptibility courses, more than one step between adjacent letter grades. Their withdrawal and failure rates are roughly half as large. Differences in median evaluation scores are statistically significant but small, all under 0.04 points on the five-point scales, with high-susceptibility courses rated slightly more interesting and slightly less demanding. Students in the two groups are nearly identical in prior ability, differing by 0.012 in the within-cohort rank. The course fixed effects in Equation~\ref{eq:main} absorb level differences of this kind, and identification comes from variation within course over time. The size of these gaps nonetheless shows why a cross-sectional comparison of high- and low-susceptibility courses would mislead.

\input{tables_and_figures_v2/table2_balance}

\subsection{Between- and Within-Course Variation in Susceptibility over Time}
\label{app:descriptives-susceptibility}

\begin{figure}[htbp]
\centering
\includegraphics[width=\textwidth]{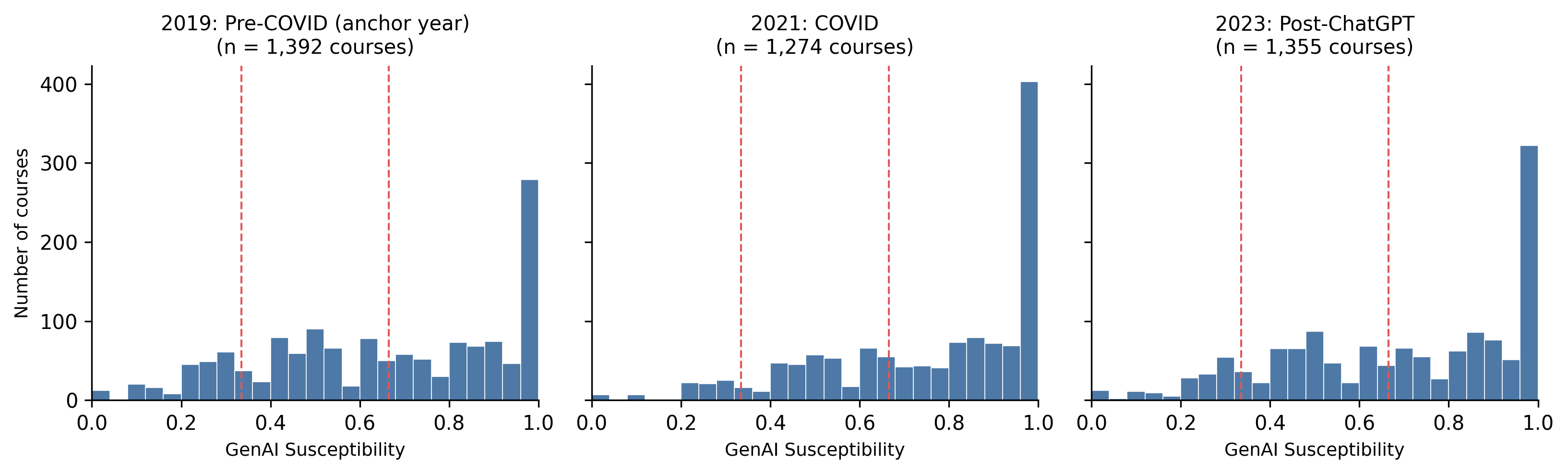}
\caption{Distribution of GenAI Susceptibility across Courses, 2019, 2021, and 2023. Each panel shows the contemporaneous susceptibility measure (the share of the final grade allocated to susceptible assessments, averaged over a course's offerings in that year) for balanced-panel courses with a syllabus in the indicated year. Dashed lines mark the cutoffs of the absolute susceptibility terciles: low ($\leq 0.335$), medium, and high ($> 0.665$).}
\label{fig:susceptibility-distribution}
\end{figure}

This subsection describes how susceptibility is distributed across courses and how it evolves within courses over time. Figure~\ref{fig:susceptibility-distribution} plots the distribution of the contemporaneous measure across courses in the balanced panel in three benchmark years: the pre-COVID anchor year (2019), the COVID period (2021), and the post-ChatGPT period (2023). In all three years the distribution is skewed toward high susceptibility, with a pronounced spike at fully susceptible courses that allocate no grade weight to real-time assessment. In 2019, around half of courses were in the high absolute tercile. COVID pushed the distribution further in that direction: in 2021, roughly two-thirds were. By 2023 the shift had partly receded, leaving a distribution more susceptible than in 2019 but well below the COVID peak. This trajectory illustrates the contamination concern in Section~\ref{sec:measure-exposure}: susceptibility measured in or near the COVID window reflects pandemic adaptation rather than stable course design.

Despite these aggregate movements, susceptibility is highly persistent within course. Table~\ref{tab:async_persistence} reports the year-to-year correlation of course susceptibility across the panel. The mean correlation between adjacent years is 0.76, and the mean correlation of every other year with 2019 is 0.73. The COVID years are the clear exception. Correlations involving 2020 or 2021 fall to between 0.42 and 0.68, while the years from 2022 onward correlate with one another at 0.71 to 0.83. This persistence justifies anchoring course GenAI susceptibility to a single pre-treatment year, and the COVID-era dip motivates the choice of 2019 in particular (Section~\ref{sec:measure-exposure}).

\input{tables_and_figures_v2/table_pct_async_persistence}

\begin{figure}[htbp]
\centering
\includegraphics[width=\textwidth]{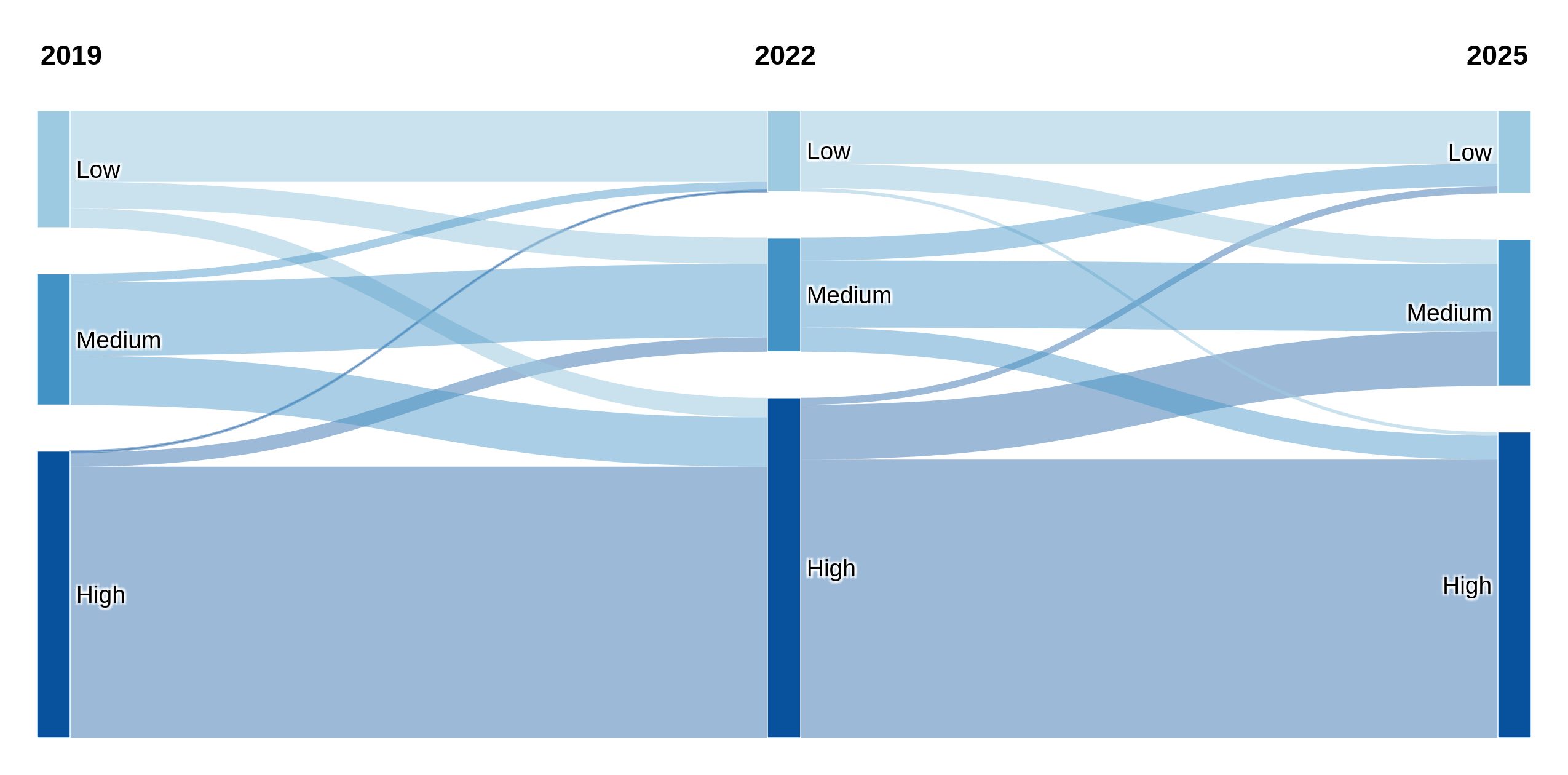}
\caption{Transitions across Susceptibility Terciles, 2019, 2022, and 2025. Flows trace 2025 offerings among the low ($\leq 0.335$), medium, and high ($> 0.665$) absolute susceptibility terciles. The 2019 and 2022 terciles are the course's anchored measures for those years, and the 2025 tercile is the offering's own measure. The sample is the 1,203 offerings in 2025 (497 courses) with terciles observed in all three years. Node heights are proportional to offering counts.}
\label{fig:susceptibility-sankey}
\end{figure}

Figure~\ref{fig:susceptibility-sankey} makes the same point at the tercile level. It traces the 1,203 offerings in 2025 (from 497 courses) whose course also has 2019 and 2022 susceptibility measures. Roughly two-thirds remain in the same tercile at all three points, and nearly three-fourths occupy the same tercile in 2025 as in 2019. Those that move tend to drift toward higher susceptibility, consistent with partially lingering effects of COVID. 

\FloatBarrier
\section{The Analytic Sample and the Course Catalog}
\label{app:archive}

Our analytic sample is not a random sample of the University's courses. A single college administers the syllabus archive, and uploading is voluntary in most cases, although administrators report that some departments require uploads as part of instructor performance reviews. Our sample restrictions narrow coverage further, so the sample may represent neither the archive nor the University's offerings as a whole. This appendix compares the analytic sample with the rest of the archive and with the full catalog. The comparison is possible because the academic career data (Section~\ref{sec:data-admin}) record class size, instruction mode, and student outcomes for every course, with or without a syllabus. The catalog includes 15,624 unique courses across 322 subjects and 105,309 offerings from fall 2016 through fall 2025. A course counts as archived if it has a syllabus from any year of the archive. Because this comparison identifies the analytic sample from enrollment records, it is slightly larger here than the regression sample (1,449 versus 1,391 courses).

\subsection{Coverage}

Table~\ref{tab:archive-coverage} reports the unique courses remaining at each step of sample selection, with their shares of all courses and enrollments. Subjects represented in the archive account for 57.2 percent of courses and 58.9 percent of enrollments. Most of the 183 unrepresented subjects are outside the college that administers the archive, including professional programs such as computer science, dentistry, nursing, business administration, and social work. Within represented subjects, the analytic sample retains about 16 percent of courses but over half of enrollments. The courses lost between these steps are therefore mostly small or irregularly offered.

\input{tables_and_figures_v2/table_archive_coverage}

\subsection{Composition}

Table~\ref{tab:archive-balance} divides the catalog into three groups: courses never archived, courses archived but outside the analytic sample, and the analytic sample. Compared with courses never archived, analytic-sample courses have larger sections (33.4 versus 18.6 students) and are more often undergraduate, lecture-based, and in person.

A further concern is selection into the archive. Instructors who upload syllabi voluntarily may invest more in course design and pedagogy, which could correlate with adapting courses to mitigate GenAI effects. Our results may then not extend to instructors who invest less, whose students' outcomes could have changed. Outcomes from fall 2016 to fall 2021 offer some evidence on this. Compared with courses never archived, analytic-sample courses have lower average grades (3.63 versus 3.70), lower failure rates (1.0 versus 2.6 percent), and higher withdrawal rates (2.2 versus 1.3 percent). We do not test these differences, and we cannot attribute them to instructors because course types also differ across the groups.

Our estimates therefore apply most directly to large undergraduate lecture courses in the subjects represented in the archive. This population is less representative of graduate and professional programs, but it is arguably of greatest interest to higher education policymakers. 

\input{tables_and_figures_v2/table_archive_balance}

\FloatBarrier
\section{LLM Annotation Pipeline Validation} 
\label{app:llm-pipeline}

\citet{chirikovAIGradeInflation2026} also uses an LLM to identify which courses ChatGPT's release was more likely to affect. Our approach differs from his task-composition measure in two respects. It captures grade-weighted assessment design rather than counts of writing and coding tasks, and it is validated directly against human coding. In his related measurement paper \citep{chirikovHowInstructorsRegulate2026}, systematic human validation covers the syllabus AI policy classification, while the task-exposure measure rests on face validity and spot checks. We treat the two as complementary measures of a latent construct, and the difference in measurement is one plausible source of our differing results. 

Our pipeline converts each syllabus PDF to Markdown using \texttt{pymupdf4llm}, then reconstructs fields with GPT-5.1 through a university-managed API gateway approved for moderately sensitive data, including intellectual property like course syllabi. The pipeline categorizes assessments into 10 categories and reconstructs their final grade weightings. Syllabi differed in their presentation of final grade weightings. Some provided clear tables with values for each assessment, others noted weights next to sections describing each assessment in detail, and still others presented final grade weightings in plain text not marked with a section header. The 10 categories are \texttt{closed\_exam}, \texttt{open\_exam}, \texttt{unknown\_exam}, \texttt{quiz\_participation}, \texttt{homework}, \texttt{papers}, \texttt{presentation}, \texttt{performance}, \texttt{project\_present}, and \texttt{project\_no\_present}. The prompts elicit step-by-step reasoning and include few-shot examples to improve categorization. Finally, each output is validated against the required format and automatically re-queried on violation.
    
To validate the pipeline, we drew a stratified sample of 525 syllabi from 2015 to 2025 that underrepresents departments with many sections of the same course each term (usually introductory courses). We developed the codebook iteratively after initial exploration of the full corpus. Seven trained research team members each coded 150 syllabi independently, so that every syllabus had two annotators. Annotators resolved all disagreements in synchronous consensus meetings, and a third coder spot-checked the final labels. These consensus labels are the gold standard for evaluating the LLM outputs. The codebook and prompts appear in Appendices~\ref{app:codebook} and~\ref{app:prompts}.

Table~\ref{tab:eval-summary} reports the mean absolute error (MAE) of the pipeline for the fields used to construct the treatment. Of the 525 syllabi, 429 contain usable final grade weightings; we exclude the remaining 96 from this validation. 

Some syllabi report weights as points, or as percentages that do not sum to 100\%, so we normalize each category's weight by the total across categories. We also merge two pairs of categories that the pipeline systematically confused. First, the pipeline consistently labeled \texttt{unknown\_exam} as \texttt{closed\_exam}, likely because closed exams are the usual default. The research team shared this intuition, but the codebook instructed annotators to label an exam unknown unless the syllabus explicitly marked it as in-class, closed-notes, and closed-book. Because both categories count as non-susceptible, merging them shows more accurately how well the pipeline serves our purpose. Second, the pipeline often confused \texttt{homework} and \texttt{papers}, which both count as susceptible, so we merge them as well.
 
After merging, the exam categories are among the most accurate (MAE of 0.030 for closed or unknown exams and 0.022 for open exams). The merged homework and papers category has the largest error (MAE = 0.085), likely because syllabi use terms such as ``assignments,'' ``problem sets,'' and ``papers'' loosely. The \texttt{performance} category is the most accurate (MAE = 0.016), reflecting clear markers for live skill demonstrations in music, arts, and fitness courses. Early pipeline iterations performed poorly on this category because ``performance'' can also describe how well a student did on other assessments, so we chose few-shot examples to clarify it. 

We construct \texttt{Susceptibility} from the normalized weights using the formula in Section~\ref{sec:measure-exposure}. The pipeline recovers it with MAE = 0.063 on the 0--1 scale ($N = 429$), which we consider adequate for treatment construction and downstream analysis.

\begin{table}[htbp]
\centering
\small
\caption{Validation of the LLM Annotation Pipeline against Human Consensus Labels}
\label{tab:eval-summary}
\begin{tabular}{llcc}
\toprule
Field & Metric & Score & N \\
\midrule
\multicolumn{4}{l}{\textit{Treatment variable}} \\
\quad Susceptibility (continuous) & MAE & 0.063 & 429 \\
\midrule
\multicolumn{4}{l}{\textit{Assessment weight extraction}} \\
\quad Closed/Unknown Exam & MAE & 0.030 & 429 \\
\quad Open Exam & MAE & 0.022 & 429 \\
\quad Homework/Papers & MAE & 0.085 & 429 \\
\quad Quiz/Participation & MAE & 0.054 & 429 \\
\quad Presentation & MAE & 0.023 & 429 \\
\quad Project (With Presentation) & MAE & 0.053 & 429 \\
\quad Project (No Presentation) & MAE & 0.053 & 429 \\
\quad Performance & MAE & 0.016 & 429 \\
\bottomrule
\end{tabular}
\begin{minipage}{\linewidth}
\vspace{0.3em}
\footnotesize
\textit{Notes:} Mean absolute error (MAE) of pipeline output against consensus labels from seven human annotators, computed on normalized weight shares (0--1 scale). Of the 525 validation syllabi, the 96 without final grade weighting information are excluded. Because of systematic misclassification, \texttt{closed\_exam} and \texttt{unknown\_exam} (both non-susceptible) are merged, as are \texttt{homework} and \texttt{papers} (both susceptible).
\end{minipage}
\end{table}

\subsection{Annotator Codebook}
\label{app:codebook}

The seven annotators explored the syllabus archive together and iteratively developed a shared understanding of what to record from each syllabus. The codebook mainly coordinates formatting and the labeling of edge cases. It covers fields ranging from course logistics to learning objectives and in-class technology policies, which the pipeline also reconstructs; validation of all fields is the subject of separate work in progress. Below we reproduce the portion relevant to this study, the final grade weightings.

\bigskip
\noindent\rule{\linewidth}{0.6pt}
\begin{quote}
\singlespacing\small

\noindent\textbf{\texttt{assessment\_type\_weighting} (dict-like):}

\begin{itemize}
    \item Use this category template and replace X with weighting (leave as X if not mentioned)

    \smallskip
    \noindent\texttt{closed\_exam: X ,\ open\_exam: X,\ unknown\_exam: X,\ homework: X,\ papers: X,\ quiz\_participation: X,\ presentation: X,\ project\_present: X,\newline\ project\_no\_present: X,\ performance: X}
    \smallskip

    \item \textbf{Percentage vs.\ Points:} Just use the number listed in syllabus, we can re-normalize later

    \item \textbf{Labs:} Anything that happens in a lab could go under other categories.
    \begin{itemize}
        \item If graded for lab report, would go under papers
        \item If graded to perform task, would go under performance
        \item If graded for showing up, would go under \texttt{quiz\_participation}
    \end{itemize}

    \item \textbf{Multiple Grading Options:} If there are multiple options for grade weighting (e.g., optional final exam), then use the average of the options.
    \begin{itemize}
        \item Do not include extra credit options.
        \item For example, if a class gives two options (e.g., with or without exams), then the weighting is like below, we take the average of the two options:
        \begin{itemize}
            \item no final: Participation: 25\%; Report: 10\%; Two papers 45\%; Exam 20\%
            \item with final: Participation: 25\%; Report: 10\%; Two papers 35\%; Exam 30\%
            \item What you put in the annotation sheet:

            \smallskip
            \noindent\texttt{closed\_exam: X ,\ open\_exam: X,\ unknown\_exam: 25,\ homework: 10,\ papers: 40,\ quiz\_participation: 25,\ presentation: X,\newline\ project\_present: X,\ project\_no\_present: X,\ performance: X}
            \smallskip
        \end{itemize}
    \end{itemize}

    \item \textbf{Category Definitions}
    \begin{itemize}
        \item \texttt{closed\_exam}: Synchronous, closed book, closed notes or asynchronous on lockdown browser; also includes final papers written in class on paper (not with laptop). Any exam where review sheets are provided/allowed still counts as closed exam, since it implies other materials are not allowed.

        \item \texttt{open\_exam}: Any exam where AI could be used, e.g., take-home, asynchronous, if laptops are allowed during the exam, or if exams are via Canvas. Exams for courses that were fully remote during COVID-19 should be assumed open.

        \item \texttt{unknown\_exam}: Exam is mentioned but not enough detail to determine whether open or closed.

        \item \texttt{homework}: Includes problem sets and reading responses, anything with a deliverable, often submitted online or with help of computer.

        \item \texttt{papers}: Includes essays and final papers written asynchronously or on a laptop.

        \item \texttt{quiz\_participation}: Includes attendance, participation, and quizzes.
        \begin{itemize}
            \item Quizzes are distinct from tests or exams, often done in class with iClicker or asynchronously on Canvas.
            \item If quiz and exam is lumped into one bucket, use exam categories.
        \end{itemize}

        \item \texttt{presentation}: Any synchronous or asynchronous presentation, could be video turned in online.

        \item \texttt{project\_present}: Any projects with presentation or oral component, or where something must be physically built (engineering or architecture or art). If it mentions how much of grade goes to presentation vs.\ deliverable, put that in Notes column.

        \item \texttt{project\_no\_present}: Any projects without presentation or oral component; only deliverable like report or digital art piece/design.

        \item \texttt{performance}:
        \begin{itemize}
            \item 1) Tasks that must be performed like music, dance, or theatre performance
            \item 2) Tasks that are performed by hand, such as collecting lab data, doing experiments at lab
            \item 3) Other live skills demonstrations like in fitness-based classes (e.g., swimming, dribbling a basketball)
        \end{itemize}
    \end{itemize}
\end{itemize}

\medskip
\noindent\textbf{Other notes}\\
Some language course syllabi are wholly or mostly in non-English languages. Annotators were instructed to take what they could gather from context and English portions, but not use a translator.

\end{quote}
\noindent\rule{\linewidth}{0.6pt}
\bigskip

\subsection{LLM Prompts}
\label{app:prompts}

Below is the full prompt used to reconstruct syllabus information, including final grade weightings by assessment category. Its instructions are more detailed than the human codebook because the team refined its shared understanding of the task through discussion. For transparency and reproducibility, we retain fields irrelevant to this study. 

\subsubsection*{Main Prompt}

The system message below instructs the model on field extraction rules, category decision logic, and a self-verification checklist. Few-shot examples (shown in the next subsection) are inserted immediately after the opening instructions.

\bigskip
\noindent\rule{\linewidth}{0.6pt}
\begin{quote}
\singlespacing\footnotesize

You are an expert at extracting structured data from university course syllabi.
You will use step-by-step reasoning for ambiguous fields before writing final values.

\medskip
\noindent IMPORTANT:
\begin{itemize}
  \item Study the examples carefully.
  \item Mentally correct any OCR errors.
  \item Extract ALL fields. Use null for fields not found.
\end{itemize}

\noindent\textit{[Few-shot examples are inserted here; see the Few-Shot Examples subsection below.]}

\medskip
\noindent\textbf{FIELDS TO EXTRACT:}

\begin{enumerate}
\item \texttt{course\_title} (str): Full course title.
\item \texttt{course\_code} (str): Course code and section number (e.g., ``STATS401-001'').
\item \texttt{all\_instructors} (list of str): Faculty instructor names ONLY. Do NOT include GSIs or IAs.
\item \texttt{instructor\_email} (list of str): Faculty instructor emails. Count should match \texttt{all\_instructors}.
\item \texttt{crosslists} (list of str): Cross-listed course codes. Format: lowercase, no spaces (e.g., ``cs101'', ``stats250-002'').
\item \texttt{prerequisites} (list of str): Prerequisite course codes. Same format. Exclude non-course prerequisites like ``Senior Standing''.
\item \texttt{num\_GSI} (int): Number of Graduate Student Instructors. GSIs hold office hours or lead discussion sections.
\item \texttt{num\_IA} (int): Number of Instructional Assistants/Graders. IAs grade and do admin work. Default to IA if role is unclear.
\item \texttt{num\_OH} (float): Total weekly office hours (faculty + GSI combined). Do NOT count ``by appointment'' hours.
\item \texttt{learning\_objs\_text} (str): Copy-paste the learning objectives section. Search for ``objectives'', ``outcomes'', ``goals'', ``aims'', or ``by the end of this course''. If none, use the course description.
\item \texttt{in\_class\_tech\_policy} (str): MUST be exactly one of: \texttt{not\_mention}, \texttt{tech\_required}, \texttt{tech\_prohibited}, \texttt{tech\_limited}, \texttt{other}.
    \begin{itemize}
        \item \texttt{tech\_required}: Technology required for class activities or exams, or fully virtual lectures. iClicker alone does NOT count.
        \item \texttt{tech\_prohibited}: All technology banned during lectures or exams.
        \item \texttt{tech\_limited}: Some tech allowed in some contexts, not others (e.g., laptops OK for notes, not exams).
        \item \texttt{other}: Tech policy exists but does not fit the categories above.
        \item \texttt{not\_mention}: No technology policy found.
    \end{itemize}
\item \texttt{assessment\_type\_weighting} (dict): Map each assessment component to EXACTLY ONE of the 10 categories. Use the exact numbers written in the syllabus --- could be percentages (e.g., 25\%) OR raw points (e.g., 150 pts). Do NOT convert or normalize. Use 0 for categories not present.

    \medskip
    \noindent CATEGORY DECISION RULES:
    \begin{itemize}
        \item \texttt{closed\_exam}: Closed-book/notes, proctored in-person, or lockdown-browser exams. Review sheets allowed = still closed. In-person exam with unspecified open/closed policy = \texttt{closed\_exam} (in-person default).
        \item \texttt{open\_exam}: Take-home exams, open-book/notes exams, exams with a due-date range (e.g., ``due Friday''), unproctored Canvas exams. RULE: if \texttt{exam\_location} = ``take\_home'', the take-home portion (typically the FINAL exam) MUST go to \texttt{open\_exam} --- \texttt{open\_exam} must be non-zero. A separate in-class midterm in the same course may still be \texttt{closed\_exam}. Never put a take-home exam in \texttt{closed\_exam} or \texttt{unknown\_exam}.
        \item \texttt{unknown\_exam}: Exam mentioned but format is truly indeterminate --- no location, no due-date, no open/closed signal. Use ONLY when you genuinely cannot tell whether in-person or take-home. Do NOT use for take-home exams.
        \item \texttt{homework}: Short, recurring assignments --- problem sets, worksheets, reading responses, online quizzes. Key: ROUTINE and individually SMALL.
        \item \texttt{papers}: Longer written work --- essays, term papers, research papers. Key: SUBSTANTIAL individual writing, often turned in once or twice.
        \item \texttt{quiz\_participation}: Attendance, in-class participation, iClicker, pop quizzes (distinct from scheduled exams). If ``quiz + exam'' are lumped without a clear split, use the exam category.
        \item \texttt{presentation}: Standalone presentations (in-person or video) not tied to a larger project.
        \item \texttt{project\_present}: Projects WITH a presentation, oral defense, or physical/performance component.
        \item \texttt{project\_no\_present}: Projects WITHOUT presentation --- deliverable only (reports, code, digital designs).
        \item \texttt{performance}: Physical or artistic performances ONLY --- music recitals, dance routines, theatrical roles, athletic skill demonstrations (e.g., Kinesiology ``Skill Performance''), hands-on lab practicals (experiments students physically perform). Examples: ``Skill 1 Performance'' (CPR demonstration), ``Lab Practicum,'' ``Piano Recital,'' ``Dance Performance.''

        CRITICAL --- do NOT use \texttt{performance} for:
            \begin{itemize}
                \item Phrases like ``exam performance'' or ``course performance'' (these describe how well a student did, NOT a type of assessment)
                \item Regular class participation, iClicker, attendance ($\rightarrow$ \texttt{quiz\_participation})
                \item Lab reports or write-ups ($\rightarrow$ \texttt{papers})
                \item Lab attendance ($\rightarrow$ \texttt{quiz\_participation})
            \end{itemize}
        \item LABS: lab writeup $\rightarrow$ \texttt{papers}; physically performing a lab task $\rightarrow$ \texttt{performance}; lab attendance $\rightarrow$ \texttt{quiz\_participation}.
    \end{itemize}

\item[12a.] \texttt{assessment\_inventory} (str): REQUIRED structured inventory --- complete ALL 4 STEPS before filling \texttt{assessment\_type\_weighting}. \textit{[See Reasoning Protocol subsection below for full text of this field.]}

\item \texttt{exam\_location} (str): MUST be exactly one of: \texttt{in\_class}, \texttt{take\_home}, \texttt{not\_listed}, or null.
    \begin{itemize}
        \item \texttt{in\_class}: Exams held in person during scheduled class time. ``Location: TBD'' implies in\_class.
        \item \texttt{take\_home}: Take-home exams OR online exams taken asynchronously (including due-date-range exams). If ANY exam has a take-home format, use ``take\_home''.
        \item \texttt{not\_listed}: Exams exist but location or format cannot be determined.
        \item null: No exams (consistent with 0 for \texttt{closed\_exam}, \texttt{open\_exam}, and \texttt{unknown\_exam}).
    \end{itemize}
    LINK TO WEIGHTS: If \texttt{exam\_location} = ``take\_home'', \texttt{open\_exam} MUST be non-zero --- the take-home component (especially the final exam) goes to \texttt{open\_exam}. An in-class midterm in the same course may stay in \texttt{closed\_exam}.

\item[13a.] \texttt{exam\_reasoning} (str): Before filling \texttt{exam\_location}, quote the exact text about where/how exams occur, then state your classification and why. \textit{[See Reasoning Protocol subsection below.]}

\item \texttt{ai\_policy\_explicit} (str): ``yes'' or ``no''. Is there an explicit policy on generative AI (ChatGPT, Claude, DALL-E, Copilot, etc.)? GenAI policies are rare in syllabi before 2023.
\item \texttt{ai\_policy\_text} (str): Copy-paste the exact AI policy text verbatim. null if \texttt{ai\_policy\_explicit} = ``no''.
\item \texttt{ai\_policy\_location} (str): MUST be exactly one of: \texttt{Own\_Section}, \texttt{Plagiarism\_Acad\_Integrity}, \texttt{Other}, or null.
    \begin{itemize}
        \item \texttt{Own\_Section}: AI policy has its own dedicated section or heading.
        \item \texttt{Plagiarism\_Acad\_Integrity}: AI policy is embedded within a plagiarism or academic integrity section.
        \item \texttt{Other}: AI policy mentioned elsewhere (e.g., in assignments section, a brief note, or learning objectives).
        \item null: if \texttt{ai\_policy\_explicit} = ``no''.
    \end{itemize}
\item \texttt{ai\_policy\_type} (list of str): Select ALL that apply from exactly these 6 labels. null if \texttt{ai\_policy\_explicit} = ``no''.
    \begin{itemize}
        \item ``\texttt{Not\_Permitted}'': AI tools are explicitly banned for all coursework.
        \item ``\texttt{Permitted\_Disclosure}'': AI is allowed but students must acknowledge or disclose usage.
        \item ``\texttt{Permitted\_WriteUp}'': AI is allowed but requires a detailed appendix describing prompts used and how outputs were incorporated.
        \item ``\texttt{Permitted\_CertainTasks}'': AI is allowed for specific tasks only (e.g., brainstorming, editing, not final drafts).
        \item ``\texttt{Fully\_Permitted}'': AI use is unrestricted.
        \item ``\texttt{Use\_Required}'': AI use is mandatory for at least one assignment.
    \end{itemize}
\item[17a.] \texttt{ai\_policy\_reasoning} (str): \textit{[See Reasoning Protocol subsection below.]}
\item \texttt{notes} (str): Unusual features only --- non-English text, Pass/No Pass grading, grade contradictions, significant OCR issues. 1--3 short phrases. null if nothing unusual.
\end{enumerate}

\medskip
\noindent\textbf{SELF-VERIFICATION} (check mentally before outputting):
\begin{itemize}
    \item Does \texttt{assessment\_type\_weighting} contain all 10 category names listed above?
    \item If \texttt{assessment\_inventory} STEP 1 is ``SOURCE: none'', are ALL 10 \texttt{assessment\_type\_weighting} values 0?
    \item Does the sum in \texttt{assessment\_inventory} STEP 3 roughly match the non-zero values in \texttt{assessment\_type\_weighting}?
    \item Is \texttt{exam\_location} consistent with whether any exam categories have non-zero weights?
    \item If \texttt{exam\_location} = ``take\_home'', is \texttt{open\_exam} non-zero? (The take-home portion must be in \texttt{open\_exam}; a separate in-class midterm may stay in \texttt{closed\_exam}.)
    \item Have I avoided using \texttt{performance} for any participation, attendance, or written assessment?
    \item If \texttt{ai\_policy\_explicit} = ``no'', are \texttt{ai\_policy\_location}, \texttt{ai\_policy\_type}, and \texttt{ai\_policy\_text} all null?
    \item Is \texttt{exam\_location} one of the 4 exact allowed values?
    \item Are all \texttt{ai\_policy\_type} labels spelled exactly as listed above?
\end{itemize}

\end{quote}
\noindent\rule{\linewidth}{0.6pt}
\bigskip

\subsubsection*{Reasoning Protocol}

Three fields in the main prompt elicit step-by-step reasoning before final values are recorded. These reasoning fields are stripped from the stored annotation after processing.

\bigskip
\noindent\rule{\linewidth}{0.6pt}
\begin{quote}
\singlespacing\footnotesize

\noindent\textbf{Field 12a.} \texttt{assessment\_inventory} (str): REQUIRED structured inventory --- complete ALL 4 STEPS before filling \texttt{assessment\_type\_weighting}.

\medskip
STEP 1 --- SOURCE TYPE: Identify how grade weights are presented in this syllabus. Write exactly one of:
\begin{itemize}
    \item ``SOURCE: table'' \quad if a clear grade table, schedule, or bulleted list of grade components exists.
    \item ``SOURCE: narrative'' \quad if weights are described only in running text or paragraphs (no table).
    \item ``SOURCE: mixed'' \quad if both a table AND narrative descriptions of grades are present.
    \item ``SOURCE: none'' \quad if NO grade breakdown or weighting is present anywhere in the syllabus.
\end{itemize}
CRITICAL: If SOURCE is ``none'', ALL 10 \texttt{assessment\_type\_weighting} values MUST be 0. Do NOT invent or infer weights that are not explicitly stated.

\medskip
STEP 2 --- COMPONENT LIST: For SOURCE = table/narrative/mixed, list EVERY graded component found. Write each component on its own line in this exact format:

\medskip
\noindent\texttt{[Component name] | [Raw value exactly as written] | [Direct quote from syllabus, $\leq$10 words] | [Category]}

\medskip
\noindent Example lines:
\begin{itemize}
    \item[] \texttt{"Midterm Exam | 25\% | 'midterm is worth 25\% of your grade' | closed\_exam"}
    \item[] \texttt{"Weekly Problem Sets | 150 pts | 'HW: 150 points total, due Fridays' | homework"}
\end{itemize}
If SOURCE is ``none'', write: ``(no components found)''

\medskip
STEP 3 --- SUM CHECK: Add up all raw values from STEP 2.
\begin{itemize}
    \item Write: ``Total: [sum]\%'' for percentage-based syllabi, or ``Total: [sum] pts'' for point-based.
    \item If total is not approximately 100\% for percentage-based syllabi (or is otherwise inconsistent), write: ``Discrepancy: [brief note]''
    \item A sum between 90--110\% is acceptable; flag anything outside that range.
\end{itemize}

STEP 4 --- AMBIGUITY NOTES: Flag any categorization decisions that required judgment. Write one line per ambiguous case, e.g.:
\begin{itemize}
    \item[] \texttt{"Take-home midterm -> open\_exam (take-home = open by rule; exam\_location = take\_home)"}
    \item[] \texttt{"Final Exam due date range listed -> open\_exam, not unknown\_exam (due date = take-home signal)"}
    \item[] \texttt{"Lab Practicum -> performance (hands-on physical task, not a written lab report)"}
    \item[] \texttt{"'Skill Performance' -> performance (Kinesiology physical skill demonstration, not a score)"}
    \item[] \texttt{"'exam performance' in intro text -> IGNORED (describes grading context, not an assessment component)"}
\end{itemize}
Write ``None'' if all assignments were straightforward to categorise.

\bigskip
\noindent\textbf{Field 13a.} \texttt{exam\_reasoning} (str): Before filling \texttt{exam\_location}, quote the exact text about where/how exams occur, then state your classification and why.

\bigskip
\noindent\textbf{Field 17a.} \texttt{ai\_policy\_reasoning} (str): If \texttt{ai\_policy\_explicit} = ``yes'', for EACH of the 6 type labels below, independently state YES or NO and give a one-sentence justification based on the quoted policy text:

\medskip
\noindent \texttt{Not\_Permitted} $|$ \texttt{Permitted\_Disclosure} $|$ \texttt{Permitted\_WriteUp} $|$ \texttt{Permitted\_CertainTasks} $|$ \texttt{Fully\_Permitted} $|$ \texttt{Use\_Required}

\medskip
If \texttt{ai\_policy\_explicit} = ``no'', set this to null.

\end{quote}
\noindent\rule{\linewidth}{0.6pt}
\bigskip

\subsubsection*{Few-Shot Examples}

The following seven examples are embedded in the system message immediately after the opening instructions. They illustrate how human annotators map assessment text to the 10 categories, with values extracted as-is (raw points or percentages, without normalization).

\bigskip
\noindent\rule{\linewidth}{0.6pt}
\begin{quote}
\singlespacing\footnotesize

\noindent\textbf{Example 1: Points-based grading (not percentages)}

\medskip
\noindent Assessment text from syllabus:
\begin{quote}
``There are a total of 375 points available in the course. Attendance 25 $|$ Participation 25 $|$ Response Notes 50 $|$ Mid-Term Exam 100 $|$ Final Exam 100 $|$ Three Revised Response Note Essays 75''
\end{quote}

\noindent Decision notes:
\begin{itemize}
    \item Attendance 25 + Participation 25 pts $\rightarrow$ \texttt{quiz\_participation}: 50
    \item Response Notes 50 pts $\rightarrow$ \texttt{homework}: 50 (recurring short written responses)
    \item Mid-Term 100 + Final 100 = 200 pts $\rightarrow$ \texttt{closed\_exam}: 200 (format not stated $\rightarrow$ \texttt{exam\_location}: not\_listed)
    \item Three Revised Essays 75 pts $\rightarrow$ \texttt{papers}: 75 (substantial revised writing projects)
    \item NOTE: Sum = 375 (raw point totals). Do NOT normalize to 100.
\end{itemize}

\noindent Correct output:
\begin{quote}\ttfamily\scriptsize
\{"assessment\_type\_weighting": \{"closed\_exam": 200, "open\_exam": 0, "unknown\_exam": 0, "homework": 50, "papers": 75, "quiz\_participation": 50, "presentation": 0, "project\_present": 0, "project\_no\_present": 0, "performance": 0\}, "exam\_location": "not\_listed"\}
\end{quote}

\bigskip
\noindent\textbf{Example 2: \texttt{homework} vs.\ \texttt{papers} distinction}

\medskip
\noindent Assessment text from syllabus:
\begin{quote}
``(1) Journal [50\%] \quad (2) Term Paper [50\%]''
\end{quote}

\noindent Decision notes:
\begin{itemize}
    \item Journal 50\% $\rightarrow$ \texttt{homework}: 50 (recurring journal entries = short, routine writing)
    \item Term Paper 50\% $\rightarrow$ \texttt{papers}: 50 (substantial single writing project)
    \item KEY RULE: ``journal'' / ``weekly response'' / ``reading notes'' = \texttt{homework}. ``term paper'' / ``essay'' / ``research paper'' = \texttt{papers}.
    \item No exams $\rightarrow$ \texttt{exam\_location}: null
\end{itemize}

\noindent Correct output:
\begin{quote}\ttfamily\scriptsize
\{"assessment\_type\_weighting": \{"closed\_exam": 0, "open\_exam": 0, "unknown\_exam": 0, "homework": 50, "papers": 50, "quiz\_participation": 0, "presentation": 0, "project\_present": 0, "project\_no\_present": 0, "performance": 0\}, "exam\_location": null\}
\end{quote}

\bigskip
\noindent\textbf{Example 3: Multi-category aggregation + exam type disambiguation}

\medskip
\noindent Assessment text from syllabus:
\begin{quote}
``Attendance \& Participation 15\% $|$ Homework (including AMDs) 15\% $|$ Essays and Rewrites 15\% $|$ Final Video Project 10\% $|$ Diagnostic Grammar Quizzes 5\% $|$ Quizzes \& Mini-Quizzes 20\% $|$ Tests including Gateway Vocabulary Test 15\% $|$ Oral Exams 5\%''
\end{quote}

\noindent Decision notes:
\begin{itemize}
    \item Attendance \& Participation 15\% + Grammar Quizzes 5\% + Quizzes \& Mini-Quizzes 20\% $\rightarrow$ \texttt{quiz\_participation}: 40
    \item Homework 15\% $\rightarrow$ \texttt{homework}: 15
    \item Essays and Rewrites 15\% $\rightarrow$ \texttt{papers}: 15
    \item Final Video Project 10\% $\rightarrow$ \texttt{project\_present}: 10 (project with a recorded video component)
    \item Tests 15\% $\rightarrow$ \texttt{open\_exam}: 15 (course uses take-home tests)
    \item Oral Exams 5\% $\rightarrow$ \texttt{closed\_exam}: 5 (in-person oral examination)
    \item \texttt{exam\_location} = ``take\_home'' (written tests are take-home)
\end{itemize}

\noindent Correct output:
\begin{quote}\ttfamily\scriptsize
\{"assessment\_type\_weighting": \{"closed\_exam": 5, "open\_exam": 15, "unknown\_exam": 0, "homework": 15, "papers": 15, "quiz\_participation": 40, "presentation": 0, "project\_present": 10, "project\_no\_present": 0, "performance": 0\}, "exam\_location": "take\_home"\}
\end{quote}

\bigskip
\noindent\textbf{Example 4: Table layout + multi-label AI policy}

\medskip
\noindent Assessment text from syllabus:
\begin{quote}
``Lab \quad Homework \quad Exam 1 \quad Exam 2 \quad Project\\
20\% \qquad 20\% \qquad\ 20\% \qquad 20\% \qquad 20\%''
\end{quote}

\noindent Decision notes:
\begin{itemize}
    \item Lab 20\% + Homework 20\% $\rightarrow$ \texttt{homework}: 40 (Lab = structured homework in a stats course)
    \item Exam 1 20\% + Exam 2 20\% $\rightarrow$ \texttt{closed\_exam}: 40 (in-class exams, proctored)
    \item Project 20\% $\rightarrow$ \texttt{project\_no\_present}: 20 (deliverable-only, no presentation)
    \item AI policy is present: \texttt{Permitted\_Disclosure} + \texttt{Fully\_Permitted} (embedded in academic integrity section)
\end{itemize}

\noindent Correct output:
\begin{quote}\ttfamily\scriptsize
\{"assessment\_type\_weighting": \{"closed\_exam": 40, "open\_exam": 0, "unknown\_exam": 0, "homework": 40, "papers": 0, "quiz\_participation": 0, "presentation": 0, "project\_present": 0, "project\_no\_present": 20, "performance": 0\}, "exam\_location": "in\_class", "ai\_policy\_explicit": "yes", "ai\_policy\_location": "Plagiarism\_Acad\_Integrity", "ai\_policy\_type": ["Permitted\_Disclosure", "Fully\_Permitted"]\}
\end{quote}

\bigskip
\noindent\textbf{Example 5: No exams + mixed formatting}

\medskip
\noindent Assessment text from syllabus:
\begin{quote}
``Participation 30\% \quad 2 in-term assignments 40\% \quad Final project 30'' \quad [no \% sign on last entry]
\end{quote}

\noindent Decision notes:
\begin{itemize}
    \item Participation 30\% $\rightarrow$ \texttt{quiz\_participation}: 30
    \item 2 in-term assignments 40\% $\rightarrow$ \texttt{homework}: 40 (recurring assignments, not a single major paper)
    \item Final project 30 (no \%) $\rightarrow$ \texttt{project\_no\_present}: 30 (deliverable-only, extract the number as-is)
    \item No exams $\rightarrow$ \texttt{exam\_location}: null
    \item NOTE: Extract ``30'' even without \% sign; the missing sign is a formatting inconsistency.
\end{itemize}

\noindent Correct output:
\begin{quote}\ttfamily\scriptsize
\{"assessment\_type\_weighting": \{"closed\_exam": 0, "open\_exam": 0, "unknown\_exam": 0, "homework": 40, "papers": 0, "quiz\_participation": 30, "presentation": 0, "project\_present": 0, "project\_no\_present": 30, "performance": 0\}, "exam\_location": null\}
\end{quote}

\bigskip
\noindent\textbf{Example 6: Take-home exams $\rightarrow$ \texttt{open\_exam} (not \texttt{closed\_exam} or \texttt{unknown\_exam})}

\medskip
\noindent Assessment text from syllabus:
\begin{quote}
``Grades: Class attendance and participation, including Canvas assignments: 20\%; Take-home midterm exam (due week 7): 20\%; One short essay (due week 12): 20\%; Take home final exam (due December 14): 40\%''
\end{quote}

\noindent Decision notes:
\begin{itemize}
    \item Class attendance + Canvas assignments 20\% $\rightarrow$ \texttt{quiz\_participation}: 20
    \item Take-home midterm 20\% + Take home final 40\% = 60\% $\rightarrow$ \texttt{open\_exam}: 60. KEY RULE: Any exam described as ``take-home'', given a due-date range, or completed outside class goes to \texttt{open\_exam} --- never \texttt{closed\_exam} or \texttt{unknown\_exam}. When \texttt{exam\_location} = take\_home, \texttt{open\_exam} must be non-zero. (A separate in-class midterm in the same course may stay in \texttt{closed\_exam}.)
    \item One short essay 20\% $\rightarrow$ \texttt{papers}: 20 (substantial written work due once)
    \item \texttt{exam\_location} = ``take\_home''
\end{itemize}

\noindent Correct output:
\begin{quote}\ttfamily\scriptsize
\{"assessment\_type\_weighting": \{"closed\_exam": 0, "open\_exam": 60, "unknown\_exam": 0, "homework": 0, "papers": 20, "quiz\_participation": 20, "presentation": 0, "project\_present": 0, "project\_no\_present": 0, "performance": 0\}, "exam\_location": "take\_home"\}
\end{quote}

\bigskip
\noindent\textbf{Example 7: Physical skill demonstrations $\rightarrow$ \texttt{performance}; take-home exam $\rightarrow$ \texttt{open\_exam}}

\medskip
\noindent Assessment text from syllabus (Department: School of Kinesiology):
\begin{quote}
``Quizzes 130 points; Skill Guide 40 points; Skill 1 Performance 30 points; Skill 2 Performance 10 points; Presentation 1 25 points; Presentation 2 25 points; Paper 50 points; Participation 30 points; Final Exam 100 points [take-home, submitted online]''
\end{quote}

\noindent Decision notes:
\begin{itemize}
    \item Quizzes 130 + Participation 30 = 160 pts $\rightarrow$ \texttt{quiz\_participation}: 160
    \item Skill Guide 40 pts $\rightarrow$ \texttt{homework}: 40 (written study guide, not the physical skill itself)
    \item Skill 1 Performance 30 + Skill 2 Performance 10 = 40 pts $\rightarrow$ \texttt{performance}: 40. KEY: `Performance' here = demonstrating a physical skill (e.g., CPR, first-aid) in front of instructor. Kinesiology skill demonstrations = \texttt{performance}. This is NOT a test score or course grade.
    \item Presentation 1 + 2 = 50 pts $\rightarrow$ \texttt{presentation}: 50
    \item Paper 50 pts $\rightarrow$ \texttt{papers}: 50
    \item Final Exam 100 pts (take-home) $\rightarrow$ \texttt{open\_exam}: 100 (take-home $\rightarrow$ \texttt{open\_exam})
    \item Total: 160 + 40 + 40 + 50 + 50 + 100 = 440 pts (point-based, do NOT normalize)
    \item \texttt{exam\_location} = ``take\_home''
\end{itemize}

\noindent Correct output:
\begin{quote}\ttfamily\scriptsize
\{"assessment\_type\_weighting": \{"closed\_exam": 0, "open\_exam": 100, "unknown\_exam": 0, "homework": 40, "papers": 50, "quiz\_participation": 160, "presentation": 50, "project\_present": 0, "project\_no\_present": 0, "performance": 40\}, "exam\_location": "take\_home"\}
\end{quote}

\end{quote}
\noindent\rule{\linewidth}{0.6pt}
\bigskip

\subsubsection*{Format Instruction and Output Validation}

The format instruction is sent as an assistant-role message immediately following the system message, prompting the model to return only valid JSON. If the output fails schema validation, a correction prompt is issued listing the specific constraint violations; the model is then asked to return only the corrected JSON.

\bigskip
\noindent\textbf{Format instruction (assistant turn):}

\noindent\rule{\linewidth}{0.6pt}
\begin{quote}
\singlespacing\footnotesize

Output ONLY valid JSON. Include \texttt{assessment\_inventory}, \texttt{exam\_reasoning}, and \texttt{ai\_policy\_reasoning} fields. Use 0 for absent assessment categories. Values are raw numbers --- do not normalize.

\medskip
\noindent Example:
\begin{quote}\ttfamily\scriptsize
\{"course\_title": "American Politics", "course\_code": "POLSCI101-001", "all\_instructors": ["Dr. Smith"], "instructor\_email": ["smith@[institution].edu"],\textsuperscript{$\dagger$} "crosslists": [], "prerequisites": [], "num\_GSI": 1, "num\_IA": 0, "num\_OH": 3.0, "learning\_objs\_text": "Students will understand...", "in\_class\_tech\_policy": "not\_mention", "assessment\_inventory": "SOURCE: narrative\textbackslash nMidterm | 25\% | 'midterm worth 25\%' | closed\_exam\textbackslash nFinal | 40\% | 'final exam 40\%' | closed\_exam\textbackslash nWeekly HW | 20\% | 'homework 20\%' | homework\textbackslash nTerm Paper | 15\% | 'paper due Week 12' | papers\textbackslash nTotal: 100\%\textbackslash nAmbiguity: None", "assessment\_type\_weighting": \{"closed\_exam": 65, "open\_exam": 0, "unknown\_exam": 0, "homework": 20, "papers": 15, "quiz\_participation": 0, "presentation": 0, "project\_present": 0, "project\_no\_present": 0, "performance": 0\}, "exam\_reasoning": "Syllabus: 'exams held in assigned classroom'. -> in\_class.", "exam\_location": "in\_class", "ai\_policy\_explicit": "no", "ai\_policy\_text": null, "ai\_policy\_location": null, "ai\_policy\_reasoning": null, "ai\_policy\_type": null, "notes": null\}
\end{quote}

\end{quote}
\noindent\rule{\linewidth}{0.6pt}

\medskip
\noindent\footnotesize{$\dagger$ The original prompt contained an institution-identifying email domain, anonymized here.}

\normalsize
\bigskip
\noindent\textbf{Correction prompt (system turn, issued on validation failure):}

\noindent\rule{\linewidth}{0.6pt}
\begin{quote}
\singlespacing\footnotesize

Your previous extraction had the following constraint violations:
\begin{itemize}
    \item[$-$] [violation list from automated schema check]
\end{itemize}

Please correct ONLY the violated fields and return the complete JSON.

\medskip
\noindent Original output: [original JSON]

\medskip
\noindent Constraints:
\begin{itemize}
    \item \texttt{assessment\_type\_weighting} must have exactly these 10 keys: \texttt{closed\_exam}, \texttt{open\_exam}, \texttt{unknown\_exam}, \texttt{homework}, \texttt{papers}, \texttt{quiz\_participation}, \texttt{presentation}, \texttt{project\_present}, \texttt{project\_no\_present}, \texttt{performance}
    \item All assessment weight values must be numeric (use 0 for absent categories)
    \item \texttt{exam\_location}: \texttt{in\_class} $|$ \texttt{take\_home} $|$ \texttt{not\_listed} $|$ null
    \item \texttt{in\_class\_tech\_policy}: \texttt{not\_mention} $|$ \texttt{tech\_required} $|$ \texttt{tech\_prohibited} $|$ \texttt{tech\_limited} $|$ \texttt{other}
    \item \texttt{ai\_policy\_location}: \texttt{Own\_Section} $|$ \texttt{Plagiarism\_Acad\_Integrity} $|$ \texttt{Other} $|$ null
    \item \texttt{ai\_policy\_type} labels (exact spelling): \texttt{Not\_Permitted}, \texttt{Permitted\_Disclosure}, \texttt{Permitted\_WriteUp}, \texttt{Permitted\_CertainTasks}, \texttt{Fully\_Permitted}, \texttt{Use\_Required}
    \item If \texttt{ai\_policy\_explicit}=``no'': \texttt{ai\_policy\_type}, \texttt{ai\_policy\_location}, \texttt{ai\_policy\_text} must all be null
\end{itemize}

\medskip
\noindent The requery assistant turn reads: ``Return ONLY the corrected JSON.''

\end{quote}
\noindent\rule{\linewidth}{0.6pt}
\bigskip

\FloatBarrier
\section{Robustness: Treatment-Measure Comparisons} 
\label{app:robustness-2019}

The main specification measures GenAI susceptibility from a course's 2019 offerings. A natural alternative is to measure it as close to ChatGPT's release as possible, from 2022 offerings. Consistent with separating the GenAI effect from COVID effects, we argue that measures drawing on COVID-affected offerings contaminate the comparison between high- and low-susceptibility courses. The 2019 measure and the pre-COVID average correlate at 0.95, while the 2022 measure and the COVID-period average correlate with the 2019 measure at only 0.74 and 0.71. Pandemic adaptation makes some courses appear more susceptible than their stable design, and we cannot know how their susceptibility would have evolved after the pandemic in the absence of GenAI. To test whether our results depend on the measure, we re-estimate Equation~\ref{eq:main} with four alternatives: 2022 offerings and averages over all offerings in the pre-COVID period (2015 to 2019), the full pre-GenAI period (2015 to 2022), and the COVID-affected period (2020 to 2022).

Table~\ref{tab:treatment_robustness} reports the average grade estimates under each measure, each on its own balanced sample. The transient-COVID estimate follows the pattern the contamination argument predicts. It is small and not significant under the two pre-COVID measures: 0.010 (SE = 0.045) for the 2019 anchor and 0.014 (SE = 0.045) for the pre-COVID average. It grows under the COVID-affected measures, to 0.078 (SE = 0.048, n.s.) for the 2022 anchor and 0.115 (SE = 0.049, $p < 0.05$) for the COVID-period average. The full pre-GenAI average lies in between at 0.047 (SE = 0.048, n.s.).

The persistent-COVID estimates fall in a narrow band from 0.055 to 0.071 across all five measures, each significant at the 5\% level. The two estimates therefore disagree on statistical significance under every measure except the COVID-period average, as in the main specification. COVID-affected measures inflate the transient-COVID estimate, while the persistent-COVID estimate is largely insensitive to the measure.

\input{tables_and_figures_v2/table_treatment_robustness}

\FloatBarrier
\section{Robustness: Prior-Ability Proxy} 
\label{app:ability-robustness}

The main heterogeneity analysis uses the course-residualized within-cohort first-term GPA rank (\texttt{AbilityRank}, Section~\ref{sec:measure-abilityrank}). Table~\ref{tab:ability_robustness} shows that the flat, null tercile pattern holds under five alternative measures of prior academic preparation. Each involves trade-offs in coverage and construct validity, but together they give a consistent picture.

First, the non-residualized within-cohort first-term GPA rank yields the same flat, non-significant pattern, so the result is not an artifact of residualization. The raw rank compares students with their entire entering cohort, while \texttt{AbilityRank} compares them with coursemates in the same offerings. 

Second, the Fall 2020 cohort's first-term grades are distorted because that semester was mostly virtual and offered a pass/no-record option (Section~\ref{sec:measure-abilityrank}). Both checks for this distortion modify the raw rank. Dropping the cohort yields null estimates in every tercile, with persistent-COVID estimates slightly larger than under the raw rank (0.060 to 0.081 versus 0.049 to 0.057, none significant at the 5\% level). Our data also record the letter grades instructors assigned to students who elected pass/no-record. Retaining the cohort and recomputing first-term GPA with these letter grades also yields null estimates in every tercile.

Third, we replace \texttt{AbilityRank} with two pre-college measures: high school GPA and SAT/ACT scores. Neither is affected by course selection at the University. High school GPA is not standardized across high schools, a limitation that test scores address. Following \citet{friedmanStandardizedTestScores2025}, we convert ACT composites to the SAT scale with the official concordance tables \citep{actGuide2018ACTSAT2018}, convert SAT totals from before the March 2016 redesign to the current scale \citep{collegeboardConcordanceTables2016}, keep each student's highest converted score, and rank it within entering cohort. Both measures cover far fewer students than first-term GPA, which every student has, because reporting high school GPA and test scores was optional in some cases. Missingness is also more common among transfer and international students, so it is not random. Estimates under these measures are also null, with one exception: the persistent-COVID estimate for the bottom high school GPA tercile is 0.098 (SE = 0.041, $p < 0.05$). One estimate significant at the 5\% level among the 30 tercile estimates from the five alternative proxies is consistent with chance, and it does not survive a Bonferroni correction. No proxy yields an equalizing effect in the bottom tercile under both COVID assumptions.

\input{tables_and_figures_v2/table_ability_robustness}

\FloatBarrier
\section{Additional Main Results Tables and Figures}
\label{app:additional-main}

This appendix collects supporting tables and event-study figures for the results in Section~\ref{sec:results}, each in order of mention.

\subsection{Tables}

Table~\ref{tab:grade_distribution} reports the grade-distribution threshold estimates discussed in Section~\ref{sec:results-grade-dist}. Each column is a linear probability model for clearing the indicated letter-grade threshold (or, in the last column, not earning at least a D), with the same fixed effects, clustering, and COVID modeling as the main outcomes. The GenAI net COVID row reports the persistent-COVID estimate.

Table~\ref{tab:het_ability_terciles} reports the average grade estimates by \texttt{AbilityRank} tercile discussed in Section~\ref{sec:results-grades-het}.

Table~\ref{tab:course_evals} reports the full course-evaluation estimates discussed in Section~\ref{sec:results-evals}. The evaluation outcomes are offering-level medians, but the regressions use student-offering rows, so each offering is weighted by its enrollment. Collapsing to one observation per offering (30,073 offerings) gives unweighted estimates that support the same conclusions. The transient-COVID estimates are 0.040 (SE = 0.047) for understanding, 0.101 (SE = 0.060) for interest, and $-0.219$ (SE = 0.091) for relative workload. The persistent-COVID estimates are $-0.018$ (SE = 0.046), $-0.038$ (SE = 0.052), and 0.024 (SE = 0.049). Unweighted, understanding and interest show no significant change under either COVID assumption at the 5\% level, and the relative workload estimates still disagree across the two assumptions.

\input{tables_and_figures_v2/table_grade_distribution}

\input{tables_and_figures_v2/table_het_ability_terciles}

\input{tables_and_figures_v2/table_course_evals}

\subsection{Figures}

The remaining event studies follow Equation~\eqref{eq:event-study} at the semester level.

\begin{figure}[htbp]
\centering
\includegraphics[width=0.78\linewidth]{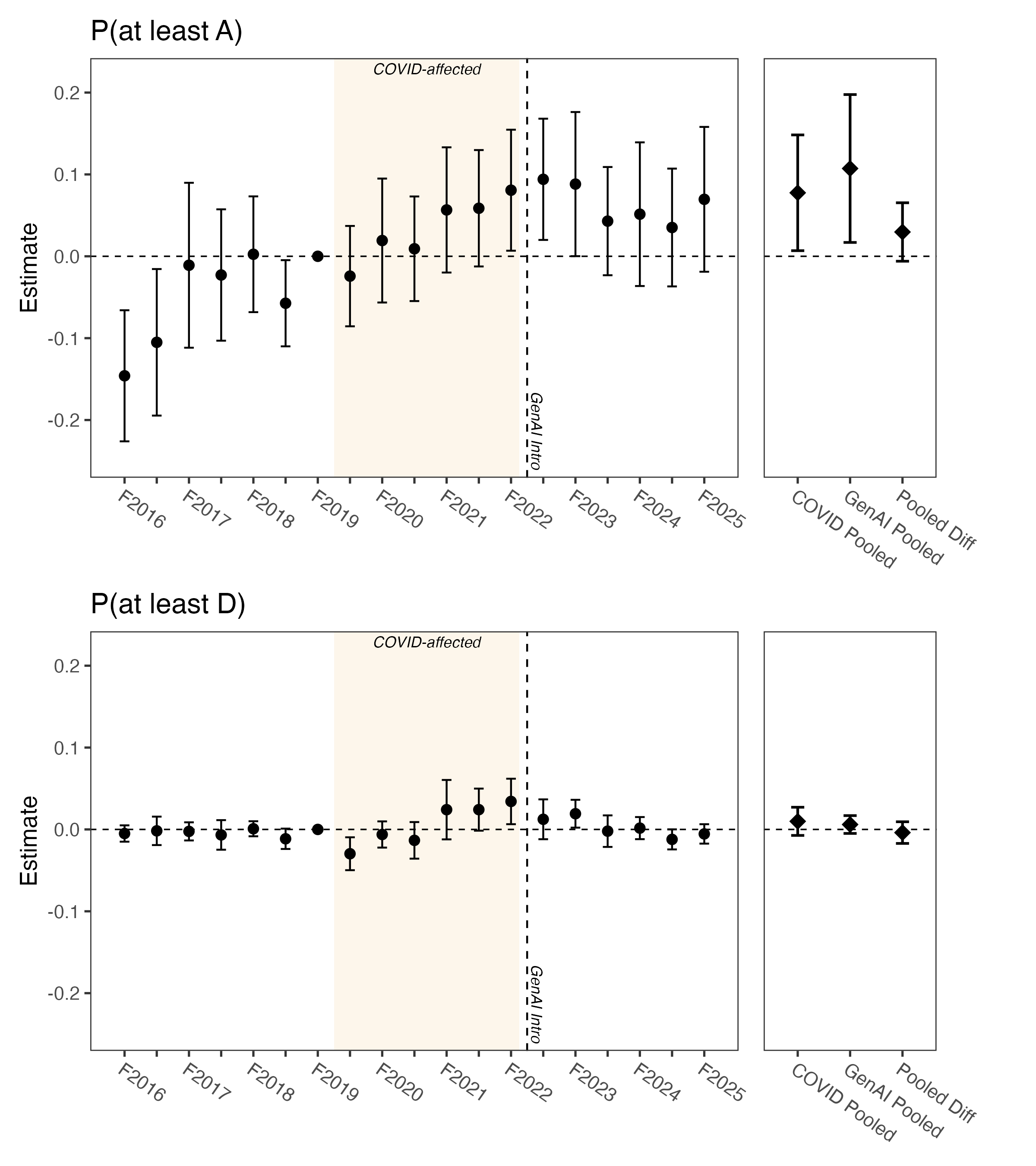}
\caption{Event Study: Grade-Distribution Thresholds. Left panels: coefficients on course susceptibility interacted with semester indicators (Equation~\ref{eq:event-study}) for earning at least an A (top) and at least a D (bottom), normalized to Fall 2019. Right panels: pooled coefficients for the COVID-affected terms (shaded band) and the post-GenAI terms from Equation~\ref{eq:main}, and their difference with a delta-method standard error. The post-GenAI coefficient is the transient-COVID estimate and the difference is the persistent-COVID estimate. Bars are 95 percent confidence intervals. Coefficients are per unit of susceptibility, the difference between fully susceptible and fully non-susceptible courses. Regressions include student, course, and semester fixed effects; standard errors are clustered by course.}
\label{fig:es-thresholds}
\end{figure}

Figure~\ref{fig:es-thresholds} plots the event studies for the top and bottom of the grade distribution: the probabilities of earning at least an A and at least a D. The A coefficients rise steadily from 2016 through the COVID period and level off after ChatGPT's release, while the D coefficients stay near zero. Because the rise at the A threshold predates ChatGPT, we interpret it as descriptive rather than causal.

\begin{figure}[htbp]
\centering
\includegraphics[width=0.78\linewidth]{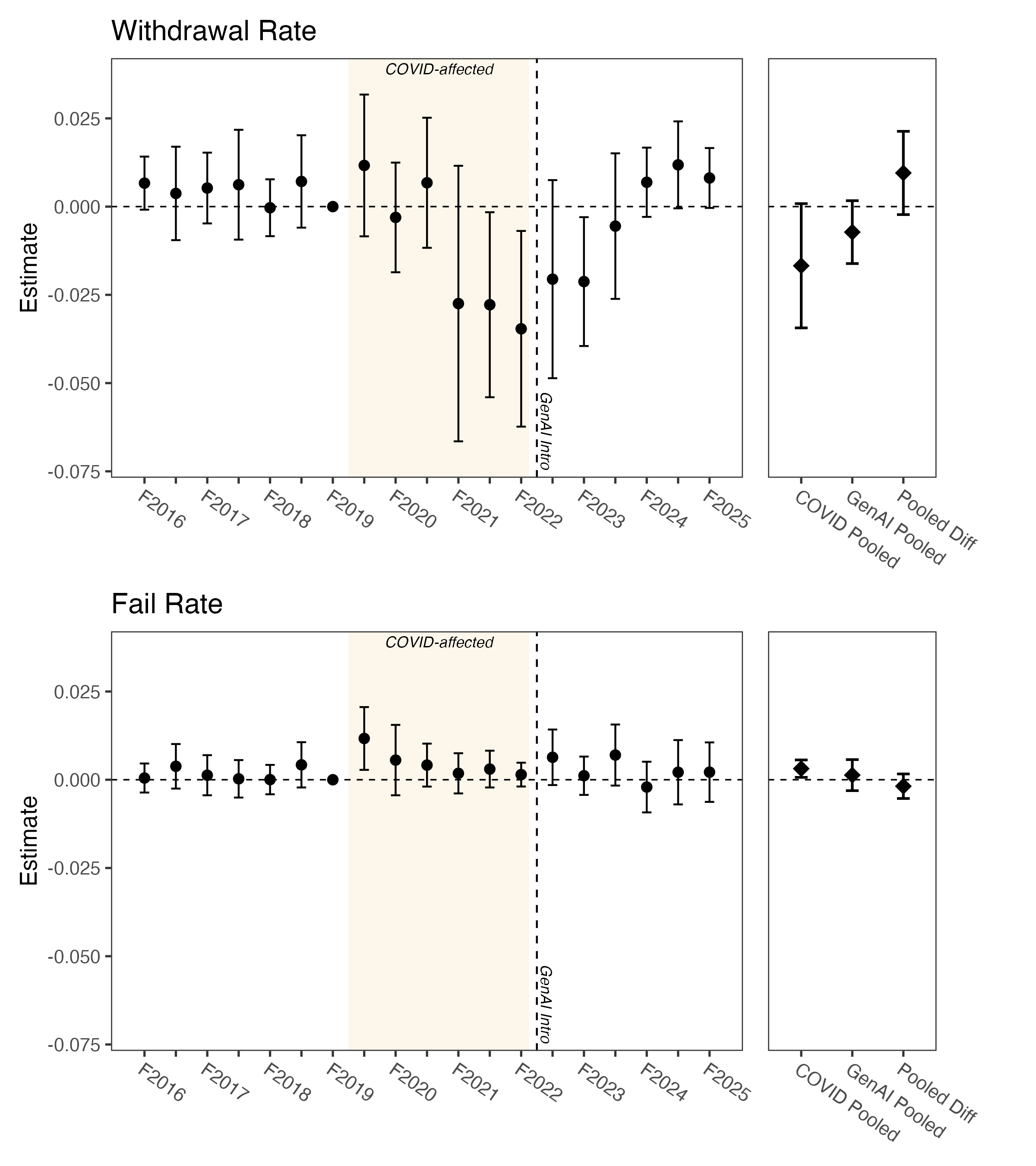}
\caption{Event Study: Withdrawal and Failure Rates. Left panels: coefficients on course susceptibility interacted with semester indicators (Equation~\ref{eq:event-study}) for withdrawal (top) and failure (bottom), normalized to Fall 2019. Right panels: pooled coefficients for the COVID-affected terms (shaded band) and the post-GenAI terms from Equation~\ref{eq:main}, and their difference with a delta-method standard error. The post-GenAI coefficient is the transient-COVID estimate and the difference is the persistent-COVID estimate. Bars are 95 percent confidence intervals. Coefficients are per unit of susceptibility, the difference between fully susceptible and fully non-susceptible courses. Regressions include student, course, and semester fixed effects; standard errors are clustered by course.}
\label{fig:es-withdraw-fail}
\end{figure}

\begin{figure}[htbp]
\centering
\includegraphics[width=0.65\linewidth]{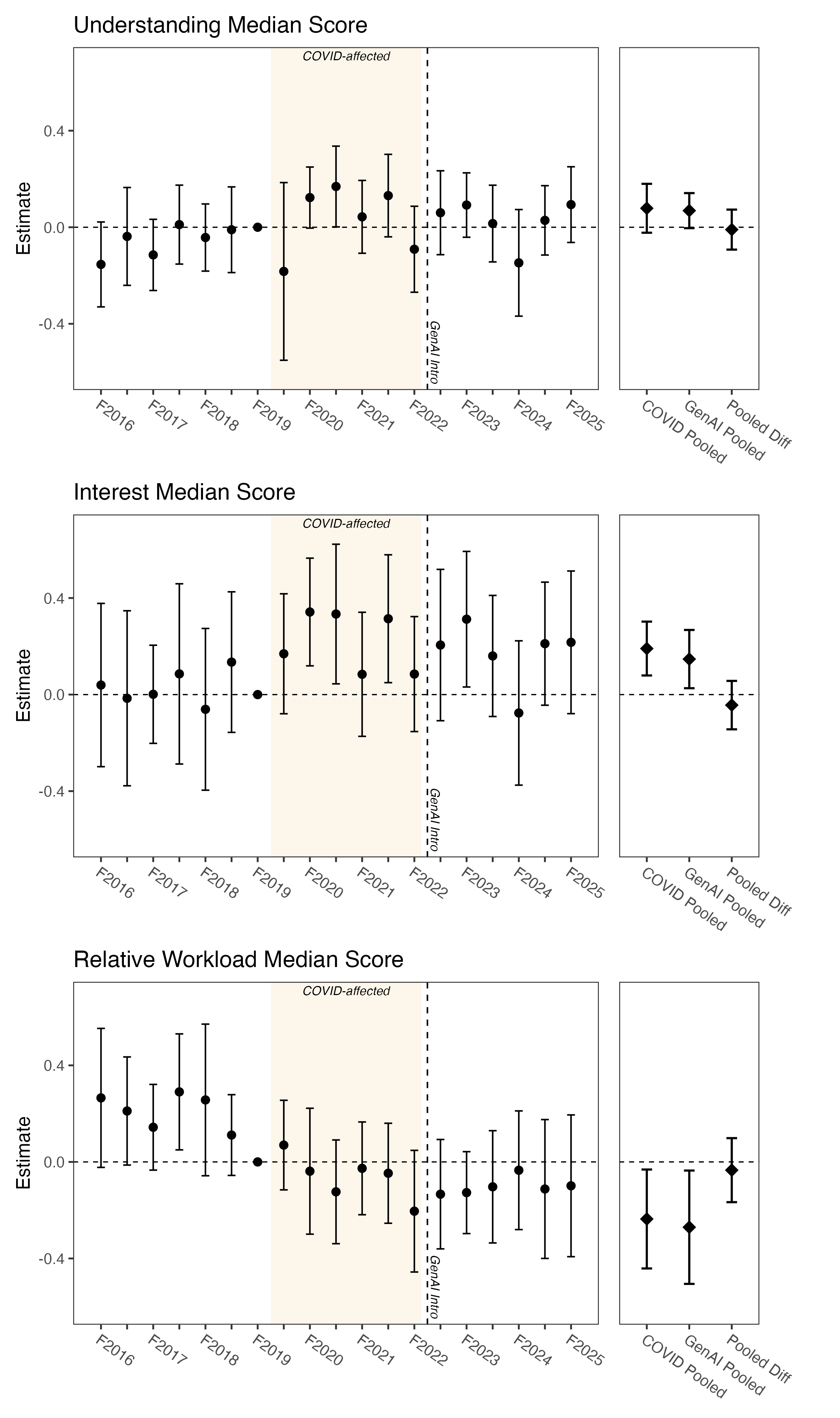}
\caption{Event Study: Course Evaluation Outcomes. Left panels: coefficients on course susceptibility interacted with semester indicators (Equation~\ref{eq:event-study}) for median self-reported understanding, interest, and relative workload (reverse-coded so that higher values indicate heavier workload), normalized to Fall 2019. Right panels: pooled coefficients for the COVID-affected terms (shaded band) and the post-GenAI terms from Equation~\ref{eq:main}, and their difference with a delta-method standard error. The post-GenAI coefficient is the transient-COVID estimate and the difference is the persistent-COVID estimate. Bars are 95 percent confidence intervals. Coefficients are per unit of susceptibility, the difference between fully susceptible and fully non-susceptible courses. Regressions include course and semester fixed effects; standard errors are clustered by course.}
\label{fig:es-evals}
\end{figure}

\FloatBarrier
\section{Sensitivity to Violations of Parallel Trends}
\label{app:honestdid}

As noted in Section~\ref{sec:methods}, tests for parallel pre-treatment trends are weak evidence for the parallel trends assumption, largely because they may be underpowered \citep{rothPretestCautionEventStudy2022}. For average grades, we assess the power of our pre-trend tests and bound the estimate under violations of parallel trends.

Following \citet{rothPretestCautionEventStudy2022}, a linear violation of parallel trends with slope 0.013 would bias our estimate by 0.121 grade points, yet our pre-trend tests would have only 50 percent power to detect it. A bias of this size matches the per-unit estimate in \citet{chirikovAIGradeInflation2026} (0.12) and exceeds the effects reported by \citet{hausmanGenerativeAIsImpact} and in this paper. 

\citet{rambachanMoreCredibleApproach2023} propose confidence sets that remain valid under bounded violations of parallel trends. Table~\ref{tab:honestdid-grade} reports them under three assumptions: no violation, a linear violation of arbitrary slope, and a violation as large as the largest observed in the pre-COVID period. To keep COVID effects out of the benchmark, we bound the violations using pre-COVID terms only. 

All three confidence sets include zero, consistent with the transient-COVID grade estimate. The sets widen as the allowed violations grow, so the evidence does not point to a precise zero. We conclude instead that there is no robust evidence that GenAI availability raised grades in more susceptible courses.

\input{tables_and_figures_v2/table_honestdid_grade}

\FloatBarrier
\section{GenAI Susceptibility as Continuous Treatment}
\label{app:contdid}

Our treatment, GenAI susceptibility, is continuous, whereas the classic difference-in-differences design assumes a binary treatment. \citet{callawayDifferenceinDifferencesContinuousTreatment2024} show that a causal interpretation of TWFE estimates with a continuous treatment requires a stronger assumption than standard parallel trends. Beyond parallel trends between treated and untreated courses, highly susceptible courses, had they been less susceptible, must have followed the trends of the less susceptible courses. This assumption fails if courses select into susceptibility levels based on outcome-related differences, such as how acceptable it is to use external resources in different subjects. 

\begin{figure}[htbp]
\centering
\includegraphics[width=0.65\linewidth]{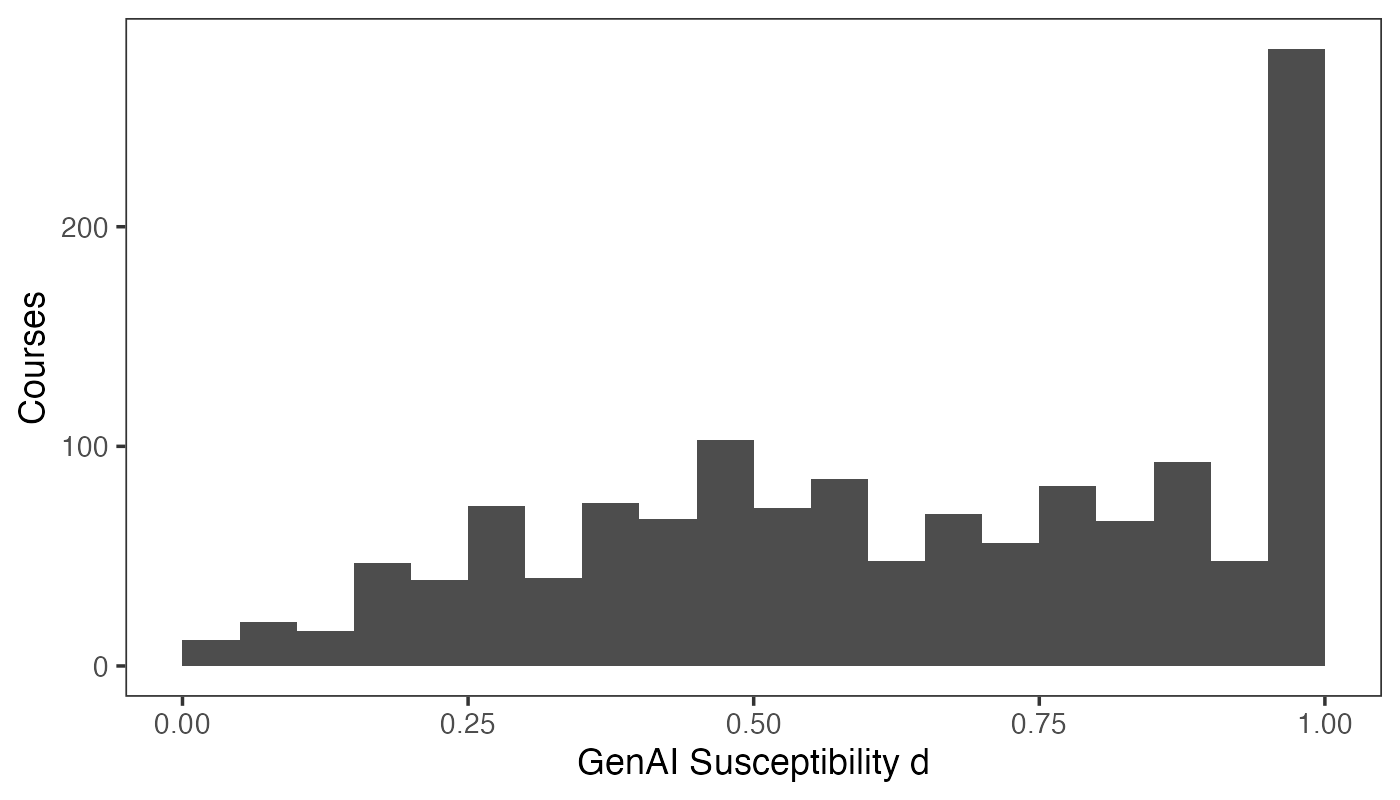}
\caption{Distribution of 2019-Anchored GenAI Susceptibility in the Regression Sample. One observation per course, measured from the course's 2019 offerings. Courses at zero form the comparison group for $\text{ATT}(d)$.}
\label{fig:dose-distribution}
\end{figure}

Using the \textit{contdid} R package accompanying \citet{callawayDifferenceinDifferencesContinuousTreatment2024}, we estimate grade effects at every level of susceptibility, comparing the post-ChatGPT period with the pre-COVID period. These estimates omit student fixed effects, so they are not directly comparable with our main results. The average treatment effect on the treated, $\text{ATT}(d)$, compares courses at each susceptibility level $d$ with courses that have no susceptible assessment. It requires only standard parallel trends between each level and that group. The average causal response on the treated, $\text{ACRT}(d)$, is the slope of the dose-response curve. Our main TWFE coefficient summarizes it, and it requires the stronger assumption. Figure~\ref{fig:dose-distribution} shows the distribution of 2019 susceptibility across the courses used in this analysis.

\begin{figure}[htbp]
\centering
\includegraphics[width=0.7\linewidth]{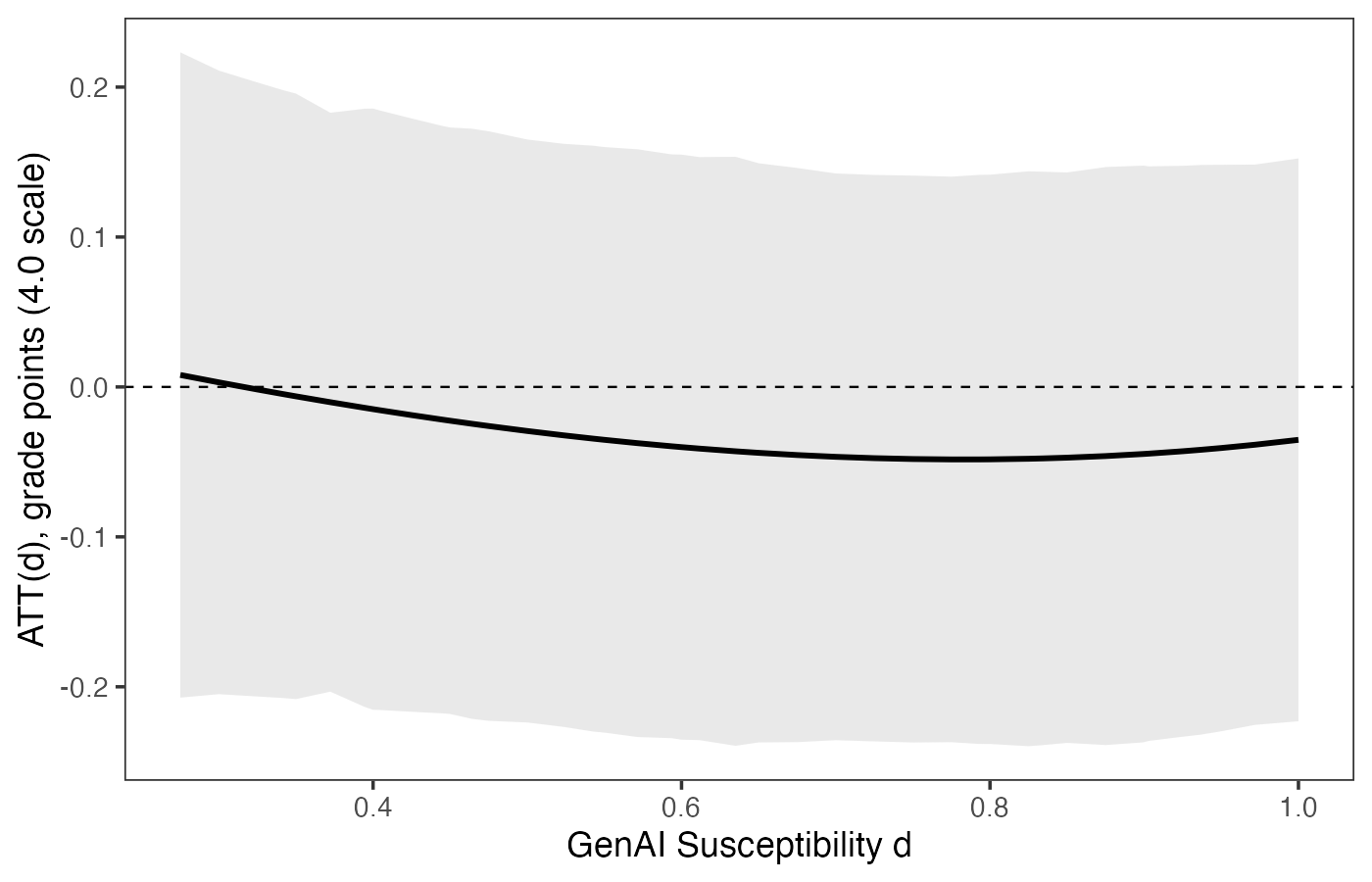}
\caption{Average Grade Effects by Level of GenAI Susceptibility, $\text{ATT}(d)$. The curve compares courses at each susceptibility level $d$ with courses that had no susceptible assessment in 2019, in the post-ChatGPT period relative to the pre-COVID period, following \citet{callawayDifferenceinDifferencesContinuousTreatment2024}. The shaded region is a uniform confidence band. Identification requires standard parallel trends.}
\label{fig:attd-grade}
\end{figure}

\begin{figure}[htbp]
\centering
\includegraphics[width=0.7\linewidth]{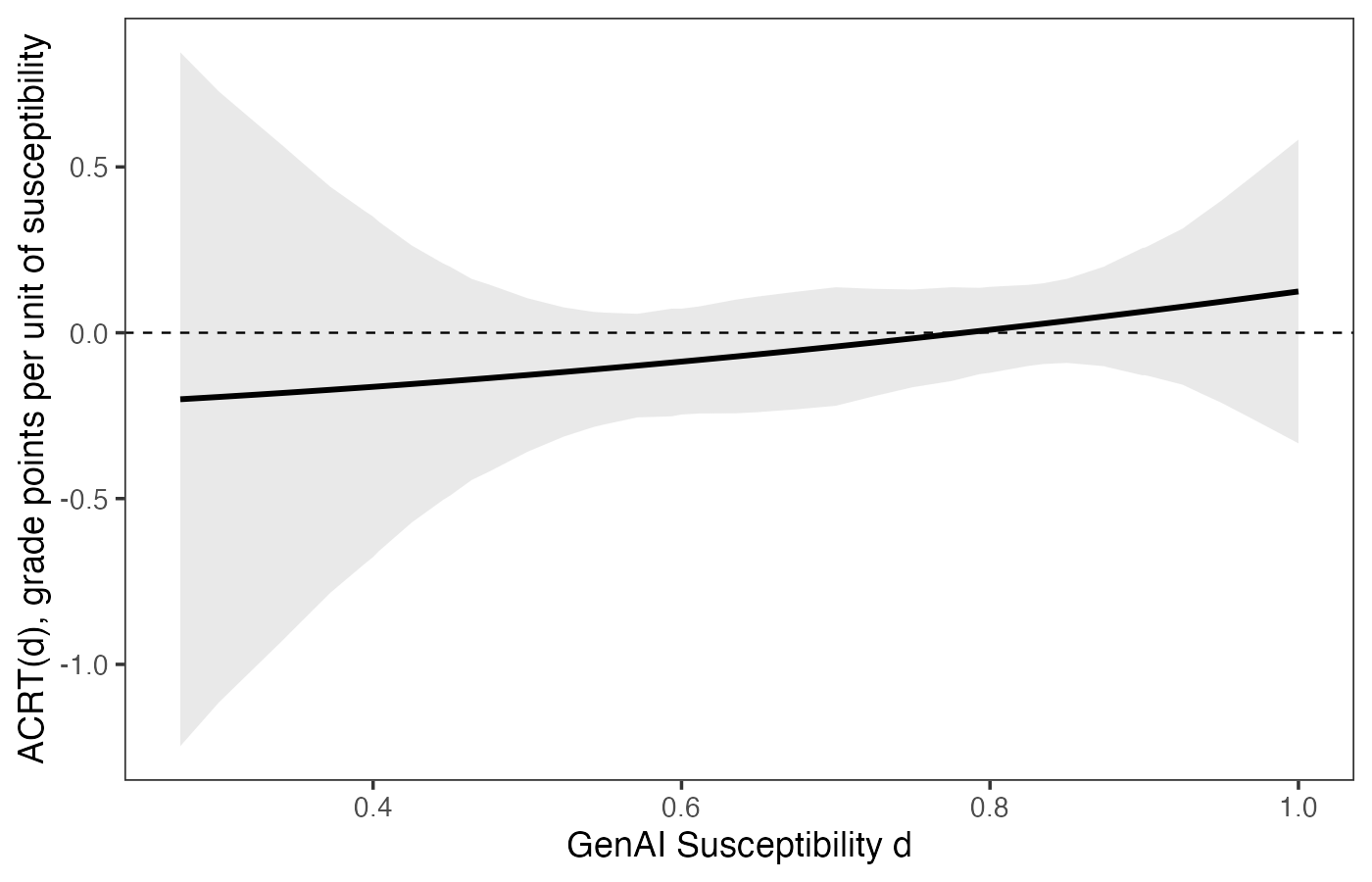}
\caption{Average Causal Response of Grades to GenAI Susceptibility, $\text{ACRT}(d)$. The curve is the slope of the dose-response function at each susceptibility level $d$, with a uniform confidence band, from the same sample and estimator as Figure~\ref{fig:attd-grade}. The two-way fixed effects coefficient summarizes this curve. Causal interpretation requires the strong parallel trends assumption.}
\label{fig:acrt-grade}
\end{figure}

Figures~\ref{fig:attd-grade} and~\ref{fig:acrt-grade} plot the two curves. Neither shows a significant effect at any level of susceptibility. The $\text{ATT}(d)$ point estimates are non-positive, near zero at the lowest levels of susceptibility and around $-0.04$ to $-0.05$ grade points at moderate-to-high levels, and the confidence band covers zero at every level. The $\text{ACRT}(d)$ band also covers zero at every level. The average ATT is $-0.038$ (SE = 0.088), and the average ACRT is $-0.007$ (SE = 0.037). We treat the ACRT as the more informative comparison to our main estimates, because our TWFE coefficient summarizes it and because the ATT relies on a very small comparison group. In both cases, the absence of a positive effect holds across levels of susceptibility, so the average does not conceal offsetting responses at different doses.

The main limitation of the ATT analysis is the scarcity of clean comparison courses: only nine had no susceptible assessment in 2019. We argue that this scarcity is common in higher education, where few course grades rest entirely on proctored assessment.

\FloatBarrier
\section{Comparing Grade Effects with Prior Studies}
\label{app:prior-comparison}

This appendix compares our grade estimates with those of \citet{hausmanGenerativeAIsImpact} and \citet{chirikovAIGradeInflation2026}. We ask two questions of each comparison. First, what is the smallest grade increase our estimate rules out? Second, can our estimate be statistically distinguished from prior ones? Table~\ref{tab:prior_comparison} reports the results.

\paragraph{Common units.} We express all effects in points on the 4.0 grade scale, or as shares for the grade distribution. Our main estimates and those of \citet{chirikovAIGradeInflation2026} are directly comparable in form. Both are coefficients on the post-ChatGPT period interacted with a 0 to 1 exposure measure, so each is the change for a course moving from no exposure to full exposure. The two exposure measures are related but distinct constructs. Ours is the proportion of the final grade from susceptible assessments, while his is the proportion of course tasks that involve writing or coding. His courses also sit lower on his scale than ours do on ours (unweighted course means of 0.24 versus 0.65; weighted by enrollment, ours is 0.53), so his per-unit coefficient extrapolates further beyond his data. For the grade distribution, his share of A grades corresponds to our probability of earning at least an A.

\paragraph{Binary contrasts.} \citet{hausmanGenerativeAIsImpact} report only a binary contrast between AI-compatible and AI-incompatible courses, so we compare it with binary contrasts. For our data we use the top ($> 0.665$) versus bottom ($\leq 0.335$) absolute tercile of 2019 susceptibility, with mean susceptibility of 0.89 and 0.22. For \citet{chirikovAIGradeInflation2026} we use his robustness estimate comparing courses with any writing or coding task to courses with none (his Table A4). \citeauthor{hausmanGenerativeAIsImpact} measure grades on a 0 to 100 scale, and no conversion to grade points is exact. We bracket it with two mappings. A linear rescaling of 0 to 100 onto 0 to 4 gives 0.04 grade points per point, while treating each letter grade as a 10-point band gives 0.10. We use their first post-ChatGPT year (0.970 points, SE = 0.289), which is measured relative to 2021--22. Their second-year estimate is larger (1.484, SE = 0.406), which would widen the gaps reported here.

\paragraph{Tests.} We report both our transient- and persistent-COVID estimates (Section~\ref{sec:covid-modeling}). For each estimate we report the upper end of its 90\% confidence interval. This is the smallest effect the estimate rules out with a one-sided test at the 5\% level. It is the relevant half of a two one-sided tests (TOST) equivalence procedure, since the question is whether grades rose by at least a given amount. We express these bounds relative to 0.3 grade points, the gap between adjacent letter grades such as B and B+. We also test whether each of our estimates equals each prior estimate in the same panel, using $z = (\hat\beta_{\text{ours}} - \hat\beta_{\text{prior}})/\sqrt{SE_{\text{ours}}^2 + SE_{\text{prior}}^2}$ and treating the studies as independent.

\paragraph{Results.} Our estimates are generally at or below the prior ones, but most differences are not statistically significant. Across our two estimates, the average grade effect rules out increases above 0.109 grade points per unit of susceptibility, or 0.36 of a letter-grade step; the transient-COVID estimate alone rules out increases above 0.084. Neither estimate is statistically distinguishable at the 5\% level from \citeauthor{chirikovAIGradeInflation2026}'s estimate of 0.12 ($p = 0.10$ transient, $p = 0.37$ persistent). The two estimates diverge at the top of the distribution. Our persistent-COVID estimate for earning at least an A (0.030) is well below \citeauthor{chirikovAIGradeInflation2026}'s 0.13 ($p < 0.01$), while our transient-COVID estimate (0.107) is closer to it ($p = 0.68$). In the binary comparisons, our estimates cannot be distinguished at the 5\% level from \citeauthor{hausmanGenerativeAIsImpact}'s under the linear conversion or from \citeauthor{chirikovAIGradeInflation2026}'s binary estimate. Only our transient-COVID estimate differs from \citeauthor{hausmanGenerativeAIsImpact}'s under the 10-point conversion ($p < 0.01$). Every upper bound we report is below half of a letter-grade step, and the binary bounds are between a tenth and a quarter of a step. The prior binary estimates are of similar size, between 0.03 and 0.10 grade points.

\paragraph{Benchmarks.} To give these magnitudes a reference point, we compare them with grade movements common to all courses. We regress final grades on course and semester fixed effects in the analytic sample and average the semester effects by period, relative to the pre-COVID semesters (Fall 2016 to Fall 2019). Within courses, grades rose by about 0.03 points per year before the pandemic. They averaged 0.14 points above the pre-COVID level both from Winter 2020 to Fall 2022 and from Winter 2023 to Fall 2025. The average student-course observation has a susceptibility of 0.53. At the largest effect our estimates allow (0.109 per unit), a course of average susceptibility would gain about 0.06 points from GenAI, about 40 percent of the common drift over the same period.

\input{tables_and_figures_v2/table_prior_comparison}

\FloatBarrier
\section{Predictive Power of Standardized Test Scores}
\label{app:grade-signal}

This appendix details the regressions in Section~\ref{sec:null-makeup} (Table~\ref{tab:grade-signal}). They ask whether, after ChatGPT's release, students' pre-college standardized test scores became less predictive of grades in more susceptible courses.

\paragraph{Measures.} The main regressor is each student's highest SAT or ACT score on the current SAT scale, in hundreds of points. ACT composites and SAT totals from before the March 2016 redesign are converted with the official concordance tables, as described in Appendix~\ref{app:ability-robustness}. The outcome is final grade value on the 0 to 4.0 scale (Appendix~\ref{app:grade-pts}). Unlike the heterogeneity analysis, which is limited to fall entrants, this analysis includes every student with a test score.

\paragraph{Slope of grades on test scores.} The object of interest is how steeply grades rise with test scores within a course-semester. We estimate how that slope responds to susceptibility with the student-level regression
\begin{equation}
\begin{split}
  g_{sct} \;=\; & \alpha_{ct} + T_s \, (\mu_c + \nu_t) \\
    & + \theta_1 \, T_s \times \text{Susceptibility}_{c} \times \text{Post}^{\text{AI}}_t \\
    & + \theta_2 \, T_s \times \text{Susceptibility}_{c} \times \text{COVID Year}_t \\
    & + \varepsilon_{sct},
\end{split}
\label{eq:grade-signal-slope}
\end{equation}
where $g_{sct}$ is the grade of student $s$ in course $c$ in semester $t$ and $T_s$ is the student's test score. Course-semester fixed effects $\alpha_{ct}$ absorb each offering's grading level, including the level effects of susceptibility in Equation~\ref{eq:main}. The terms $\mu_c$ and $\nu_t$ give each course and each semester its own score slope. The coefficients $\theta_1$ and $\theta_2$ therefore compare changes in the slope across courses that differ in susceptibility. As in Equation~\ref{eq:main}, $\theta_1$ is the transient-COVID estimate and $\theta_1 - \theta_2$ is the persistent-COVID estimate.

Every regressor other than the fixed effects is the test score multiplied by a variable that is constant within a course-semester. By the Frisch-Waugh-Lovell theorem, the coefficients and their clustered standard errors equal those from a course-semester regression. That regression relates each course-semester's own slope to course and semester fixed effects and the two susceptibility interactions, weighting by the course-semester's sum of squared score deviations. We estimate the model this way, excluding course-semesters with fewer than 10 students with a score and clustering standard errors by course.

\paragraph{Bounds.} As for average grades (Appendix~\ref{app:prior-comparison}), the largest decline the estimates rule out is the lower end of a 90\% confidence interval, a one-sided test at the 5\% level. In column (1) these limits are $-0.007 - 1.645 \times 0.017 = -0.035$ for the transient-COVID estimate and $-0.016 - 1.645 \times 0.010 = -0.033$ for the persistent-COVID estimate. Declines larger than 0.035, about a sixth of the pre-COVID slope of 0.21, are therefore ruled out.

\paragraph{Test-optional selection.} In June 2020, the University made standardized test scores optional for applicants. Students who did not submit scores earned lower first-term grades (Section~\ref{sec:null-makeup}), which raises selection concerns. Column (1) of Table~\ref{tab:grade-signal} uses every student with a score. Column (2) keeps students who entered in Fall 2020 or earlier, when scores were required. The restriction has a cost: in the restricted sample, the post-GenAI period consists mostly of upper-level students.

\FloatBarrier
\section{Unadjusted Outcome Trends by Susceptibility Group}
\label{app:raw-dynamics}

\begin{figure}[htbp]
\centering
\begin{subfigure}[t]{0.62\textwidth}
\centering
\includegraphics[width=\linewidth]{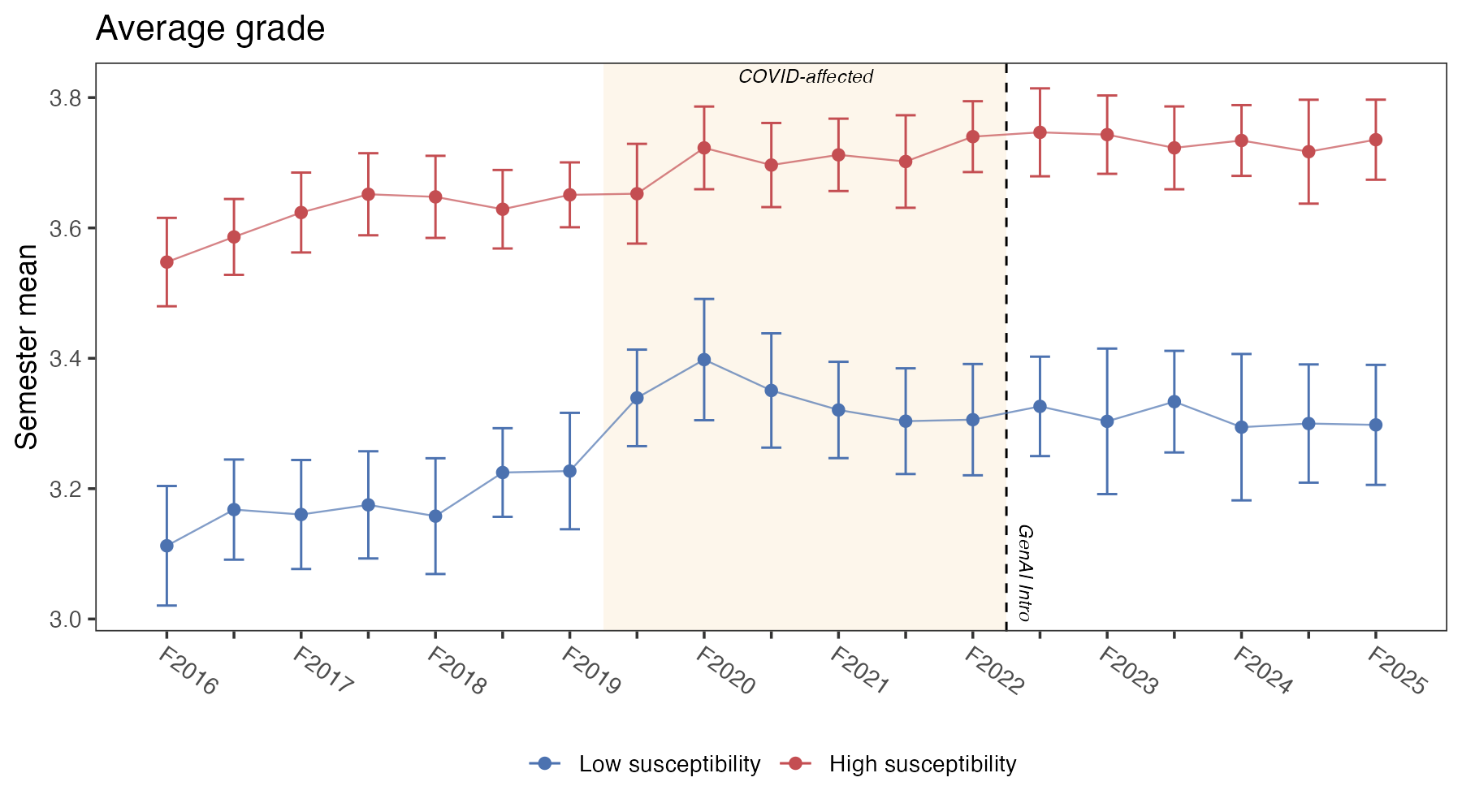}
\caption{Average grade}
\label{fig:raw-grade}
\end{subfigure}

\vspace{0.6em}

\begin{subfigure}[t]{0.48\textwidth}
\centering
\includegraphics[width=\linewidth]{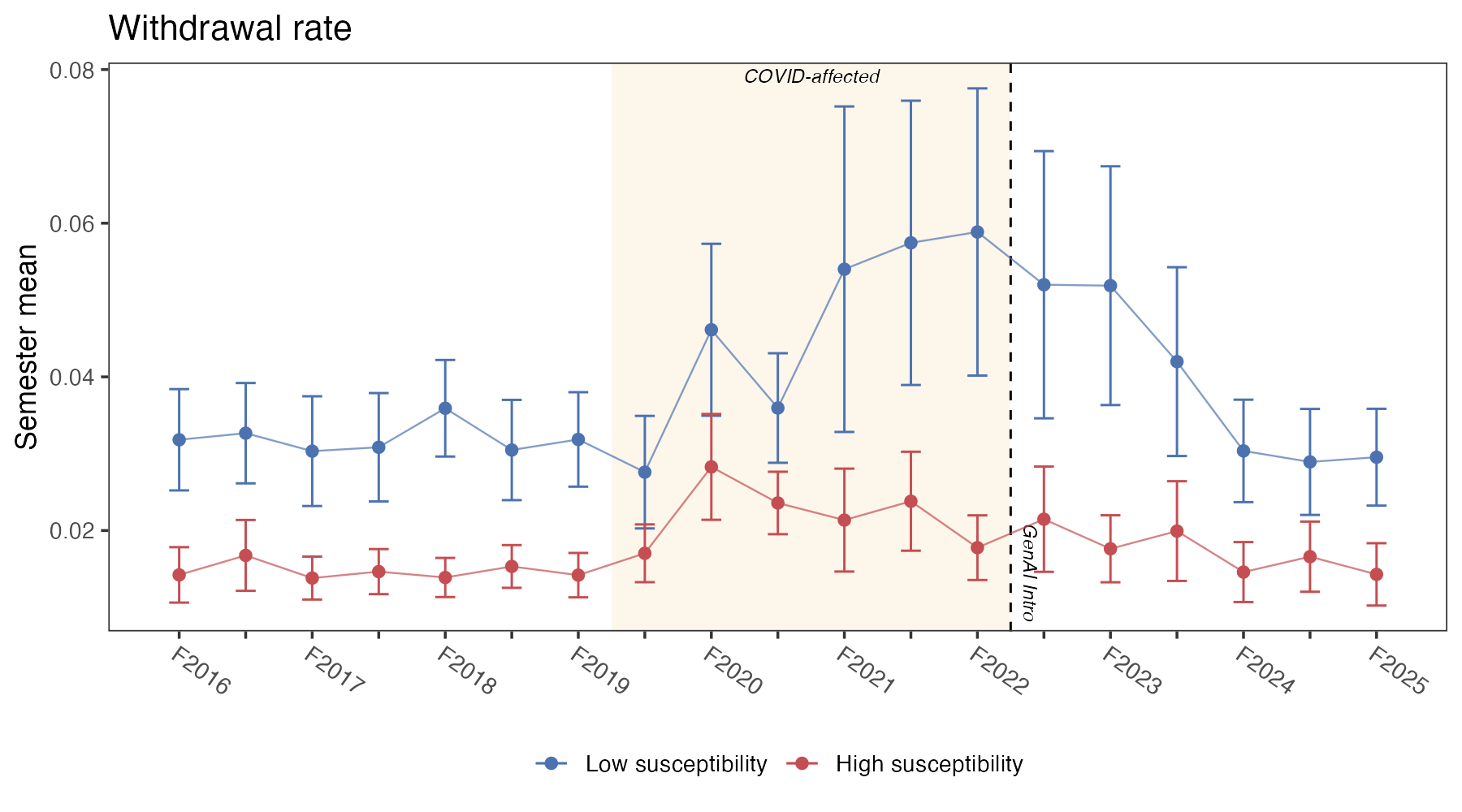}
\caption{Withdrawal rate}
\label{fig:raw-withdrawal}
\end{subfigure}\hfill
\begin{subfigure}[t]{0.48\textwidth}
\centering
\includegraphics[width=\linewidth]{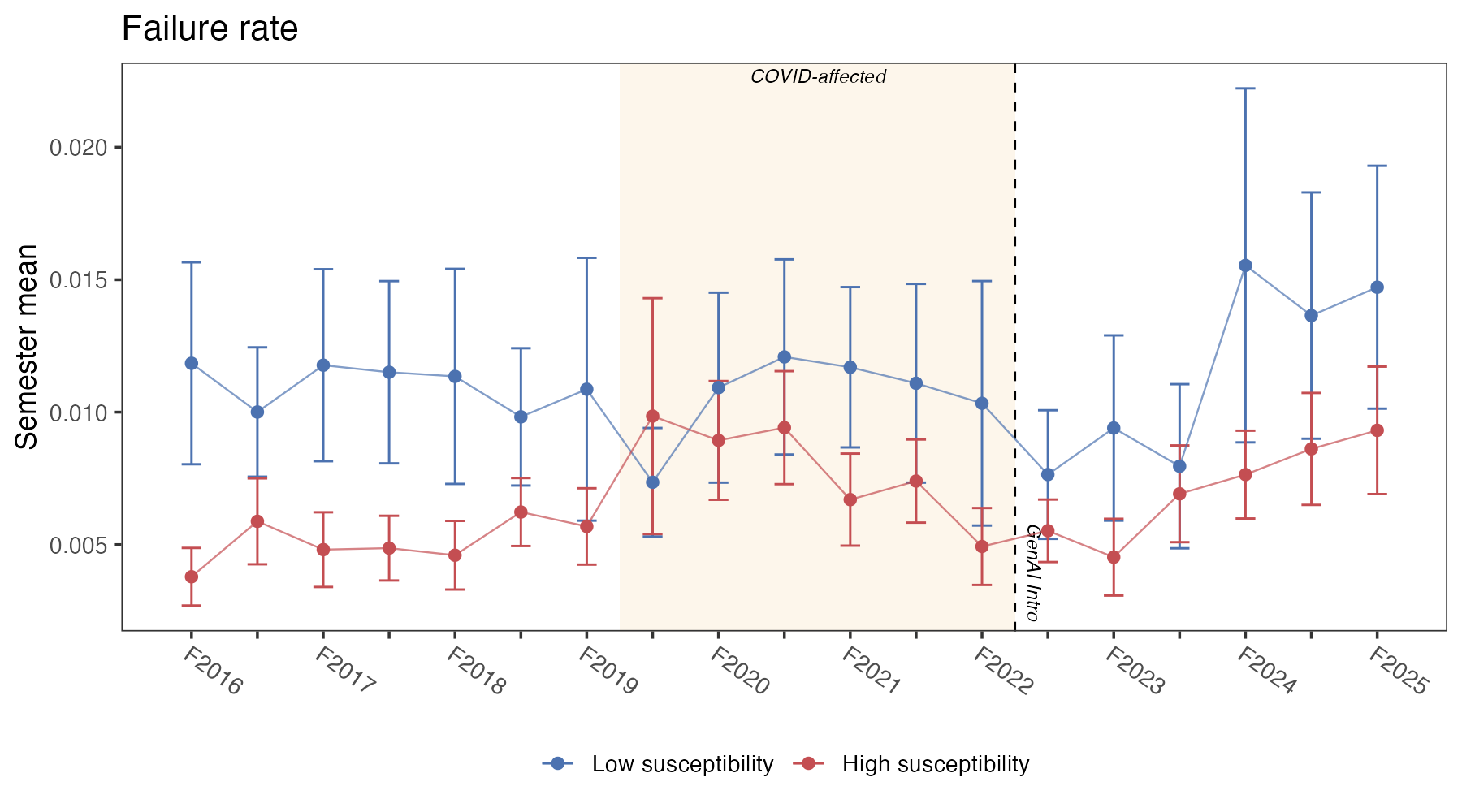}
\caption{Failure rate}
\label{fig:raw-failure}
\end{subfigure}
\caption{Unadjusted Student Outcomes by Susceptibility Group. Points are semester means for low-susceptibility ($\leq 0.335$) and high-susceptibility ($> 0.665$) courses, by 2019-anchored susceptibility. Bars are 95 percent confidence intervals with standard errors clustered by course. The shaded band marks the COVID-affected terms (Winter 2020 to Fall 2022), and the dashed line marks ChatGPT's release.}
\label{fig:raw-student-outcomes}
\end{figure}

\begin{figure}[htbp]
\centering
\begin{subfigure}[t]{0.62\textwidth}
\centering
\includegraphics[width=\linewidth]{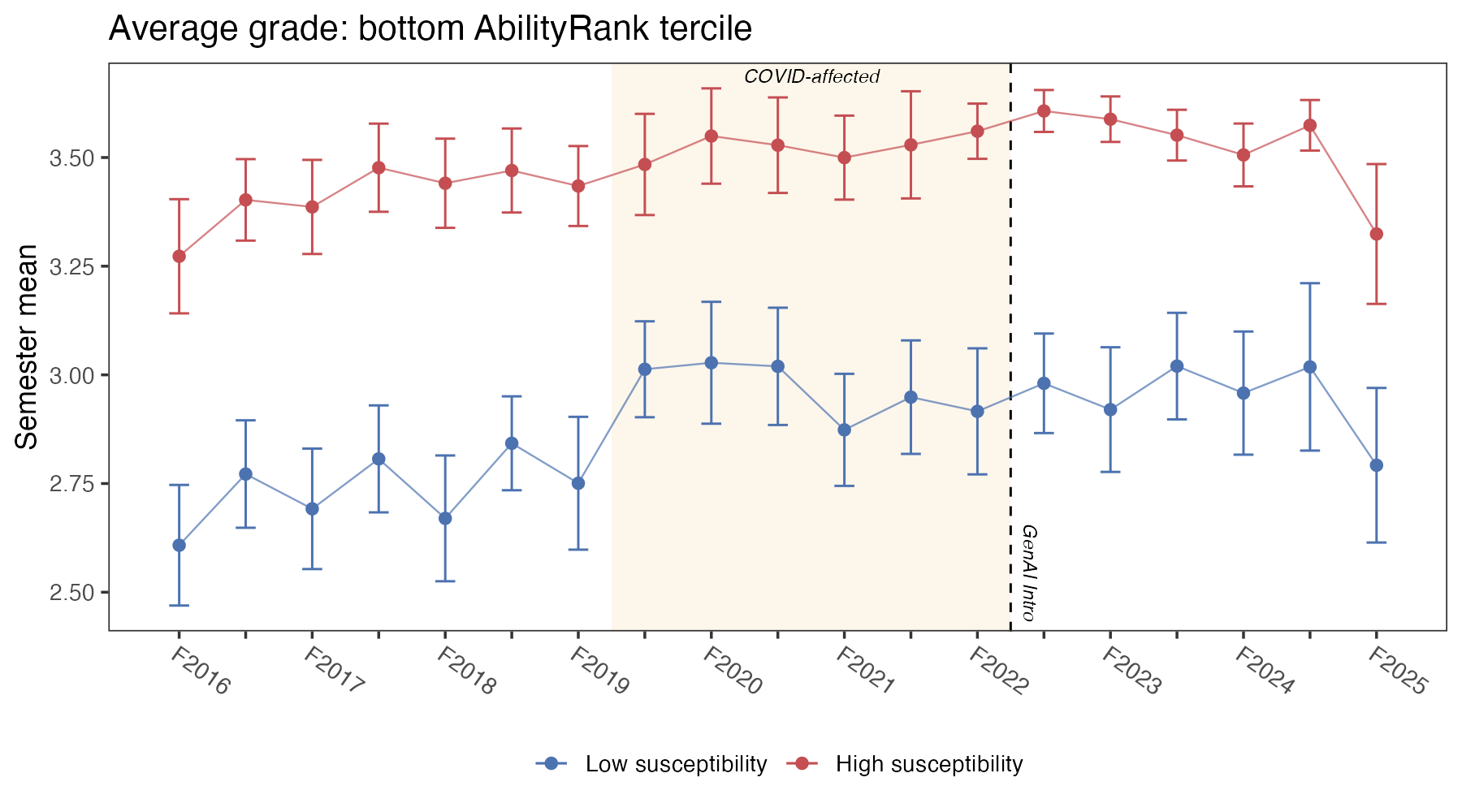}
\caption{Bottom AbilityRank tercile}
\label{fig:raw-gpa-low}
\end{subfigure}

\vspace{0.6em}

\begin{subfigure}[t]{0.48\textwidth}
\centering
\includegraphics[width=\linewidth]{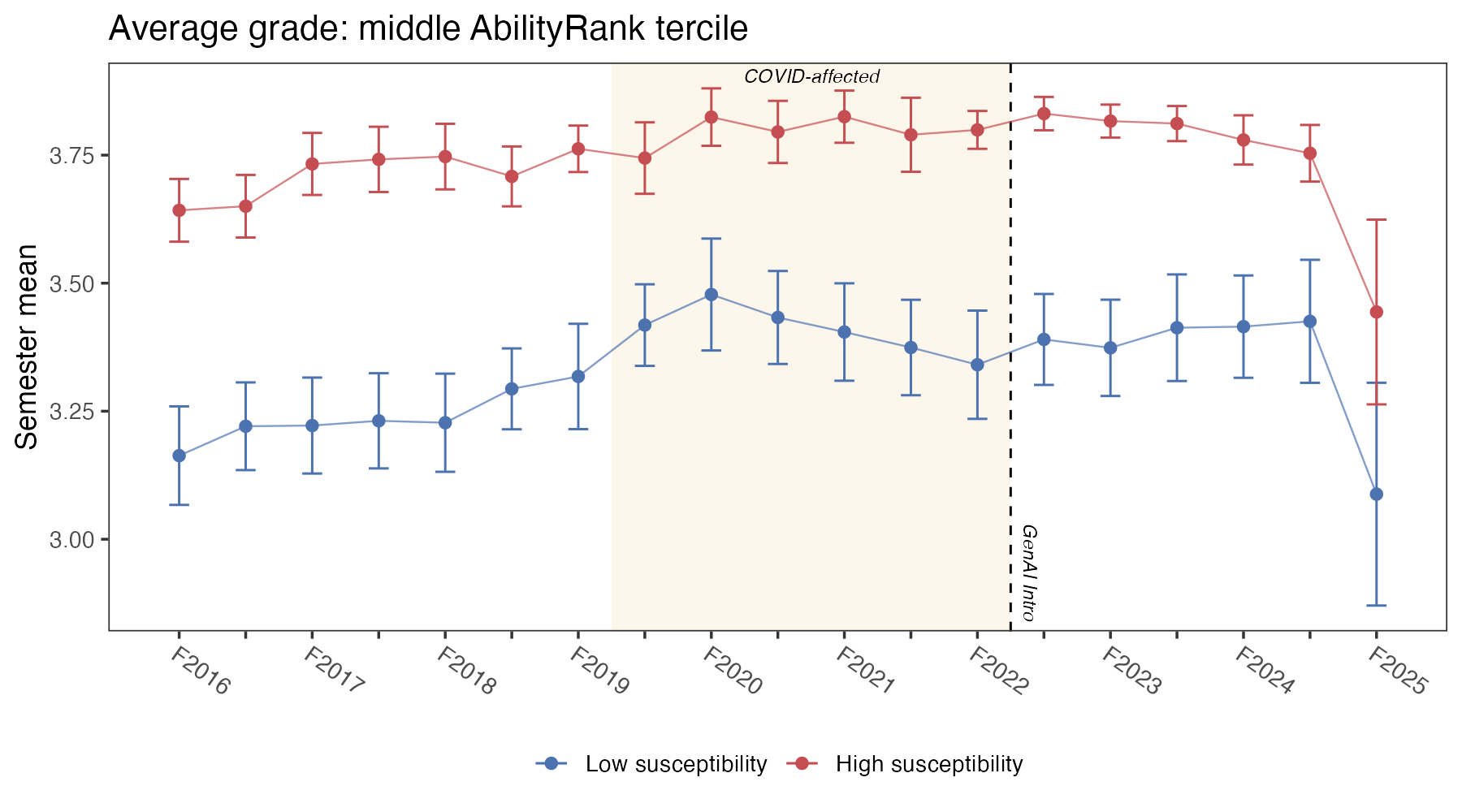}
\caption{Middle AbilityRank tercile}
\label{fig:raw-gpa-mid}
\end{subfigure}\hfill
\begin{subfigure}[t]{0.48\textwidth}
\centering
\includegraphics[width=\linewidth]{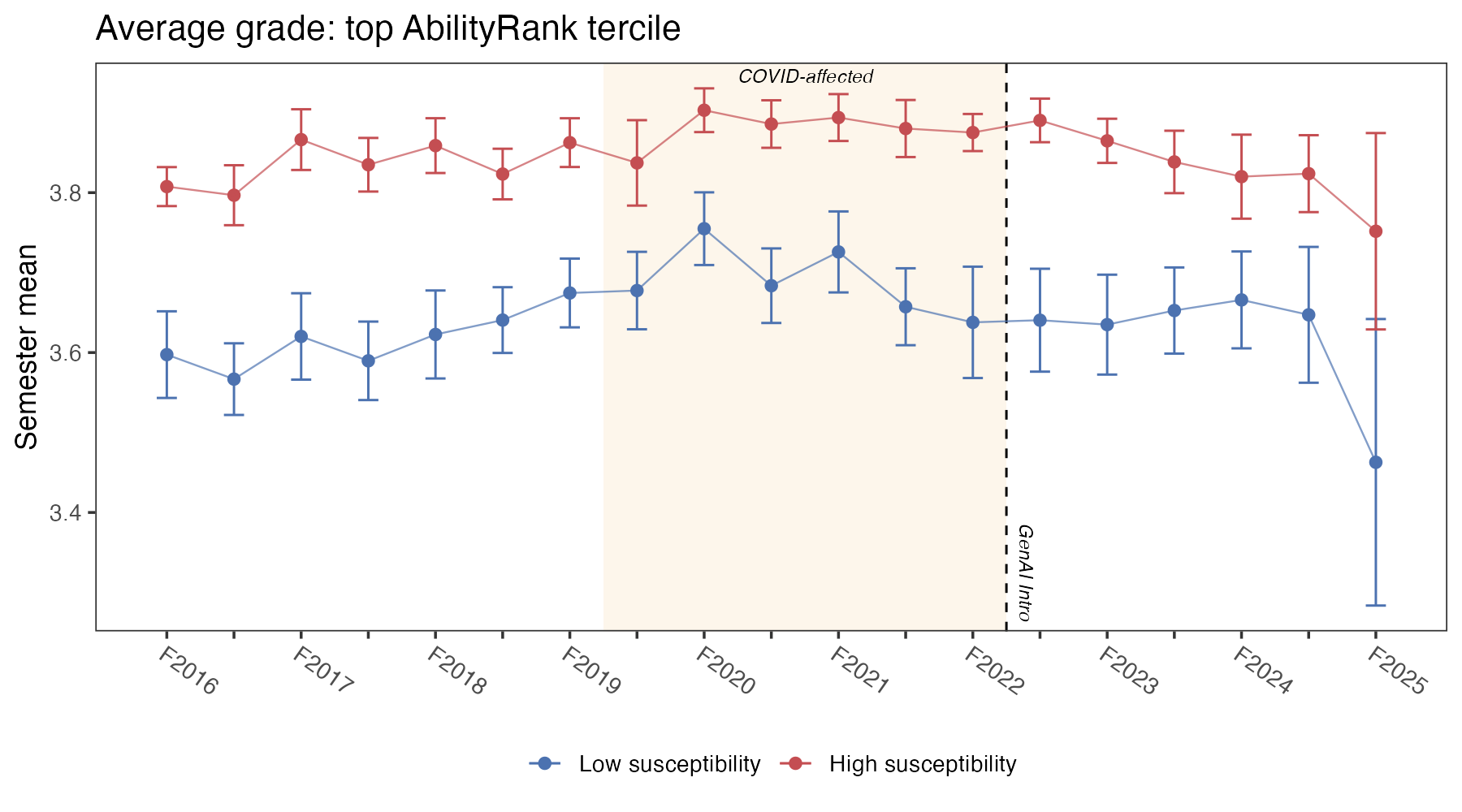}
\caption{Top AbilityRank tercile}
\label{fig:raw-gpa-high}
\end{subfigure}
\caption{Unadjusted Average Grade by Susceptibility Group and \texttt{AbilityRank} Tercile. Points are semester means for low-susceptibility ($\leq 0.335$) and high-susceptibility ($> 0.665$) courses, by 2019-anchored susceptibility. Bars are 95 percent confidence intervals with standard errors clustered by course. The shaded band marks the COVID-affected terms (Winter 2020 to Fall 2022), and the dashed line marks ChatGPT's release.}
\label{fig:raw-gpa-het}
\end{figure}

\begin{figure}[htbp]
\centering
\begin{subfigure}[t]{0.62\textwidth}
\centering
\includegraphics[width=\linewidth]{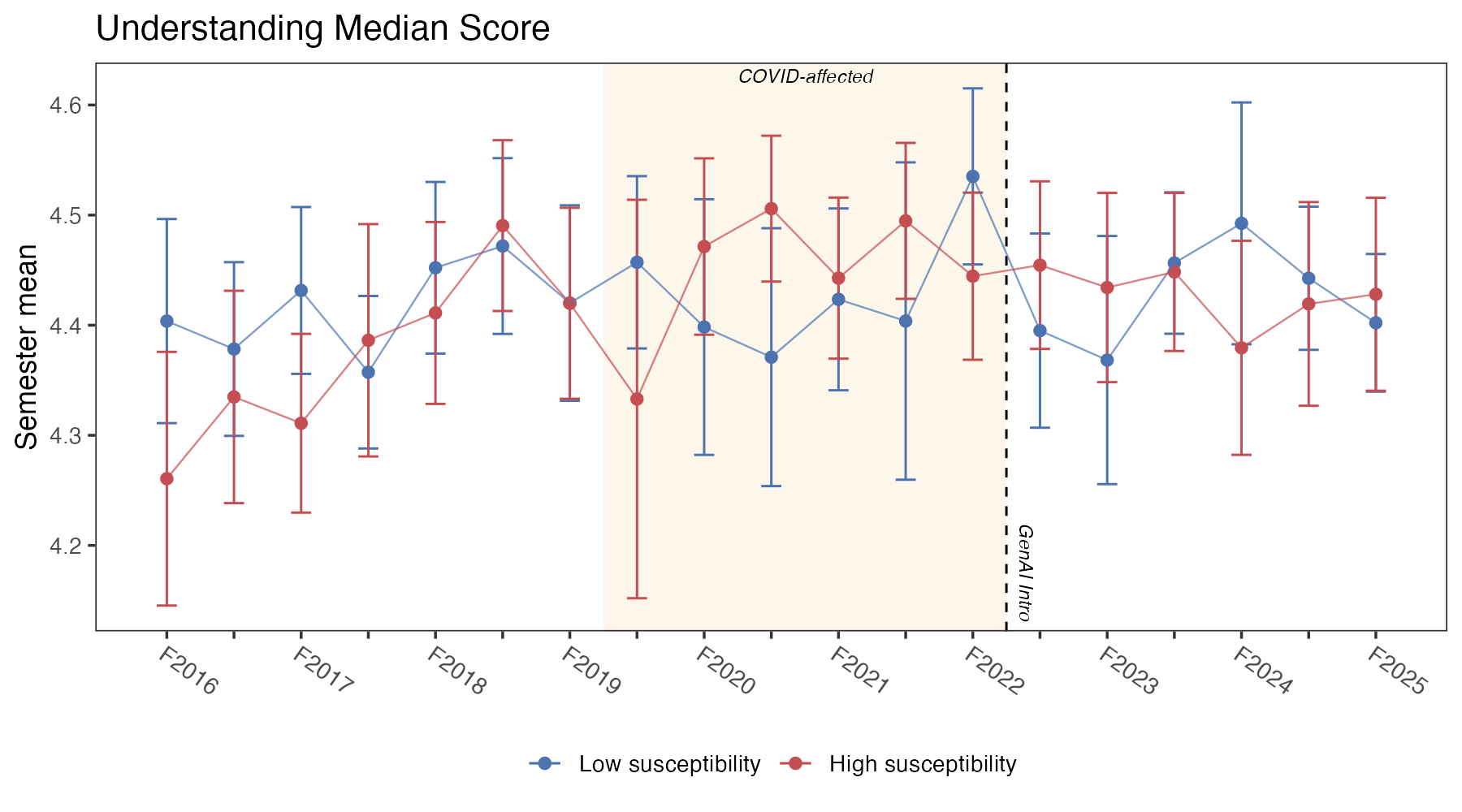}
\caption{Self-reported understanding}
\label{fig:raw-eval-understanding}
\end{subfigure}

\vspace{0.6em}

\begin{subfigure}[t]{0.48\textwidth}
\centering
\includegraphics[width=\linewidth]{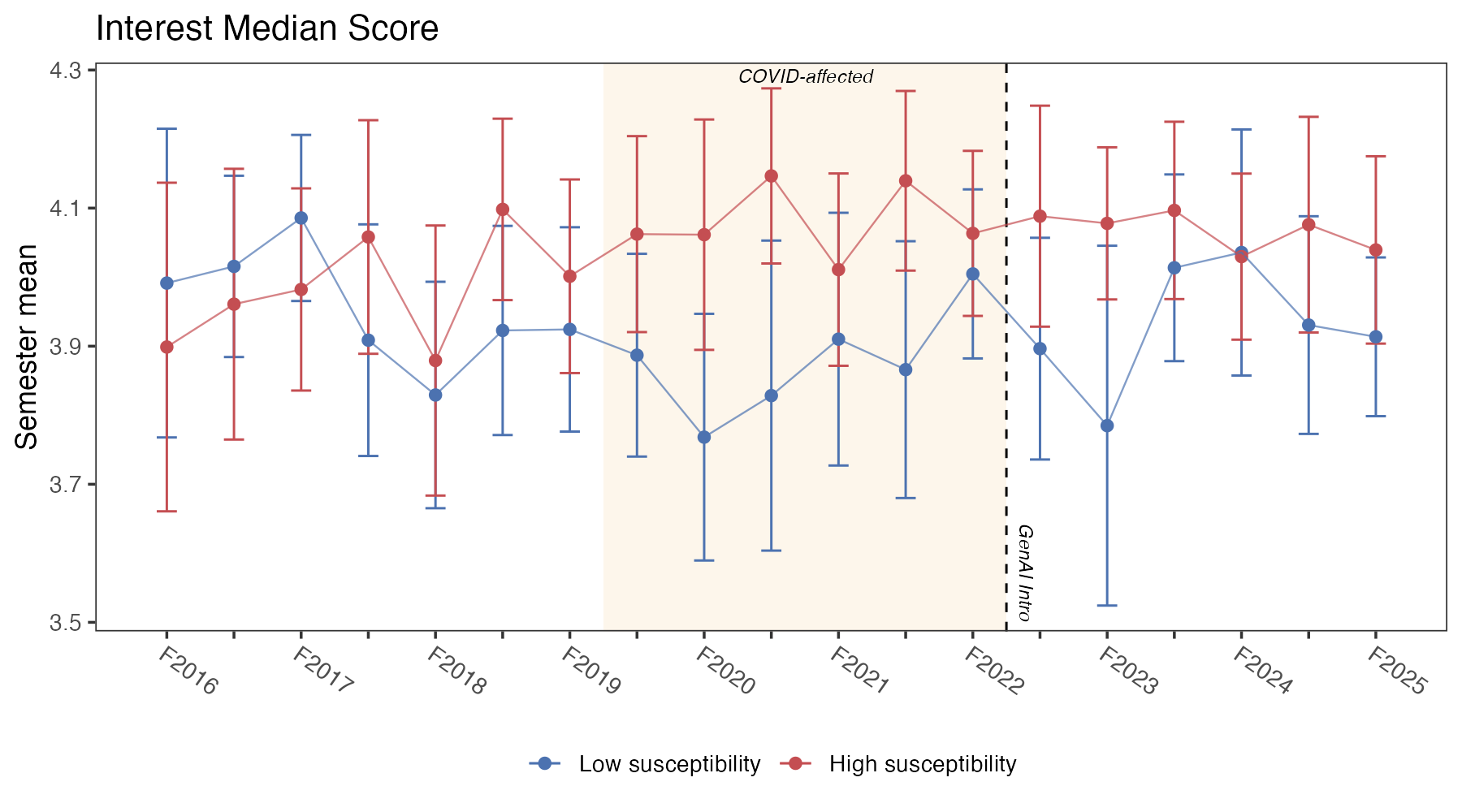}
\caption{Self-reported interest}
\label{fig:raw-eval-interest}
\end{subfigure}\hfill
\begin{subfigure}[t]{0.48\textwidth}
\centering
\includegraphics[width=\linewidth]{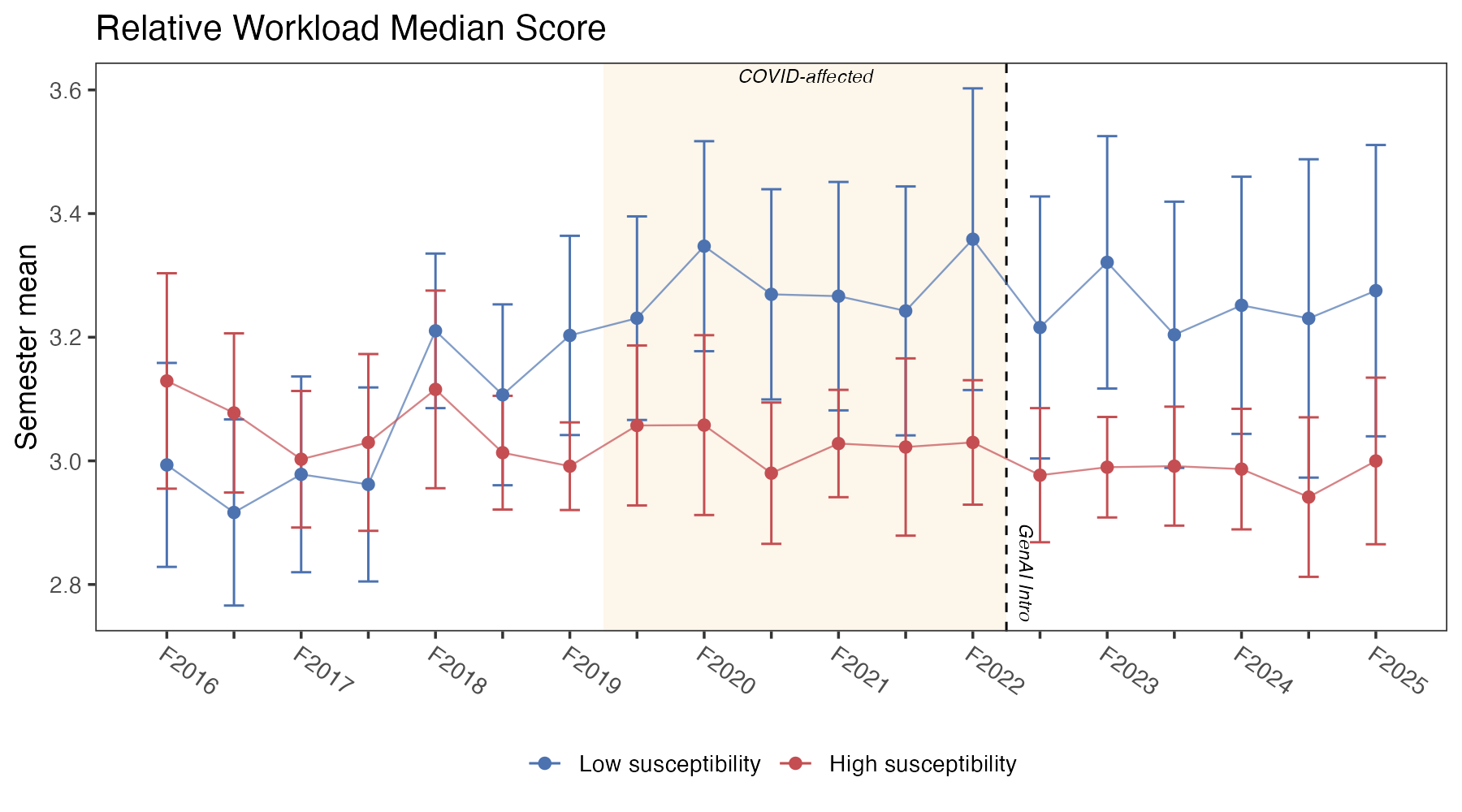}
\caption{Perceived relative workload (reverse-coded)}
\label{fig:raw-eval-workload}
\end{subfigure}
\caption{Unadjusted Course Evaluation Outcomes by Susceptibility Group. Points are semester means of median ratings for low-susceptibility ($\leq 0.335$) and high-susceptibility ($> 0.665$) courses, by 2019-anchored susceptibility. Relative workload is reverse-coded so that higher values indicate heavier workload. Bars are 95 percent confidence intervals with standard errors clustered by course. The shaded band marks the COVID-affected terms (Winter 2020 to Fall 2022), and the dashed line marks ChatGPT's release.}
\label{fig:raw-course-evals}
\end{figure}

Figures~\ref{fig:raw-student-outcomes}, \ref{fig:raw-gpa-het}, and~\ref{fig:raw-course-evals} plot unadjusted semester means of each outcome for low- and high-susceptibility courses, complementing the continuous event studies in Section~\ref{sec:results} and Appendix~\ref{app:additional-main}.


\end{document}

%% file: tables_and_figures_v2/table_student_outcomes.tex
\begin{landscape}
\begin{table}
\centering
\begin{threeparttable}
\caption{Difference-in-Differences Estimates for Student Outcomes}
\label{tab:student_outcomes}
\small
\setlength{\tabcolsep}{4pt}
\begin{tabular}{llllllllll}
& GPA & GPA & GPA & Withdraw & Withdraw & Withdraw & Fail & Fail & Fail \\
Susceptibility $\times$ Post-AI & \num{0.039} & \num{0.040} & \num{0.010} & \num{-0.000} & \num{0.001} & \num{-0.007} & \num{0.001} & \num{-0.000} & \num{0.001} \\
& (\num{0.037}) & (\num{0.031}) & (\num{0.045}) & (\num{0.003}) & (\num{0.003}) & (\num{0.005}) & (\num{0.002}) & (\num{0.002}) & (\num{0.002}) \\
Susceptibility $\times$ COVID Year &  &  & \num{-0.060}* &  &  & \num{-0.017}* &  &  & \num{0.003}** \\
&  &  & (\num{0.036}) &  &  & (\num{0.009}) &  &  & (\num{0.001}) \\
Susceptibility & \num{0.577}*** &  &  & \num{-0.032}*** &  &  & \num{-0.007}*** &  &  \\
& (\num{0.054}) &  &  & (\num{0.007}) &  &  & (\num{0.003}) &  &  \\
Post-AI & \num{0.058}** &  &  & \num{-0.001} &  &  & \num{0.000} &  &  \\
& (\num{0.026}) &  &  & (\num{0.002}) &  &  & (\num{0.001}) &  &  \\
N & \num{1174344} & \num{1162801} & \num{1162801} & \num{1220330} & \num{1208492} & \num{1208492} & \num{1220330} & \num{1208492} & \num{1208492} \\
\midrule
GenAI net COVID &  &  & 0.071*** &  &  & 0.010 &  &  & -0.002 \\
&  &  & (0.024) &  &  & (0.006) &  &  & (0.002) \\
Student FE &  & $\checkmark$ & $\checkmark$ &  & $\checkmark$ & $\checkmark$ &  & $\checkmark$ & $\checkmark$ \\
Course FE &  & $\checkmark$ & $\checkmark$ &  & $\checkmark$ & $\checkmark$ &  & $\checkmark$ & $\checkmark$ \\
Semester FE &  & $\checkmark$ & $\checkmark$ &  & $\checkmark$ & $\checkmark$ &  & $\checkmark$ & $\checkmark$ \\
\end{tabular}
\begin{tablenotes}[para,flushleft]
\small
\item \textit{Notes:} Standard errors clustered by course in parentheses. * $p<0.1$, ** $p<0.05$, *** $p<0.01$. For each outcome, the first column is OLS without fixed effects, the second adds student, course, and semester fixed effects, and the third is the main specification (Equation~\ref{eq:main}), which adds the Susceptibility $\times$ COVID Year interaction. In the third column, the Susceptibility $\times$ Post-AI coefficient $\hat\beta_1$ is the transient-COVID estimate, measured against the pre-COVID baseline. The ``GenAI net COVID'' row reports the persistent-COVID estimate $\hat\beta_1 - \hat\beta_2$ with a delta-method standard error, which assumes that the COVID-era differential persisted into the post-AI period. GPA is final grade value on the 4.0 scale and excludes withdrawals.
\end{tablenotes}
\end{threeparttable}
\end{table}
\end{landscape}

%% file: tables_and_figures_v2/table_grade_signal.tex
\begin{table}[htbp]
\centering
\begin{threeparttable}
\caption{Slope of Grades on Standardized Test Scores by GenAI Susceptibility}
\label{tab:grade-signal}
\small
\begin{tabular}{lcc}
\toprule
& All scores & Required era \\
& (1) & (2) \\
\midrule
Susceptibility $\times$ Post-AI & -0.007 & 0.009 \\
 & (0.017) & (0.020) \\
Susceptibility $\times$ COVID Year & 0.009 & 0.019 \\
 & (0.015) & (0.014) \\
GenAI net COVID & -0.016 & -0.010 \\
 & (0.010) & (0.019) \\
Pre-COVID mean & 0.208 & 0.208 \\
Course-semesters & 10,528 & 8,038 \\
\bottomrule
\end{tabular}
\begin{tablenotes}[para,flushleft]
\small
\item \textit{Notes:} Descriptive regressions; see Appendix~\ref{app:grade-signal} for the specification. The outcome is the slope of grade points on the SAT or ACT score (converted to the current SAT scale, per 100 points) within each course-semester, estimated from Equation~\ref{eq:grade-signal-slope}. Column (1) uses every student with a score, including students without an entering cohort. Column (2) keeps students whose first term at the University was Fall 2020 or earlier, who applied when scores were required. Course-semesters with fewer than 10 students with a score are excluded. Pre-COVID mean is the weighted mean slope from Fall 2016 to Fall 2019. Standard errors clustered by course in parentheses. * $p<0.1$, ** $p<0.05$, *** $p<0.01$.
\end{tablenotes}
\end{threeparttable}
\end{table}

%% file: tables_and_figures_v2/table1_summary_stats.tex
\begin{table}[h!]
\centering
\caption{Summary Statistics for the Balanced Panel}
\label{tab:summary_stats}
\begin{tabular}{lccc}
\toprule
Variable & Mean & SD & $N$ \\
\midrule
\multicolumn{4}{l}{\textit{Treatment}} \\
  Susceptibility (2019 anchor) & 0.526 & 0.304 & 1,448,677 \\
\midrule
\multicolumn{4}{l}{\textit{Student Outcomes}} \\
  Final Grade Value (0--4.0) & 3.485 & 0.723 & 1,782,773 \\
  Withdrawal (indicator) & 0.024 & 0.154 & 1,887,320 \\
  Failure (indicator, grade below D-) & 0.008 & 0.090 & 1,887,320 \\
\midrule
\multicolumn{4}{l}{\textit{Course Evaluations}} \\
  Eval: Understanding of Subject (1--5) & 4.450 & 0.349 & 1,114,278 \\
  Eval: Interest in Subject (1--5) & 4.032 & 0.571 & 1,114,333 \\
  Eval: Relative Workload  (1--5) & 3.071 & 0.546 & 1,194,555 \\
\midrule
\multicolumn{4}{l}{\textit{Student Demographics}} \\
  First-Term GPA (0--4.0) & 3.489 & 0.581 & 1,886,662 \\
\bottomrule
\end{tabular}
\smallskip
\begin{minipage}{\linewidth}
\small \textit{Notes:} Balanced panel: all terms from 2015 to 2025, restricted to courses observed both in 2015--2019 and in 2023--2025. The regression sample is narrower (Section~\ref{sec:data-sample}). The unit of observation is the student-offering, so course-level variables (treatment and evaluations) are implicitly enrollment-weighted. Workload is reverse-coded from the original survey scale so that higher values indicate heavier workload.
\end{minipage}
\end{table}

%% file: tables_and_figures_v2/table_top_subjects.tex
\begin{table}[h!]
\centering
\caption{Ten Largest Subjects in the Balanced Panel by Courses, Offerings, and Enrollments}
\label{tab:top_subjects}
\begin{tabular}{rlr@{\hspace{1.5em}}rlr@{\hspace{1.5em}}rlr}
\toprule
\multicolumn{3}{c}{Unique Courses} &
\multicolumn{3}{c}{Course-Term Offerings} &
\multicolumn{3}{c}{Student Enrollments} \\
\cmidrule(lr){1-3}\cmidrule(lr){4-6}\cmidrule(l){7-9}
Rank & Subject & $N$ & Rank & Subject & $N$ & Rank & Subject & $N$ \\
\midrule
  1 & ASIANLAN & 91 &  1 & MATH & 1,363 &  1 & CHEM & 170,886 \\
  2 & MATH & 78 &  2 & PSYCH & 1,138 &  2 & MATH & 130,742 \\
  3 & PSYCH & 75 &  3 & ASIANLAN & 978 &  3 & PSYCH & 126,155 \\
  4 & HISTORY & 57 &  4 & CHEM & 878 &  4 & ECON & 111,871 \\
  5 & CHEM & 53 &  5 & ENGLISH & 822 &  5 & ENGLISH & 83,692 \\
  6 & ENGLISH & 46 &  6 & ECON & 705 &  6 & BIOLOGY & 80,043 \\
  7 & SOC & 45 &  7 & IOE & 605 &  7 & STATS & 60,901 \\
  8 & IOE & 44 &  8 & SOC & 589 &  8 & SPANISH & 52,594 \\
  9 & ECON & 41 &  9 & HISTORY & 575 &  9 & EARTH & 42,708 \\
  10 & POLSCI & 37 &  10 & SPANISH & 532 &  10 & PHYSICS & 41,490 \\
\bottomrule
\end{tabular}
\smallskip
\begin{minipage}{\linewidth}
\small \textit{Notes:} Subject codes follow the University's classification. Counts use the balanced panel: all terms from 2015 to 2025, restricted to courses observed both in 2015--2019 and in 2023--2025. Each ranking is computed separately, so a subject may appear in more than one.
\end{minipage}
\end{table}

%% file: tables_and_figures_v2/table2_balance.tex
\begin{table}[h!]
\centering
\caption{Pre-COVID Means for Low- and High-Susceptibility Courses}
\label{tab:balance}
\begin{tabular}{lcccc}
\toprule
Variable & Low Suscep. & High Suscep. & Difference & $p$-value \\
& ($\leq 0.335$) & ($> 0.665$) & (High $-$ Low) & \\
\midrule
\multicolumn{5}{l}{\textit{Student Outcomes}} \\
  Final Grade Value & 3.175 & 3.622 & +0.447 & 0.000$^{***}$ \\
  Withdrawal Rate & 0.032 & 0.015 & -0.017 & 0.000$^{***}$ \\
  Failure Rate & 0.011 & 0.005 & -0.006 & 0.000$^{***}$ \\
\midrule
\multicolumn{5}{l}{\textit{Course Evaluations}} \\
  Eval: Understanding & 4.419 & 4.381 & -0.037 & 0.000$^{***}$ \\
  Eval: Interest  & 3.946 & 3.985 & +0.039 & 0.000$^{***}$ \\
  Eval: Relative Workload & 3.065 & 3.046 & -0.020 & 0.000$^{***}$ \\
\midrule
\multicolumn{5}{l}{\textit{Student Demographics}} \\
  Prior Ability (raw within-cohort rank) & 0.470 & 0.482 & +0.012 & 0.000$^{***}$ \\
\bottomrule
\end{tabular}
\smallskip
\begin{minipage}{\linewidth}
\small \textit{Notes:} Fall and winter terms from fall 2016 through fall 2019 in the balanced panel. Low and high susceptibility are the bottom ($\leq 0.335$) and top ($> 0.665$) absolute terciles of the 2019-anchored susceptibility measure; the middle tercile is excluded. $p$-values are from Welch two-sample $t$-tests. $^{*}$ $p<0.1$, $^{**}$ $p<0.05$, $^{***}$ $p<0.01$.
\end{minipage}
\end{table}

%% file: tables_and_figures_v2/table_pct_async_persistence.tex
\begin{table}[htbp]
\centering
\begin{threeparttable}
\caption{Year-to-Year Correlations of Course GenAI Susceptibility}
\label{tab:async_persistence}
\small
\setlength{\tabcolsep}{3.5pt}
\begin{tabular}{lccccccccccc}
\toprule
 & '15 & '16 & '17 & '18 & '19 & '20 & '21 & '22 & '23 & '24 & '25 \\
\midrule
'15 & 1.00 &  &  &  &  &  &  &  &  &  &  \\
'16 & 0.84 & 1.00 &  &  &  &  &  &  &  &  &  \\
'17 & 0.82 & 0.85 & 1.00 &  &  &  &  &  &  &  &  \\
'18 & 0.78 & 0.83 & 0.87 & 1.00 &  &  &  &  &  &  &  \\
'19 & 0.80 & 0.79 & 0.83 & 0.89 & 1.00 &  &  &  &  &  &  \\
'20 & 0.50 & 0.55 & 0.50 & 0.57 & 0.61 & 1.00 &  &  &  &  &  \\
'21 & 0.50 & 0.49 & 0.54 & 0.53 & 0.56 & 0.42 & 1.00 &  &  &  &  \\
'22 & 0.65 & 0.66 & 0.68 & 0.70 & 0.73 & 0.56 & 0.68 & 1.00 &  &  &  \\
'23 & 0.68 & 0.67 & 0.69 & 0.71 & 0.73 & 0.56 & 0.64 & 0.81 & 1.00 &  &  \\
'24 & 0.65 & 0.66 & 0.69 & 0.67 & 0.70 & 0.51 & 0.58 & 0.76 & 0.82 & 1.00 &  \\
'25 & 0.73 & 0.70 & 0.70 & 0.71 & 0.68 & 0.51 & 0.58 & 0.71 & 0.81 & 0.83 & 1.00 \\
\bottomrule
\end{tabular}
\begin{tablenotes}[para,flushleft]\small
\item \textit{Notes:} Each cell is the Pearson correlation of course susceptibility between the two indicated calendar years, across courses observed in both years (pairwise complete, minimum 30 shared courses per cell). A course-year value averages susceptibility over the enrolled students in that year, so offerings are weighted by enrollment. The sample is the balanced panel (Appendix~\ref{app:descriptives}). The mean adjacent-year correlation is 0.76, and the mean correlation of each other year with 2019 is 0.73.
\end{tablenotes}
\end{threeparttable}
\end{table}

%% file: tables_and_figures_v2/table_archive_coverage.tex
\begin{table}[htbp]
\centering
\begin{threeparttable}
\caption{Coverage of the Syllabus Archive and the Analytic Sample}
\label{tab:archive-coverage}
\small
\begin{tabular}{lrrr}
\toprule
 & & Share of & Share of \\
Conditioning step & Courses & courses & enrollments \\
\midrule
Full University catalog & 15,624 & 1.000 & 1.000 \\
Subject represented in archive & 8,931 & 0.572 & 0.589 \\
Courses with archived syllabus & 4,906 & 0.314 & 0.450 \\
Has 2019 syllabus (susceptibility measured) & 1,768 & 0.113 & 0.317 \\
Final analytic sample (observed before and after fall 2022) & 1,449 & 0.093 & 0.308 \\
\bottomrule
\end{tabular}
\begin{tablenotes}[para,flushleft]\small
\item \textit{Notes:} The starting point is all graded undergraduate and graduate offerings in the administrative records from fall 2016 through fall 2025 (15,624 unique courses across 322 subjects). Archive rows count a syllabus from any year of the archive, not only 2015--2025. Each row applies one further sample restriction, and shares are relative to the full catalog. The largest source of attrition is the 183 subjects not represented in the syllabus archive, which account for 41.1\% of all enrollments.
\end{tablenotes}
\end{threeparttable}
\end{table}

%% file: tables_and_figures_v2/table_archive_balance.tex
\begin{table}[htbp]
\centering
\begin{threeparttable}
\caption{Course Characteristics and Outcomes: Analytic Sample, Rest of Archive, and Never-Archived Courses}
\label{tab:archive-balance}
\begin{tabular}{lccc}
\toprule
 & Never & Archived, not & Analytic \\
 & archived & in analytic sample & sample \\
\midrule
\multicolumn{4}{l}{\textit{Scope}} \\
  Unique courses & 10,718 & 3,457 & 1,449 \\
  Share of courses & 0.686 & 0.221 & 0.093 \\
  Share of enrollments & 0.550 & 0.142 & 0.308 \\
  Subjects represented & 320 & 137 & 89 \\
\midrule
\multicolumn{4}{l}{\textit{Course Characteristics}} \\
  Mean section size & 18.6 & 18.6 & 33.4 \\
  Undergraduate share & 0.421 & 0.677 & 0.794 \\
  Lecture share & 0.385 & 0.469 & 0.532 \\
  Graded-basis share & 0.737 & 0.823 & 0.838 \\
  Online share & 0.028 & 0.005 & 0.005 \\
  Mean credit units & 2.90 & 3.03 & 3.25 \\
\midrule
\multicolumn{4}{l}{\textit{Student Outcomes, Fall 2016 to Fall 2021}} \\
  Average grade & 3.697 & 3.688 & 3.627 \\
  Withdrawal rate & 0.013 & 0.020 & 0.022 \\
  Failure rate & 0.026 & 0.014 & 0.010 \\
\bottomrule
\end{tabular}
\begin{tablenotes}[para,flushleft]\small
\item \textit{Notes:} The columns divide the 15,624 unique courses in the University catalog from fall 2016 through fall 2025 into courses never archived (no syllabus in any year of the archive), courses archived but outside the analytic sample, and analytic-sample courses. Characteristics and outcomes are course-level means. 
\end{tablenotes}
\end{threeparttable}
\end{table}

%% file: tables_and_figures_v2/table_treatment_robustness.tex
\begin{landscape}
\begin{table}
\centering
\begin{threeparttable}
\caption{Average Grade Estimates under Alternative Measures of GenAI Susceptibility}
\label{tab:treatment_robustness}
\small
\setlength{\tabcolsep}{4pt}
\begin{tabular}{llllll}
& 2019 & Pre-COVID & Pre-GenAI & 2022 & COVID period \\
& (1) & (2) & (3) & (4) & (5) \\
Susceptibility $\times$ Post-AI & \num{0.010} & \num{0.014} & \num{0.047} & \num{0.078} & \num{0.115}** \\
& (\num{0.045}) & (\num{0.045}) & (\num{0.048}) & (\num{0.048}) & (\num{0.049}) \\
Susceptibility $\times$ COVID Year & \num{-0.060}* & \num{-0.041} & \num{-0.011} & \num{0.019} & \num{0.055} \\
& (\num{0.036}) & (\num{0.040}) & (\num{0.043}) & (\num{0.043}) & (\num{0.051}) \\
N & \num{1162801} & \num{1341383} & \num{1459896} & \num{1181433} & \num{1344274} \\
\midrule
GenAI net COVID & 0.071*** & 0.055** & 0.058** & 0.060** & 0.060** \\
& (0.024) & (0.022) & (0.024) & (0.025) & (0.025) \\
Student FE & $\checkmark$ & $\checkmark$ & $\checkmark$ & $\checkmark$ & $\checkmark$ \\
Course FE & $\checkmark$ & $\checkmark$ & $\checkmark$ & $\checkmark$ & $\checkmark$ \\
Semester FE & $\checkmark$ & $\checkmark$ & $\checkmark$ & $\checkmark$ & $\checkmark$ \\
\end{tabular}
\begin{tablenotes}[para,flushleft]
\small
\item \textit{Notes:} Standard errors clustered by course in parentheses. * $p<0.1$, ** $p<0.05$, *** $p<0.01$. Each column re-estimates the main specification (Equation~\ref{eq:main}) for average grade with a different measure of course GenAI susceptibility: 2019 offerings (col.\ 1, the main measure), and averages over all offerings in the pre-COVID period 2015--2019 (col.\ 2), the full pre-GenAI period 2015--2022 (col.\ 3), 2022 offerings (col.\ 4), and the COVID-affected period 2020--2022 (col.\ 5). Each column uses its own balanced sample (courses with the measure and with observations both before and after 2022), so $N$ differs across columns. The Susceptibility $\times$ Post-AI coefficient $\hat\beta_1$ is the transient-COVID estimate. The ``GenAI net COVID'' row reports the persistent-COVID estimate $\hat\beta_1 - \hat\beta_2$ with a delta-method standard error.
\end{tablenotes}
\end{threeparttable}
\end{table}
\end{landscape}

%% file: tables_and_figures_v2/table_ability_robustness.tex
\begin{table}
\centering
\begin{threeparttable}
\caption{Average Grade Estimates by Tercile under Alternative Measures of Prior Academic Preparation}
\label{tab:ability_robustness}
\small
\setlength{\tabcolsep}{4pt}
\begin{tabular}{lllllll}
& AbilityRank & Raw rank & Drop Fall 2020 & P/F-adjusted & HS GPA & SAT/ACT \\
\midrule
\multicolumn{7}{l}{\textit{Susceptibility $\times$ Post-AI (transient-COVID estimate)}} \\
Bottom tercile & -0.062 & -0.061 & -0.038 & -0.050 & 0.016 & -0.061 \\
 & (0.062) & (0.061) & (0.066) & (0.062) & (0.056) & (0.060) \\
Middle tercile & -0.020 & -0.011 & -0.007 & -0.012 & 0.036 & -0.070 \\
 & (0.052) & (0.053) & (0.056) & (0.054) & (0.060) & (0.054) \\
Top tercile & 0.032 & 0.070 & 0.069 & 0.065 & 0.048 & -0.037 \\
 & (0.048) & (0.050) & (0.053) & (0.050) & (0.056) & (0.043) \\
\midrule
\multicolumn{7}{l}{\textit{GenAI net COVID (persistent-COVID estimate)}} \\
Bottom tercile & 0.053 & 0.055 & 0.081* & 0.049 & 0.098** & 0.046 \\
 & (0.046) & (0.044) & (0.047) & (0.044) & (0.041) & (0.045) \\
Middle tercile & 0.047 & 0.049 & 0.060 & 0.046 & 0.079 & 0.025 \\
 & (0.039) & (0.042) & (0.046) & (0.042) & (0.048) & (0.043) \\
Top tercile & 0.046 & 0.057 & 0.079* & 0.070* & 0.070* & 0.024 \\
 & (0.036) & (0.036) & (0.041) & (0.036) & (0.039) & (0.032) \\
\end{tabular}
\begin{tablenotes}[para,flushleft]
\small
\item \textit{Notes:} Standard errors clustered by course in parentheses. * $p<0.1$, ** $p<0.05$, *** $p<0.01$. Each cell estimates the main specification (Equation~\ref{eq:main}) for average grade within the indicated tercile of the column's measure, with student, course, and semester fixed effects. The top panel reports the transient-COVID estimate $\hat\beta_1$; the bottom panel reports the persistent-COVID estimate $\hat\beta_1 - \hat\beta_2$ with a delta-method standard error. Columns: the main course-residualized within-cohort first-term GPA rank (\texttt{AbilityRank}); the raw (non-residualized) within-cohort rank; the raw rank excluding the Fall 2020 entering cohort; the raw rank with first-term GPA recomputed from instructor letter grades for pandemic pass/no-record elections; within-cohort high school GPA (first-year entrants only); and within-cohort SAT/ACT scores converted to a common scale. Samples differ across columns.
\end{tablenotes}
\end{threeparttable}
\end{table}

%% file: tables_and_figures_v2/table_grade_distribution.tex
\begin{landscape}
\begin{table}
\centering
\begin{threeparttable}
\caption{Difference-in-Differences Estimates by Letter-Grade Threshold}
\label{tab:grade_distribution}
\small
\setlength{\tabcolsep}{4pt}
\begin{tabular}{llllll}
& $\geq$ A & $\geq$ B & $\geq$ C & $\geq$ D & Not $\geq$ D \\
Susceptibility $\times$ Post-AI & \num{0.107}** & \num{-0.020} & \num{-0.003} & \num{0.006} & \num{-0.006} \\
& (\num{0.046}) & (\num{0.021}) & (\num{0.011}) & (\num{0.006}) & (\num{0.006}) \\
Susceptibility $\times$ COVID Year & \num{0.078}** & \num{-0.050}*** & \num{-0.010} & \num{0.010} & \num{-0.010} \\
& (\num{0.036}) & (\num{0.018}) & (\num{0.011}) & (\num{0.009}) & (\num{0.009}) \\
N & \num{1208492} & \num{1208492} & \num{1208492} & \num{1208492} & \num{1208492} \\
\midrule
GenAI net COVID & 0.030 & 0.030*** & 0.007 & -0.004 & 0.004 \\
& (0.018) & (0.011) & (0.006) & (0.007) & (0.007) \\
Student FE & $\checkmark$ & $\checkmark$ & $\checkmark$ & $\checkmark$ & $\checkmark$ \\
Course FE & $\checkmark$ & $\checkmark$ & $\checkmark$ & $\checkmark$ & $\checkmark$ \\
Semester FE & $\checkmark$ & $\checkmark$ & $\checkmark$ & $\checkmark$ & $\checkmark$ \\
\end{tabular}
\begin{tablenotes}[para,flushleft]
\small
\item \textit{Notes:} Standard errors clustered by course in parentheses. * $p<0.1$, ** $p<0.05$, *** $p<0.01$. Each column is a linear probability model, estimated with the main specification (Equation~\ref{eq:main}), for clearing the indicated letter-grade threshold; the final column is the complement of the fourth, not earning at least a D. As in \citet{chirikovAIGradeInflation2026}, every enrollment enters the denominator, so withdrawals, incompletes, and non-letter grades count as not clearing any threshold. The Susceptibility $\times$ Post-AI coefficient $\hat\beta_1$ is the transient-COVID estimate. The ``GenAI net COVID'' row reports the persistent-COVID estimate $\hat\beta_1 - \hat\beta_2$ with a delta-method standard error.
\end{tablenotes}
\end{threeparttable}
\end{table}
\end{landscape}

%% file: tables_and_figures_v2/table_het_ability_terciles.tex
\begin{landscape}
\begin{table}
\centering
\begin{threeparttable}
\caption{Average Grade Estimates by \texttt{AbilityRank} Tercile}
\label{tab:het_ability_terciles}
\small
\setlength{\tabcolsep}{4pt}
\begin{tabular}{llll}
& Bottom & Middle & Top \\
Susceptibility $\times$ Post-AI & \num{-0.062} & \num{-0.020} & \num{0.032} \\
& (\num{0.062}) & (\num{0.052}) & (\num{0.048}) \\
Susceptibility $\times$ COVID Year & \num{-0.115}** & \num{-0.067}* & \num{-0.014} \\
& (\num{0.048}) & (\num{0.035}) & (\num{0.030}) \\
N & \num{266174} & \num{248771} & \num{238140} \\
\midrule
GenAI net COVID & 0.053 & 0.047 & 0.046 \\
& (0.046) & (0.039) & (0.036) \\
Student FE & $\checkmark$ & $\checkmark$ & $\checkmark$ \\
Course FE & $\checkmark$ & $\checkmark$ & $\checkmark$ \\
Semester FE & $\checkmark$ & $\checkmark$ & $\checkmark$ \\
\end{tabular}
\begin{tablenotes}[para,flushleft]
\small
\item \textit{Notes:} Standard errors clustered by course in parentheses. * $p<0.1$, ** $p<0.05$, *** $p<0.01$. Each column estimates the main specification (Equation~\ref{eq:main}) for average grade within one tercile of \texttt{AbilityRank}, the course-residualized within-cohort first-term GPA rank (Section~\ref{sec:measure-abilityrank}). The sample is restricted to undergraduate, fall-entry students from cohorts that entered before ChatGPT's release, for whom the proxy is measured pre-treatment. The Susceptibility $\times$ Post-AI coefficient $\hat\beta_1$ is the transient-COVID estimate. The ``GenAI net COVID'' row reports the persistent-COVID estimate $\hat\beta_1 - \hat\beta_2$ with a delta-method standard error.
\end{tablenotes}
\end{threeparttable}
\end{table}
\end{landscape}

%% file: tables_and_figures_v2/table_course_evals.tex
\begin{landscape}
\begin{table}
\centering
\begin{threeparttable}
\caption{Difference-in-Differences Estimates for Course Evaluation Outcomes}
\label{tab:course_evals}
\small
\setlength{\tabcolsep}{4pt}
\begin{tabular}{llllllllll}
& understand & understand & understand & interest & interest & interest & workload & workload & workload \\
Susceptibility $\times$ Post-AI & \num{0.020} & \num{0.031} & \num{0.069}* & \num{0.044} & \num{0.056} & \num{0.147}** & \num{-0.213}** & \num{-0.158}* & \num{-0.270}** \\
& (\num{0.033}) & (\num{0.030}) & (\num{0.037}) & (\num{0.049}) & (\num{0.049}) & (\num{0.062}) & (\num{0.092}) & (\num{0.084}) & (\num{0.120}) \\
Susceptibility $\times$ COVID Year &  &  & \num{0.079} &  &  & \num{0.191}*** &  &  & \num{-0.236}** \\
&  &  & (\num{0.052}) &  &  & (\num{0.057}) &  &  & (\num{0.105}) \\
Susceptibility & \num{-0.052} &  &  & \num{0.098} &  &  & \num{-0.212}* &  &  \\
& (\num{0.059}) &  &  & (\num{0.111}) &  &  & (\num{0.116}) &  &  \\
Post-AI & \num{-0.005} &  &  & \num{0.004} &  &  & \num{0.103} &  &  \\
& (\num{0.020}) &  &  & (\num{0.035}) &  &  & (\num{0.066}) &  &  \\
N & \num{938602} & \num{938601} & \num{938601} & \num{938593} & \num{938592} & \num{938592} & \num{938593} & \num{938592} & \num{938592} \\
\midrule
GenAI net COVID &  &  & -0.010 &  &  & -0.044 &  &  & -0.034 \\
&  &  & (0.042) &  &  & (0.051) &  &  & (0.068) \\
Course FE &  & $\checkmark$ & $\checkmark$ &  & $\checkmark$ & $\checkmark$ &  & $\checkmark$ & $\checkmark$ \\
Semester FE &  & $\checkmark$ & $\checkmark$ &  & $\checkmark$ & $\checkmark$ &  & $\checkmark$ & $\checkmark$ \\
\end{tabular}
\begin{tablenotes}[para,flushleft]
\small
\item \textit{Notes:} Standard errors clustered by course in parentheses. * $p<0.1$, ** $p<0.05$, *** $p<0.01$. Outcomes are offering-level median ratings on five-point scales; workload is reverse-coded so that higher values indicate heavier workload. Regressions are estimated on student-offering rows, so each offering is weighted by its enrollment. For each outcome, the first column is OLS without fixed effects, the second adds course and semester fixed effects, and the third is the main specification (Equation~\ref{eq:main}), which adds the Susceptibility $\times$ COVID Year interaction. In the third column, the Susceptibility $\times$ Post-AI coefficient $\hat\beta_1$ is the transient-COVID estimate, measured against the pre-COVID baseline. The ``GenAI net COVID'' row reports the persistent-COVID estimate $\hat\beta_1 - \hat\beta_2$ with a delta-method standard error, which assumes that the COVID-era differential persisted into the post-AI period.
\end{tablenotes}
\end{threeparttable}
\end{table}
\end{landscape}

%% file: tables_and_figures_v2/table_honestdid_grade.tex
\begin{table}[htbp]
\centering
\begin{threeparttable}
\caption{Robust Confidence Sets for the Average Post-ChatGPT Grade Effect}
\label{tab:honestdid-grade}
\begin{tabular}{lc}
\toprule
Assumed violation of parallel trends & Robust confidence set \\
\midrule
None (conventional interval) & $[-0.080,\ 0.088]$ \\
Linear violation of any slope ($M = 0$) & $[-0.118,\ 0.057]$ \\
Violation up to the largest pre-COVID deviation ($\bar{M} = 1$) & $[-0.528,\ 0.580]$ \\
\bottomrule
\end{tabular}
\begin{tablenotes}[para,flushleft]\small
\item \textit{Notes:} Confidence sets for the average post-ChatGPT effect on final grades, in grade points per unit of GenAI susceptibility, from the semester event study, following \citet{rambachanMoreCredibleApproach2023}. The first row assumes no parallel trends violation. The second assumes a linear violation of arbitrary slope. The third assumes a violation as large as observed in the pre-COVID period. Violations are bounded using pre-COVID terms only.
\end{tablenotes}
\end{threeparttable}
\end{table}

%% file: tables_and_figures_v2/table_prior_comparison.tex
\begin{table}[htbp]
\centering
\begin{threeparttable}
\caption{Grade Effects in This Paper and Prior Studies on a Common Scale}
\label{tab:prior_comparison}
\footnotesize
\setlength{\tabcolsep}{2pt}
\begin{tabular}{lcccc}
\toprule
& Estimate & 95\% CI & \shortstack{Upper 90\% bound\\(grade steps)} & \shortstack{Difference\\$p$} \\
\midrule
\multicolumn{5}{l}{\textit{Panel A. Grade points, per unit of exposure share}} \\
\hspace{0.5em}Chirikov (2026) & 0.120 (0.050) & [0.022, 0.218] &  &  \\
\hspace{0.5em}This paper, transient COVID & 0.010 (0.045) & [$-$0.078, 0.098] & 0.084 (0.28) & 0.10 \\
\hspace{0.5em}This paper, persistent COVID & 0.071 (0.024) & [0.025, 0.117] & 0.109 (0.36) & 0.37 \\
\addlinespace
\multicolumn{5}{l}{\textit{Panel B. Share earning an A, per unit of exposure share}} \\
\hspace{0.5em}Chirikov (2026) & 0.130 (0.030) & [0.071, 0.189] &  &  \\
\hspace{0.5em}This paper, transient COVID & 0.107 (0.046) & [0.017, 0.198] & 0.183 & 0.68 \\
\hspace{0.5em}This paper, persistent COVID & 0.030 (0.018) & [$-$0.006, 0.065] & 0.060 & $<$0.01 \\
\addlinespace
\multicolumn{5}{l}{\textit{Panel C. Grade points, high versus low exposure}} \\
\hspace{0.5em}Hausman et al. (2026), 0.04 per point & 0.039 (0.012) & [0.016, 0.061] &  &  \\
\hspace{0.5em}Hausman et al. (2026), 0.10 per point & 0.097 (0.029) & [0.040, 0.154] &  &  \\
\hspace{0.5em}Chirikov (2026), binary exposure & 0.030 (0.020) & [$-$0.009, 0.069] &  &  \\
\hspace{0.5em}This paper, transient COVID & $-$0.016 (0.030) & [$-$0.074, 0.042] & 0.033 (0.11) & 0.08, $<$0.01, 0.20 \\
\hspace{0.5em}This paper, persistent COVID & 0.049 (0.016) & [0.018, 0.080] & 0.075 (0.25) & 0.60, 0.15, 0.45 \\
\bottomrule
\end{tabular}
\begin{tablenotes}[para,flushleft]
\small
\item \textit{Notes:} Estimates with standard errors in parentheses. Panels A and B report coefficients on Post-AI $\times$ Exposure, where exposure is a 0 to 1 share: the share of the final grade from unproctored assessment in this paper, and the share of course tasks that involve writing or coding in \citet{chirikovAIGradeInflation2026}. Each coefficient is the change for a course moving from 0 to 1 on its own share. Panel B compares \citeauthor{chirikovAIGradeInflation2026}'s share of A grades (his university awards no A+) with our probability of earning at least an A. Panel C compares binary contrasts: \citet{hausmanGenerativeAIsImpact}'s AI-compatible versus AI-incompatible courses (their first post-ChatGPT year, 0.970 points on a 0 to 100 scale, SE 0.289, converted at 0.04 or 0.10 grade points per point), \citeauthor{chirikovAIGradeInflation2026}'s courses with any writing or coding task versus none (his Table A4), and our top versus bottom susceptibility tercile (mean susceptibility 0.89 versus 0.22). ``Transient COVID'' is the post-AI coefficient relative to the pre-COVID baseline; ``persistent COVID'' subtracts the COVID-period coefficient (the GenAI net COVID row of the main tables). The upper 90\% bound is the smallest effect our estimate rules out with a one-sided 5\% test; in parentheses, that bound divided by 0.3 grade points, the gap between adjacent letter grades such as B and B+. Difference $p$ is a two-sided test of equality with each prior estimate in the same panel, treating the studies as independent; in Panel C the three values compare against Hausman et al.\ at 0.04 and 0.10 grade points per point and against \citeauthor{chirikovAIGradeInflation2026}'s binary estimate, in that order. See Appendix~\ref{app:prior-comparison}.
\end{tablenotes}
\end{threeparttable}
\end{table}